\documentclass[seceq]{ptptex}

\usepackage{graphicx,bm}
\usepackage{wrapft}

\title{
Introduction to Nonequilibrium Statistical Mechanics \\
with Quantum Field Theory%
}

\author{
Takafumi \textsc{Kita}%
}

\inst{
Department of Physics, Hokkaido University, Sapporo 060-0810, Japan
}

\recdate{December 30, 2009; revised February 27, 2010}
\notypesetlogo

\abst{
In this article, we present a concise and self-contained introduction to nonequilibrium statistical mechanics with quantum field theory by considering an ensemble of interacting identical bosons or fermions as an example. Readers are assumed to be familiar with the Matsubara formalism of equilibrium statistical mechanics such as Feynman diagrams, the proper self-energy, and Dyson's equation. The aims are threefold: (i) to explain the fundamentals of nonequilibrium quantum field theory as simple as possible on the basis of the knowledge of the equilibrium counterpart; (ii) to elucidate the hierarchy in describing nonequilibrium systems from Dyson's equation on the Keldysh contour to the Navier-Stokes equation in fluid mechanics via quantum transport equations and the Boltzmann equation; (iii) to derive an expression of nonequilibrium entropy that evolves with time. In stage (i), we introduce nonequilibrium Green's function and the self-energy uniquely on the round-trip Keldysh contour, thereby avoiding possible confusions that may arise from defining multiple Green's functions at the very beginning. We try to present the Feynman rules for the perturbation expansion as simple as possible. In particular, we focus on the self-consistent perturbation expansion with the Luttinger-Ward thermodynamic functional, i.e., Baym's $\Phi$-derivable approximation, which has a crucial property for nonequilibrium systems of obeying various conservation laws automatically. We also show how the two-particle correlations can be calculated within the $\Phi$-derivable approximation, i.e., an issue of how to handle the ``Bogoliubov-Born-Green-Kirkwood-Yvons (BBGKY) hierarchy''. Aim (ii) is performed through successive reductions of relevant variables with the Wigner transformation, the gradient expansion based on the Groenewold-Moyal product, and Enskog's expansion from local equilibrium. This part may be helpful for convincing readers that nonequilibrium systems can be handled microscopically with quantum field theory, including fluctuations. We also discuss a derivation of the quantum transport equations for electrons in electromagnetic fields based on the gauge-invariant Wigner transformation so that the Lorentz force is reproduced naturally.
As for (iii), the Gibbs entropy of equilibrium statistical mechanics suffers from the flaw that it does not evolve in time. We show here that a microscopic expression of nonequilibrium dynamical entropy can be derived from the quantum transport equations so as to be compatible with the law of increase in entropy as well as equilibrium statistical mechanics.
}

\PTPindex{052, 056, 062, 356, 512}  

\begin{document}

\maketitle

\tableofcontents

\section{Introduction}

Nonequilibrium quantum field theory on the real-time Keldysh contour\cite{Keldysh64} has been used extensively in recent years
to describe a wide range of dynamical phenomena in condensed matter physics, nuclear matter, and high-energy physics.\cite {Bonitz00,BS03,BF06} 
The method enables us to handle a wide range of many-body systems with quantum effects and/or strong correlations microscopically from first principles. There are already excellent review articles\cite{Langreth76,Danielewicz84,CSHY85,RS86,Berges04} and textbooks\cite{HJ98,Rammer07} on the topic. The approach is also explained briefly in some standard textbooks on statistical many-body theory.\cite{LP,Mahan}
Thus, adding another review here on the topic may require some justifications for doing so.
The distinct features of this article may be summarized as follows.

(i) Considering specifically a system of identical bosons or fermions with a two-body interaction, we try to explain the basics of nonequilibrium quantum field theory concisely and clearly on the basis of the knowledge of the equilibrium counterpart. For example, Green's function and the self-energy are introduced uniquely on the round-trip Keldysh contour in exactly the same way as those on the imaginary-time Matsubara contour. We thereby avoid defining multiple real-time Green's functions at the very beginning so as not to cause confusion. We also try to present Feynman rules for the perturbation expansion as simple as possible. One of the extra factors here is how to carry out the summations over the forward and backward paths of the Keldysh contour.  Several variant rules have been proposed for it such as the one by Keldysh\cite{Keldysh64} and another by Langreth (i.e., the Langreth theorem).\cite{Langreth76}  We present here an alternative rule for the summation, which we think is the simplest possible. It is worth pointing out that the consideration here can be extended easily to other systems, e.g., electrons in solids with a periodic potential, impurities, phonons, etc. Indeed, we only need to modify the basic Hamiltonian to incorporate those effects. This may be realized by looking at the variety of applications of the method ranging from condensed matter physics to high-energy physics.\cite{Bonitz00,BS03,BF06}

(ii) We specifically focus on the self-consistent perturbation expansion with the Luttinger-Ward thermodynamic functional,\cite{LW60} which reproduces the Hartree-Fock theory as the lowest-order approximation.\cite{Luttinger60} Its essence lies in determining Green's function $G$ and the self-energy $\Sigma$ self-consistently using Dyson's equation $(G_{0}^{-1}-\Sigma)G=1$ and the relation $\Sigma\propto \delta\Phi/\delta G$ based on a given functional $\Phi=\Phi[G]$. As shown by Baym,\cite{Baym62} the approximation scheme has a unique property of obeying various conservation laws automatically, i.e., a property indispensable for describing nonequilibrium systems but not satisfied by the simple perturbation expansion. Despite this crucial feature, the self-consistent perturbation expansion has not been paid sufficient attention in those review articles and textbooks. We give it a full account together with a complete proof of the conservation laws.  We also derive the Bethe-Salpeter equation for two-particle correlation functions in the $\Phi$-derivable approximation; thus, choosing a definite $\Phi=\Phi[G]$ will be shown to amount to determining the whole Bogoliubov-Born-Green-Kirkwood-Yvons (BBGKY) hierarchy.\cite{Cercignani88}

(iii) We try to elucidate the hierarchy in describing nonequilibrium systems from the microscopic Dyson's equation on the Keldysh contour to the macroscopic Navier-Stokes equation; the latter is shown to be reached from the former by successive reductions of relevant variables through quantum transport equations and the Boltzmann equation. The Navier-Stokes equation is an archetype of nonlinear evolution equations on which a wide variety of nonequilibrium phenomena (e.g., turbulence, chaos, and pattern formation) have been discussed extensively.\cite{CH93,CG09} A manifest derivation of it from Dyson's equation may convince readers that those phenomena can be handled microscopically from first principles and enable them to incorporate effects beyond the phenomenological approach such as fluctuations.

(iv) We derive a microscopic expression of nonequilibrium entropy, which evolves with time. Entropy $S$ is the central quantity in thermodynamics as embodied in the Clausius inequality $dS\geq TdQ$, where $T$ and $Q$ denote temperature and heat, respectively. It tells us that entropy for any isolated system should increase monotonically with time. On the other hand, it was shown by Jaynes\cite{Jaynes57} that every equilibrium statistical ensemble can be identified as the maximum of the Gibbs (or information) entropy 
\begin{equation}
S = -k_{B}\sum_{\nu}p_{\nu}\ln p_{\nu} ,\hspace{10mm}\left\{\begin{array}{ll} k_{\rm B} & : \mbox{Boltzmann constant}\\
 p_{\nu} & : \mbox{probability of the state $\nu$}
\end{array}\right. ,
\label{S-Gibbs}
\end{equation} 
under some constraints. However, the Gibbs entropy
suffers from the lack of dynamics, i.e., it cannot describe the law of increase in entropy.\cite{LL80} Except for the Boltzmann entropy for dilute classical gases\cite{Boltzmann72,Cercignani88} and its extensions to classical denser systems,\cite{Resibois78,Lebowitz04} we do not have a widely accepted statistical-mechanical expression for nonequilibrium entropy to represent the Clausius inequality microscopically. Following Ivanov et al.,\cite{Ivanov00} we show here that such an expression can be derived from the quantum transport equations mentioned above so as to be compatible with the law of increase in entropy and also to embrace the Boltzmann entropy as a limit.

(v) We extend the derivation of quantum transport equations to electrons in electromagnetic fields, where special care is necessary for the gauge invariance of the equation to reproduce the Lorentz force adequately. 

(vi) To supplement contents above, we include in Appendix A a full account of the second quantization method as an equivalent alternative to the description with many-body wave functions in the configuration space. We also describe the equilibrium self-consistent perturbation expansion with $\Phi$ in Appendix B, the Luttinger-Ward thermodynamic functional in Appendix C, and a derivation of an expression of equilibrium entropy in Appendix D.

Now, we briefly summarize the relevant history of quantum field theory in statistical mechanics. The application of the method to equilibrium many-body phenomena was pioneered by Matsubara in 1955 based on the simple perturbation expansion with the imaginary-time Matsubara Green's function.\cite{Matsubara55} The finite-temperature version of Wick's theorem,\cite{Wick50} which forms the basis for the perturbation expansion with the Matsubara Green's function, was first proved by Bloch and de Dominicis\cite{BdD59} in 1959 and refined by Gaudin\cite{Gaudin60} shortly. In 1960, Luttinger and Ward\cite{LW60} provided a formally exact expression of the thermodynamic potential as a functional of the fully renormalized Green's function $G$, which contains the functional $\Phi[G]$ mentioned above in the form of the skeleton diagram expansion. It was subsequently used by Luttinger\cite{Luttinger60} to clarify some general properties of interacting normal fermions, e.g., the Fermi-surface sum rule anticipated by Landau in his Fermi-liquid theory.\cite{Landau56} Luttinger\cite{Luttinger60} also noted that the skeleton diagram expansion forms a basis for a self-consistent perturbation expansion where the Hartree-Fock theory is reproduced as the lowest-order approximation. Shortly, equilibrium quantum field theory was made quite popular among theorists through the well-known textbook by Abrikosov, Gor'kov, and Dzyaloshinski,\cite{AGD63} which even contains its applications to superconductivity and Bose-Einstein condensation as well as a microscopic justification of the Landau Fermi-liquid theory. Meanwhile, the method had been used fairly extensively  for describing nonequilibrium transport phenomena by tracing the time evolution of the density matrix.\cite{Bogoliubov62} A crucial step was made along this line by Kadanoff and Baym in 1962,\cite{KB62} who pioneered the use of Green's function for this purpose. However, their treatment was still based on the analytic continuation from the imaginary-time Matsubara Green's function. Baym\cite{Baym62} successively clarified a sufficient condition for various conservation laws to be satisfied automatically in terms of $\Phi[G]$ and called it ``$\Phi$-derivable approximation''. The approximation scheme is essentially identical in content with the self-consistent approximation of Luttinger but superior to the latter in its explicit usage of $\Phi$.\cite{Luttinger60} On the other hand, Schwinger\cite{Schwinger61} introduced Green's functions on a closed time-path contour in 1961 to calculate time-dependent expectation values of physical quantities for a Brownian motion. In 1964, Keldysh\cite{Keldysh64,comment1} clearly recognized that quantum field theory with Green's function can treat general nonequilibrium phenomena in the same way as equilibrium phenomena to develop a perturbation expansion with respect to the interaction on the real-time round-trip contour of $-\infty\leq t\leq \infty$. The theory made an epoch in nonequilibrium statistical mechanics with quantum field theory.
However, its great potential for describing nonequilibrium phenomena seems to have been appreciated rather slowly.
Some of its earliest applications include: a calculation of the tunneling current through a metal-insulator-metal junction by Caroli et al.;\cite{CCNS71} a derivation of a transport equation for a superconductor by Aronov and Grevich;\cite{AG75} a derivation of nonequilibrium quasiclassical equations of superconductivity by Larkin and Ovchinnikov.\cite{LO75}
Later developments may be found in Refs.\ \citen{Bonitz00,BS03,BF06,Langreth76,Danielewicz84,CSHY85,RS86,Berges04,HJ98,Rammer07}.

\section{Representations and expectation value\label{sec:rep}}

We first introduce the Schr\"odinger, Heisenberg, and interaction representations for a time-dependent Hamiltonian. We then derive an expression for expectation values of physical quantities in terms of the latter two representations.

\subsection{Schr\"odinger and Heisenberg representations}

The Schr\"odinger and Heisenberg representations are usually introduced in terms of a time-independent Hamiltonian. Since we treat nonequilibrium dynamical phenomena, we start by defining those representations explicitly for a time-dependent Hamiltonian.

We consider a many-body system described with a time-dependent Hamiltonian $\hat{\cal H}(t)$. By adopting the second quantization metnod, the Schr\"odinger equation reads
\begin{equation}
i\hbar \frac{d|\Psi_{\nu} (t)\rangle }{dt}  = \hat{\cal H}(t)|\Psi_{\nu} (t)\rangle ,
\label{Schrodinger}
\end{equation}
where $|\Psi_{\nu}(t)\rangle$ denotes a many-body wave function specified by quantum number $\nu$ with $\langle \Psi_{\nu} (t)|\Psi_{\nu} (t)\rangle=1$.
See Appendix A for the notations of second quantization as well as for a proof that the method is completely equivalent to the description with many-body wave functions in the configuration space.
Equation (\ref{Schrodinger}) has the formal solution:
\begin{equation}
|\Psi_{\nu} (t)\rangle = \hat{\cal U}(t,t_{0})|\Psi_{\nu}(t_{0}) \rangle ,
\label{Psi(t)}
\end{equation}
where $t_{0}$ is some initial time and $\hat{\cal U}(t,t_{0})$ is defined by
\begin{subequations}
\label{U-def}
\begin{equation}
\hat{\cal U}(t,t_{0})
\equiv 1+\sum_{n=1}^{\infty}\left(-\frac{i}{\hbar}\right)^{\!\! n} \int_{t_{0}}^{t}dt_{n}
\cdots\int_{t_{0}}^{t_{3}}dt_{2} \int_{t_{0}}^{t_{2}}dt_{1}\hat{\cal H}(t_{n})
\cdots \hat{\cal H}(t_{2}) \hat{\cal H}(t_{1}) .
\label{U-def1}
\end{equation}
It is seen easily that Eq.\ (\ref{Psi(t)}) with Eq.\ (\ref{U-def1}) satisfies Eq.\ (\ref{Schrodinger}).
Although $t\geq t_{0}$ may have been assumed implicitly, definition (\ref{U-def}) is also effective for $t<t_{0}$.
We then observe that, for $t>t_{0}$ ($t<t_{0}$), the Hamiltonians on the right-hand side are arranged in chronological order from right to left (left to right). Hence, it follows that the operator $\hat{\cal U}(t,t_{0})$ can also be put into an exponential form as
\begin{equation}
\hat{\cal U}(t,t_{0})=
\left\{\begin{array}{ll}
\vspace{1mm}
\displaystyle
{\cal T}\exp\left[-\frac{i}{\hbar}\int_{t_{0}}^{t}\hat{\cal H}(t_{1})dt_{1} \right] & : t\geq t_{0} 
\\
\displaystyle
\tilde {\cal T}\exp\left[-\frac{i}{\hbar}\int_{t_{0}}^{t}\hat{\cal H}(t_{1})dt_{1} \right] & : t< t_{0} 
\end{array}\right. ,
\label{U-def2}
\end{equation}
\end{subequations}
where ${\cal T}$ ($\tilde{\cal T}$) is the time-ordering (anti-time-ordering) operator 
to arrange the Hamiltonians into the chronological order from right to left (left to right)
with an extra sign change upon every transposition of fermion field operators;\cite{AGD63,FW71,HJ98} this sign change is irrelevant here as $\hat{\cal H}(t)$ is usually composed of an even number of field operators. See, e.g., p.\ 57 of Ref.\ \citen{FW71} for the equivalence between Eq.\ (\ref{U-def1}) and Eq.\ (\ref{U-def2}).

One can show easily that $\hat{\cal U}(t,t_{0})$ obeys the equation of motion:
\begin{subequations}
\label{U-eq-motion}
\begin{equation}
i\hbar \frac{\partial\hat{\cal U}(t,t_{0}) }{\partial t}  = \hat{\cal H}(t)\hat{\cal U}(t,t_{0}) .
\label{U-eq-motion1}
\end{equation}
Expressing the integrals in Eq.\ (\ref{U-def}) as $\int_{t_{0}}^{t}dt_{1}\int_{t_{1}}^{t}dt_{2}
\cdots \int_{t_{n-1}}^{t}dt_{n}$, we also obtain
\begin{equation}
i\hbar \frac{\partial\hat{\cal U}(t,t_{0}) }{\partial t_{0}}  = -\hat{\cal U}(t,t_{0}) \hat{\cal H}(t_{0}).
\label{U-eq-motion2}
\end{equation}
\end{subequations}
The operator has the properties:
\begin{subequations}
\label{U-prop}
\begin{equation}
\hat{\cal U}(t,t_{1})\hat{\cal U}(t_{1},t_{0})=\hat{\cal U}(t,t_{0}),
\label{U-prop1}
\end{equation}
\begin{equation}
\hat{\cal U}(t_{0},t_{0})=1.
\label{U-prop2}
\end{equation}
\end{subequations}
Equation (\ref{U-prop2}) is obvious from Eq.\ (\ref{U-def1}), whereas Eq.\ (\ref{U-prop1}) can be proved as follows. We first observe that $\hat{\cal V}(t,t_{0})\equiv
\hat{\cal U}(t,t_{1})\hat{\cal U}(t_{1},t_{0})$ satisfies the same first-order differential equation with respect to $t$ as $\hat{\cal U}(t,t_{0})$. We also notice with Eq.\ (\ref{U-prop2}) that the initial values at $t=t_{1}$ are the same between the two operators, i.e., $\hat{\cal V}(t,t_{0})=\hat{\cal U}(t,t_{0})$. Hence, we  arrive at Eq.\ (\ref{U-prop1}).

Setting $t=t_{0}$ in Eq.\ (\ref{U-prop1}) enables us to identify $\hat{\cal U}^{-1}(t,t_{0})=\hat{\cal U}(t_{0},t)$.
Moreover, taking the Hermitian conjugate of Eq.\ (\ref{U-eq-motion1}) and comparing the result with Eq.\  (\ref{U-eq-motion2}), 
we obtain $\hat{\cal U}^{\dagger}(t,t_{0})=\hat{\cal U}(t_{0},t)$. These results are summarized as
\begin{equation}
\hat{\cal U}^{-1}(t,t_{0})=\hat{\cal U}^{\dagger}(t,t_{0})=\hat{\cal U}(t_{0},t).
\label{U^-1}
\end{equation}

With $\hat{\cal U}$, the expectation value of an Hermitian operator $\hat{\cal O}(t)$ in terms of wave function (\ref{Psi(t)}) can be expressed alternatively as
\begin{equation}
\langle \Psi_{\nu}(t)|\hat{\cal O}(t)|\Psi_{\nu}(t)\rangle =
\langle \Psi_{\nu}(t_{0}) | \hat{\cal O}_{\rm H}(t) | \Psi_{\nu}(t_{0})\rangle,
\label{<O>}
\end{equation}
where $\hat{\cal O}_{\rm H}(t)$ denotes the Heisenberg representation:
\begin{equation}
\hat{\cal O}_{\rm H}(t)\equiv \hat{\cal U}^{\dagger}(t,t_{0}) \hat{\cal O}(t) \,\hat{\cal U}(t,t_{0}).
\label{O_H-def}
\end{equation}
Note that $\hat{\cal O}_{\rm H}(t)$ is Hermitian with $t_{0}$ dependence through $\hat{\cal U}$; the latter dependence is dropped here for convenience. 
The left (right) side of Eq.\ (\ref{<O>}) corresponds to the Schr\"odinger (Heisenberg) picture.
We realize with $\hat{\cal U}^{\dagger}(t,t_{0})=\hat{\cal U}(t_{0},t)$ that the evaluation of Eq.\ (\ref{<O>}) in the Heisenberg picture
proceeds along the closed time path of Fig.\ 1.

\begin{figure}[t]
\begin{center}
  \includegraphics[width=0.25\linewidth]{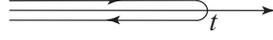}
\end{center}
\vspace{0mm}
\caption{Integration contour of Eq.\ (\protect\ref{<O>}). Both the forward and backward paths are actually on the real axis but shifted slightly upwards and downwards, respectively, to distinguish them clearly.\label{fig:0}}
\end{figure}

\subsection{Interaction representation\label{subsec:int-rep}}

We now consider those cases where $\hat{\cal H}(t)$
can be split into two parts as $\hat{\cal H}(t)=\hat{\cal H}_{0}(t)+\hat{\cal H}'(t)$.
Following Eq.\ (\ref{U-def}), we first define a time-evolution operator with $\hat{\cal H}_{0}(t)$ by
\begin{equation}
\hat{\cal U}_{0}(t,t_{0})
\equiv 1+\sum_{n=1}^{\infty}\left(-\frac{i}{\hbar}\right)^{\!\! n} \int_{t_{0}}^{t}dt_{n}
\cdots\int_{t_{0}}^{t_{3}}dt_{2} \int_{t_{0}}^{t_{2}}dt_{1}\hat{\cal H}_{0}(t_{n})
\cdots \hat{\cal H}_{0}(t_{2}) \hat{\cal H}_{0}(t_{1}) ,
\label{U_0-def}
\end{equation}
which clearly has the same properties in terms of $\hat{\cal H}_{0}$ as Eqs.\ (\ref{U-eq-motion})--(\ref{U^-1}) of $\hat{\cal U}$ with $\hat{\cal H}$.
We next introduce the so-called scattering matrix:
\begin{equation}
\hat{\cal S}(t,t_{0})\equiv \hat{\cal U}_{0}(t_{0},t)\hat{\cal U}(t,t_{0}).
\label{S-def}
\end{equation}
Its inverse and Hermitian conjugate are easily identified with Eq.\ (\ref{U^-1}) as
\begin{equation}
\hat{\cal S}^{-1}(t,t_{0})=\hat{\cal S}^{\dagger}(t,t_{0})=\hat{\cal U}(t_{0},t)\hat{\cal U}_{0}(t,t_{0}).
\label{S^-1}
\end{equation}
Note that $\hat{\cal S}^{-1}(t,t_{0})\neq \hat{\cal S}(t_{0},t)$ generally.
Using Eqs.\ (\ref{U-eq-motion}) and (\ref{U-prop}), one can show easily that 
$\hat{\cal S}(t,t_{0})$ satisfies
\begin{equation}
i\hbar\frac{\partial \hat{\cal S}(t,t_{0})}{\partial t} 
=\hat{\cal H}_{\rm I}'(t)\hat{\cal S}(t,t_{0}),
\label{S-eq-motion}
\end{equation}
where $\hat{\cal H}_{\rm I}'(t)$ denotes the interaction representation of $\hat{\cal H}'(t)$; the interaction representation is
defined for an arbitrary Hermitian operator $\hat{\cal O}(t)$ by
\begin{equation}
\hat{\cal O}_{\rm I}(t)\equiv \hat{\cal U}_{0}(t_{0},t)
\hat{\cal O}(t)\hat{\cal U}_{0}(t,t_{0}).
\label{O_I-def}
\end{equation}
Note that $\hat{\cal O}_{\rm I}(t)$ is also Hermitian with $t_{0}$ dependence through $\hat{\cal U}_{0}$; 
the latter dependence is also dropped here for convenience. 
Noting Eqs.\ (\ref{U-def1}) and (\ref{U-eq-motion1}), we can immediately write down the solution to Eq.\ (\ref{S-eq-motion}) as
\begin{equation}
\hat{\cal S}(t,t_{0})
= 1+\sum_{n=1}^{\infty}\left(-\frac{i}{\hbar}\right)^{\!\! n} \int_{t_{0}}^{t}dt_{n}
\cdots\int_{t_{0}}^{t_{3}}dt_{2} \int_{t_{0}}^{t_{2}}dt_{1}\hat{\cal H}_{\rm I}'(t_{n})
\cdots \hat{\cal H}_{\rm I}'(t_{2}) \hat{\cal H}_{\rm I}'(t_{1}) .
\label{S-int}
\end{equation}
It may also be put into an exponential form as Eq.\ (\ref{U-def2}).

Let us substitute $\hat{\cal U}(t,t_{0})=\hat{\cal U}_{0}(t,t_{0})\hat{\cal S}(t,t_{0})$ from Eq.\ (\ref{S-def}) into Eq.\ (\ref{<O>}). We thereby obtain an alternative expression for the expectation value in terms of Eqs.\ (\ref{O_I-def}) and (\ref{S-int}) as
\begin{equation}
\langle \Psi_{\nu}(t)|\hat{\cal O}(t)|\Psi_{\nu}(t)\rangle =
\langle \Psi_{\nu}(t_{0}) | \hat{\cal S}^{\dagger}(t,t_{0})\hat{\cal O}_{\rm I}(t) \hat{\cal S}(t,t_{0})| \Psi_{\nu}(t_{0})\rangle .
\label{<O_I>}
\end{equation}

\subsection{Keldysh contour}

From now on, we set $t_{0}\!=\!-\infty$. 
Using $\hat{\cal S}^{\dagger}(\infty,t)\hat{\cal S}(\infty,t)=1$ and $\hat{\cal S}(t_{2},t_{1})\hat{\cal S}(t_{1},t_{0})\!=\hat{\cal S}(t_{2},t_{0})$, we can transform Eq.\ (\ref{<O_I>}) as
\begin{eqnarray}
&&\hspace{-10mm}
\langle \Psi_{\nu}(t)|\hat{\cal O}|\Psi_{\nu}(t)\rangle =
\langle \Psi_{\nu}(-\infty) | \hat{\cal S}^{\dagger}(t,-\infty)\hat{\cal O}_{\rm I}(t) \hat{\cal S}(t,-\infty)| \Psi_{\nu}(-\infty)\rangle 
\nonumber \\
&&\hspace{16.5mm}
= \langle \Psi_{\nu}(-\infty) | \hat{\cal S}^{\dagger}(\infty,-\infty)\hat{\cal S}(\infty,t)\hat{\cal O}_{\rm I}(t) \hat{\cal S}(t,-\infty)| \Psi_{\nu}(-\infty)\rangle 
\nonumber \\
&&\hspace{16.5mm}
= \langle \Psi_{\nu}(-\infty) | \hat{\cal S}^{\dagger}(t,-\infty)\hat{\cal O}_{\rm I}(t) \hat{\cal S}^{\dagger}(\infty,t)\hat{\cal S}(\infty,-\infty)| \Psi_{\nu}(-\infty)\rangle 
.
\label{<O_I2>}
\end{eqnarray}
We also notice that Eq.\ (\ref{S-int}) with $t_{0}=-\infty$ is equivalent to
\begin{subequations}
\label{S-int-T}
\begin{equation}
\hat{\cal S}(t,-\infty)
= {\cal T} \exp\left[-\frac{i}{\hbar} \int_{-\infty}^{t}\hat{\cal H}_{\rm I}'(t_{1}) dt_{1}\right] .
\label{S-int-T1}
\end{equation}
The Hermitian conjugate of Eq.\ (\ref{S-int}) can be expressed
with the anti-time-ordering operator $\tilde{\cal T}$ as 
\begin{equation}
\hat{\cal S}^{\dagger}(t,-\infty)
= \tilde {\cal T} \exp\left[-\frac{i}{\hbar} \int_{t}^{-\infty}\hat{\cal H}_{\rm I}'(t_{1})dt_{1}\right] .
\label{S-int-T2}
\end{equation}
\end{subequations}
Substituting these expressions into it, we realize that
the evaluation of Eq.\ (\ref{<O_I2>}) proceeds on the round-trip Keldysh contour $C$ of Fig.\ 2 starting from $t_{0}=-\infty$ towards $t=\infty$, and $\hat{\cal O}_{\rm I}(t)$ can be evaluated on either of the forward or backward path equivalently.

\begin{figure}[b]
\begin{center}
  \includegraphics[width=0.25\linewidth]{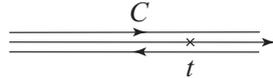}
\end{center}
\caption{Keldysh contour.\label{fig:1}}
\end{figure}

The fact just mentioned enables us to provide a concise expression to Eq.\ (\ref{<O_I2>}). 
Let us introduce the operator $\hat{\cal S}_{C}$ on $C$ by
\begin{equation}
\hat{\cal S}_{C}={\cal T}_{C}\exp\left[-\frac{i}{\hbar} \int_{C}\hat{\cal H}_{\rm I}'(t^{C})dt^{C}\right] ,
\label{S_C}
\end{equation}
where $t^{C}$ denotes a point on $C$, and ${\cal T}_{C}$ arranges operators according to their chronological order on $C$ from right to left. 
By using $\hat{\cal S}_{C}$, Eq.\ (\ref{<O_I2>}) can be expressed as
\begin{equation}
\langle \Psi_{\nu}(t)|\hat{\cal O}|\Psi_{\nu}(t)\rangle =
\langle \Psi_{\nu}(-\infty) | {\cal T}_{C}\hat{\cal S}_{C}\hat{\cal O}_{\rm I}(t)| \Psi_{\nu}(-\infty)\rangle 
= \frac{\langle \Psi_{\nu}(-\infty) | {\cal T}_{C}\hat{\cal S}_{C}\hat{\cal O}_{\rm I}(t)| \Psi_{\nu}(-\infty)\rangle}
{\langle \Psi_{\nu}(-\infty) | \hat{\cal S}_{C}| \Psi_{\nu}(-\infty)\rangle} ,
\label{<O_I3>}
\end{equation}
where $\hat{\cal O}_{\rm I}(t)$ can be on either the forward or backward path, and we have used $\langle \Psi_{\nu}(-\infty) | \hat{\cal S}_{C}| \Psi_{\nu}(-\infty)\rangle=1$ in the last equality.
One can see clearly from the last expression that only those diagrams that are connected with 
$\hat{\cal O}_{\rm I}$ contribute to the expectation value.

\section{Nonequilibrium perturbation expansion\label{sec:noneq}}

With the above preliminaries, we now embark on formulating nonequilibrium statistical mechanics for
an ensemble of  identical bosons or fermions with quantum field theory. There have been basically two approaches regarding the initial conditions at $t_{0}=-\infty$. The first one is that of Keldysh,\cite{Keldysh64} where  the initial state is void of interaction  to switch it on adiabatically together with external fields. The second one is the grand canonical ensemble with full interaction; however, the initial correlations along the imaginary-time path are neglected subsequently.\cite{RS86,HJ98} We adopt here the approach by Keldysh as it is free from any approximations and, hence, transparent. Indeed, the effect of the initial correlations may be handled using Keldysh's approach by waiting thermalization due to interaction before applying external fields.

\subsection{Hamiltonian, density matrix, and expectation value}

We consider a system of identical noninteracting bosons or fermions at $t_{0}=-\infty$. Let us apply an external field $U({\bm r},t)$ for $t>-\infty$ and also introduce the interaction $\hat{\cal H}_{\rm int}$ adiabatically.
Thus, the total Hamiltonian at time $t$ is given by 
\begin{equation}
\hat{\cal H}(t) = \hat{\cal H}_{0}(t)+\hat{\cal H}_{\rm int}a(t)  .
\label{H}
\end{equation}
Here, $\hat{\cal H}_{0}$ denotes noninteracting Hamiltonian including $U$, $\hat{\cal H}_{\rm int}$
describes the two-body interaction, and $a(t)$ is some adiabatic factor. The last quantity may be expressed with the step function  $\theta(t)$ as $a(t)\equiv\theta(-t)e^{0_{+} t}\!+\!\theta(t)$, for example,
with $0_{+}$ denoting an infinitesimal positive constant.
By neglecting the spin degrees of freedom for a while,
$\hat{\cal H}_{0}$ and $\hat{\cal H}_{\rm int}$ can be expressed as
\begin{subequations}
\label{H_0int}
\begin{eqnarray}
\hat{\cal H}_{0}&=&
\int 
\hat{\psi}^{\dagger}({\bm r}_{1})\hat{K}_{1}\hat{\psi}({\bm r}_{1}) \,d^{3}r_{1},
\label{H_0}
\\
\hat{\cal H}_{\rm int}&=&\frac{1}{2}
\int d^{3}r_{1}\int d^{3}r_{1}'\,
\hat{\psi}^{\dagger}({\bm r}_{1})\hat{\psi}^{\dagger}({\bm r}_{1}')
V({\bm r}_{1}\!-\!{\bm r}_{1}')
\hat{\psi}({\bm r}_{1}')\hat{\psi}({\bm r}_{1}) .
\label{H_int}
\end{eqnarray}
\end{subequations}
Here, $\hat{K}_{1}$ denotes
\begin{equation}
\hat{K}_{1}\equiv -\frac{\hbar^{2}}{2m}\nabla_{1}^{2}
+U({\bm r}_{1}t_{1})-\mu,
\label{K_1}
\end{equation}
with $m$ and $\mu$ as the particle mass and chemical potential, respectively;
the field operators $\hat{\psi}$ and $\hat{\psi}^{\dagger}$ satisfy
\begin{subequations}
\label{commun}
\begin{equation}
\hat{\psi}({\bm r}_{1})\hat{\psi}^{\dagger}({\bm r}_{2})\mp\hat{\psi}^{\dagger}({\bm r}_{2})\hat{\psi}({\bm r}_{1})=\delta({\bm r}_{1}-{\bm r}_{2}),
\label{commun1}
\end{equation}
\begin{equation}
\hat{\psi}({\bm r}_{1})\hat{\psi}({\bm r}_{2})\mp\hat{\psi}({\bm r}_{2})\hat{\psi}({\bm r}_{1})=0,
\label{commun2}
\end{equation}
\end{subequations}
with the upper (lower) sign corresponding to bosons (fermions),
and $V({\bm r}_{1}-{\bm r}_{1}')$ is the interaction potential with
$V(-{\bm r})=V({\bm r})$, which may be expanded as
\begin{equation}
V({\bm r})=\int \frac{d^{3}p}{(2\pi \hbar)^{3}}V_{\bm p}\,e^{i{\bm p}
\cdot{\bm r}/\hbar} ,
\label{V-Fourier}
\end{equation}
with $V_{-{\bm p}}=V_{\bm p}$.

Let us assume that the system at $t_{0}=-\infty$ is in thermodynamic equilibrium described with the grand canonical ensemble.
The corresponding density matrix $\hat{\rho}_{0}$ can be expressed with the eigenvalue $E_{\nu}(-\infty)$ and the eigenstate $|\Psi_{\nu}(-\infty)\rangle$ of $\hat{\cal H}_{0}(-\infty)$ as
\begin{equation}
\hat{\rho}_{0}=\frac{1}{Z_{0}}\sum_{\nu}e^{-E_{\nu}(-\infty)/k_{\rm B}T_{0}}|\Psi_{\nu}(-\infty)\rangle 
\langle \Psi_{\nu}(-\infty) |,
\hspace{10mm}Z_{0}\equiv \sum_{\nu}e^{-E_{\nu}(-\infty)/k_{\rm B}T_{0}},
\end{equation}
where $T_{0}$ and $k_{\rm B}$ are the initial temperature and Boltzmann constant, respectively.
We now average Eq.\ (\ref{<O>}) over the initial probability distribution $Z_{0}^{-1}e^{-E_{\nu}(-\infty)/k_{\rm B}T_{0}}$. 
Then, the expectation value of $\hat{\cal O}_{\rm H}(t)$ can be expressed as
\begin{equation}
\langle\hat{\cal O}_{\rm H}(t)\rangle \equiv {\rm Tr}\hat{\rho}_{0}  \hat{\cal O}_{\rm H}(t) =
\frac{\langle {\cal T}_{C}\hat{\cal S}_{C} \hat{\cal O}_{\rm I}(t) \rangle}
{\langle\hat{\cal S}_{C} \rangle},
\label{O(t)-I2}
\end{equation}
where we have used Eq.\ (\ref{<O_I3>}) in the second equality.
Since the initial state is a noninteracting grand canonical ensemble, we can use 
the Bloch-de Dominicis theorem \cite{BdD59,Gaudin60}, i.e., the finite-temperature version of Wick's theorem \cite{Wick50}, to evaluate the second expression of Eq.\ (\ref{O(t)-I2}) perturbatively in terms of $\hat{\cal H}_{\rm int}$. It then follows that only those Feynman diagrams connected with
$\hat{\cal O}_{\rm I}(t)$ contribute to the expectation value due to the denominator in Eq.\ (\ref{O(t)-I2}).

A few comments are in order. First, the original postulate of the noninteracting grand canonical ensemble
in proving the Bloch-de Dominicis theorem can be relaxed, as already pointed out by Danielewicz \cite{Danielewicz84}. To be specific, the Wick decomposition is justified for any density matrix of the form:
\begin{equation}
\hat{\rho}_{0}=\prod_{k}
\sum_{n_{k}}|n_{k}\rangle p_{k}^{n_{k}}\langle n_{k}|,
\hspace{10mm}|n_{k}\rangle = \frac{(\hat{c}_{k}^{\dagger})^{n_{k}}}{\sqrt{n_{k}!}}
|0\rangle_{k} ,
\label{rho_0}
\end{equation}
where $p_{k}$ denotes the probability that there is a particle in the single-particle state $k$,
$n_{k}$ is the number of particles occupying $k$, $\hat{c}_{k}^{\dagger}$ is the creation operator,
and $|0\rangle_{k}$ is the vacuum of $k$; see also Appendix \ref{subsec:DM} for the equilibrium density matrix. 
This $\hat{\rho}_{0}$ is easily shown to satisfy
\begin{equation}
\hat{c}_{k}\hat{\rho}_{0}=p_{k}\hat{\rho}_{0}\hat{c}_{k},
\hspace{10mm}
\hat{c}_{k}^{\dagger}\hat{\rho}_{0}=\frac{1}{p_{k}}\hat{\rho}_{0}\hat{c}_{k}^{\dagger}.
\label{rho_0-identity}
\end{equation}
Using these relations, one may repeat the concise proof by Gaudin \cite{Gaudin60,FW71}
to convince oneself that the Bloch-de Dominicis theorem holds 
even for $p_{k}\neq e^{-(\epsilon_{k}-\mu)/k_{\rm B}T_{0}}$, where $\epsilon_{k}$ is the single-particle energy of the state $k$.
Thus, we can evolve the system also from noninteracting nonequilibrium states of a wide class. 

Second, the effects of so-called ``initial correlations'' may be incorporated into the formalism 
by waiting for thermalization due to interaction before applying external fields.
Indeed, thermalization is achieved in this formalism by considering terms
of the second (and higher) order in the perturbation expansion, which describe collisions between particles.
In this context, it may seem puzzling why thermalization is achieved only by evolving the system mechanically.
Equivalently, what is the origin of the arrow of time in this formalism?
It is instructive for answering the question to see the results of a molecular dynamics simulation on a classical dilute hard-sphere gas  by Orban and Bellemans.\cite{OB67} They show that the entropy of the system (i.e., the Boltzmann entropy that is applicable here) does increase for the overwhelming majority of initial conditions
without any contradictions to the time-reversal symmetry in the mechanical laws.
Suggested by the results and following the viewpoint of Boltzmann,\cite{Lebowitz93} we trace the origin of the arrow of time in many-body systems 
to the overwhelmingness of those initial conditions.
The equation of motion for the nonequilibrium Green's function below will be regarded as corresponding to those overwhelming
majority of initial conditions.
Thermalization will also be achieved for systems in contact with thermal baths.
The concept of temperature is not necessary at all in this formalism; it naturally emerges as one of the properties 
in the thermalized distribution function.

\subsection{Green's function}

We now consider the effects of the interaction
\begin{equation}
\hat{\cal H}'(t)\equiv \hat{\cal H}_{\rm int}a(t)
\end{equation}
perturbatively.
A key quantity to this end is Green's function:
\begin{equation}
 G(1^{C},2^{C})
\equiv-\frac{i}{\hbar}
\langle {\cal T}_{C} \hat{\psi}_{\rm H}(1^{C})\hat{\psi}^{\dagger}_{\rm H}(2^{C})\rangle
=-\frac{i}{\hbar}
\langle {\cal T}_{C}\hat{\cal S}_{C}\hat{\psi}_{\rm I}(1^{C})\hat{\psi}^{\dagger}_{\rm I}(2^{C})\rangle_{\rm c}.
\label{G-def}
\end{equation}
Here, the superscript $^{C}$ in $1^{C}\equiv {\bm r}_{1}t^{C}_{1}$ distinguishes the forward and backward paths of the Keldysh contour,
and the subscript $_{\rm c}$ in the last expression denotes retaining only connected diagrams.
We drop the subscript $_{\rm I}$ in $\hat{\psi}_{\rm I}$ and $\hat{\psi}_{\rm I}^{\dagger}$ below; we also omit the superscript $^{C}$ whenever it is unnecessary such as those in the argument of $V$.

The perturbation expansion on $C$ can be carried out in the same manner as that of the equilibrium Matsubara formalism on the imaginary-time contour $t\in [0,-i\hbar/k_{\rm B}T]$ with the difference lying only in the contours; see Appendix \ref{sec:SCPE} for the Matsubara formalism.
Hence, it follows that Eq.\ (\ref{G-def}) satisfies Dyson's equation:
\begin{equation}
\left(\!i\hbar \frac{\partial }{\partial t_{1}^{C}}-\hat{K}_{1}\!\right)
G(1^{C},2^{C})-
\int \Sigma(1^{C},3^{C})
G(3^{C},2^{C})\,d3^{C} =\delta(1^{C},2^{C}),
\label{Dyson}
\end{equation}
where $\hat{K}_{1}$ is given by Eq.\ (\ref{K_1}).

Following Keldysh,\cite{Keldysh64} we next transform every integration on $C$
into that of the ordinary time $t\in [-\infty,\infty]$.
To this end, we express $C$ as a sum of the forward path $C_{1}$ and the backward path $C_{2}$
as
\begin{equation}
\int_{C}dt^{C}=\int_{C_{1}}dt^{1}+\int_{C_{2}}dt^{2}
=\int_{-\infty (C_{1})}^{\infty}dt^{1}-\int_{-\infty (C_{2})}^{\infty}dt^{2}.
\label{int_C}
\end{equation}
We next introduce $1^{i}\equiv {\bm r}_{1}t_{1}^{i}$ $(i=1,2)$ to distinguish times on $C_{1}$ and $C_{2}$ clearly.
We then define $G^{ij}(1,2)\equiv G(1^{i},2^{j})$ to provide them with the matrix representation:
\begin{equation}
\check{G}(1,2)\equiv 
\left[\begin{array}{cc}
G^{11}(1,2) & G^{12}(1,2)\\
G^{21}(1,2) & G^{22}(1,2)
\end{array}\right] \! .
\label{G-check-def}
\end{equation}
We express every $2\times 2$ matrix due to the path decomposition with $\,\check{ }\,$
on top of it like Eq.\ (\ref{G-check-def}). Explicit expressions of $G^{ij}$ are given by
\begin{subequations}
\label{Gcomp}
\begin{eqnarray}
&&\hspace{-10mm}
G^{21}(1,2)=
- \frac{i}{\hbar}\langle
\hat{\psi}_{\rm H}(1)\hat{\psi}_{\rm H}^{\dagger}(2)\rangle,
\label{Gcomp21}
\\
&&\hspace{-10mm}
G^{12}(1,2)=
\mp \frac{i}{\hbar}\langle\hat{\psi}_{\rm H}^{\dagger}(2)
\hat{\psi}_{\rm H}(1)\rangle,
\label{Gcomp12}
\\
&&\hspace{-10mm}
G^{11}(1,2)=-\frac{i}{\hbar}\bigl[
\theta(t_{1}\!-\!t_{2})\langle\hat{\psi}_{\rm H}(1)\hat{\psi}_{\rm H}^{\dagger}(2)\rangle
\pm\theta(t_{2}\!-\!t_{1}) \langle\hat{\psi}_{\rm H}^{\dagger}(2)\hat{\psi}_{\rm H}(1)\rangle\bigr]
\nonumber \\
&&\hspace{5.8mm}=
\theta(t_{1}\!-\!t_{2})G^{21}(1,2)
+\theta(t_{2}\!-\!t_{1})G^{12}(1,2) ,
\label{Gcomp11}
\\
&&\hspace{-10mm}
G^{22}(1,2)=-\frac{i}{\hbar}\bigl[
\theta(t_{2}\!-\!t_{1})\langle\hat{\psi}_{\rm H}(1)\hat{\psi}_{\rm H}^{\dagger}(2)\rangle
\pm\theta(t_{1}\!-\!t_{2}) \langle\hat{\psi}_{\rm H}^{\dagger}(2)\hat{\psi}_{\rm H}(1)\rangle\bigr]
\nonumber \\
&&\hspace{5.8mm}=
\theta(t_{2}\!-\!t_{1})G^{21}(1,2)
+\theta(t_{1}\!-\!t_{2})G^{12}(1,2) ,
\label{Gcomp22}
\end{eqnarray}
\end{subequations}
with
\begin{equation}
\theta (t)\equiv \left\{ \begin{array}{ll} 1 & :t\geq 0 \\
0 & : t<0
\end{array}\right. .
\label{step-fn}
\end{equation}
The off-diagonal elements $G^{12}$ and $G^{21}$ correspond to $G^{<}$ and $G^{>}$
of Kadanoff and Baym,\cite{KB62} respectively.
The four elements have the properties:
\begin{subequations}
\label{Gcomp-symm}
\begin{equation}
[G^{12}(1,2)]^{*}= -G^{12}(2,1), \hspace{3mm}
[G^{21}(1,2)]^{*}= -G^{21}(2,1), \hspace{3mm}
[G^{11}(1,2)]^{*}=- G^{22}(2,1), 
\label{Gcomp-symm1}
\end{equation}
\begin{equation}
G^{11}(1,2)+ G^{22}(1,2) = 
G^{12}(1,2)+ G^{21}(1,2) .
\label{Gcomp-symm2}
\end{equation}
\end{subequations}
Equation (\ref{Gcomp-symm}) can be expressed concisely in terms of
$\check{G}$ as
\begin{equation}
\check{G}(1,2)=-\check{\tau}_{1}\check{G}^{\dagger}(2,1)\check{\tau}_{1},
\hspace{10mm}{\rm Tr}\,\check{G}(1,2)={\rm Tr}\,\check{\tau}_{1}\check{G}(1,2) ,
\label{checkG-symm}
\end{equation}
where $\check{\tau}_{i}$ ($i=1,2,3$) are the Pauli matrices.
It is also convenient to define the self-energy matrix:
\begin{equation}
\check{\Sigma}(1,2)\equiv 
\left[\begin{array}{cc}
\Sigma^{11}(1,2) & \Sigma^{12}(1,2)\\
\Sigma^{21}(1,2) & \Sigma^{22}(1,2)
\end{array}\right] .
\label{Sigma-check-def}
\end{equation}
Now, we can express Dyson's equation (\ref{Dyson})
in terms of $\check{G}$ and $\check{\Sigma}$ as
\begin{equation}
\left(i\hbar \frac{\partial }{\partial t_{1}}-\hat{K}_{1}\right)
\check{G}(1,2)-
\int \check{\Sigma}(1,3)\check{\tau}_{3}
\check{G}(3,2)\,d3 =\check{\tau}_{3}
\delta(1,2) .
\label{Dyson-check}
\end{equation}
Here, $\check{\tau}_{3}$ between
$\check{G}$ and $\check{\Sigma}$ results from the transformation:
\begin{eqnarray*}
\int_{C}\Sigma(1^{i},3^{C})G(3^{C},2^{j})\, d3^{C}=
\int d3 \left[
\Sigma^{i1}(1,3)G^{1j}(3,2)
-\Sigma^{i2}(1,3)G^{2j}(3,2)\right] ,
\end{eqnarray*}
and $\check{\tau}_{3}$ on the right-hand side of 
Eq.\ (\ref{Dyson-check}) originates from the opposite signs in the arguments of $\theta$
in Eqs.\ (\ref{Gcomp11}) and (\ref{Gcomp22}).

Let us introduce
\begin{equation}
\check{G}^{(0)}(1,2)\equiv 
\left(i\hbar \frac{\partial }{\partial t_{1}}-\hat{K}_{1}\right)^{\!\! -1}
\check{\tau}_{3} \delta(1,2) ,
\label{G_0}
\end{equation}
which is the solution to Eq.\ (\ref{Dyson-check}) without $\check{\Sigma}$, i.e., Green's function without interaction.
Multiplying it by $\check{G}^{(0)}\check{\tau}_{3}$ from the right,
Eq.\ (\ref{Dyson-check}) can also be put into an integral form as
\begin{equation}
\check{G}(1,2)=\check{G}^{(0)}(1,2)+
\int d3\int \! d4\, \check{G}^{(0)}(1,3)\check{\tau}_{3}
\check{\Sigma}(3,4)\check{\tau}_{3}
\check{G}(4,2) .
\label{Dyson-check2}
\end{equation}
\begin{figure}[b]
\begin{center}
  \includegraphics[width=0.15\linewidth]{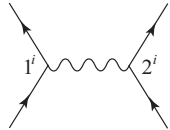}
\end{center}
\caption{Feynman diagram for the two-body interaction.\label{fig:Hint}}
\end{figure}
\subsection{Perturbation expansion of $\check{G}$}

Let us write down the Feynman rules for the perturbation expansion of $\check{G}$
to see how they can be obtained from those of the equilibrium Matsubara formalism.
Using Eq.\ (\ref{int_C}), we first express $\hat{\cal S}_{C}$ of Eq.\ (\ref{S_C}) as
\begin{eqnarray}
&&\hspace{-9mm}
\hat{\cal S}_{C}={\cal T}_{C}\exp\left[-\frac{i}{\hbar}\sum_{i=1}^{2}(-1)^{i-1}\int_{-\infty}^{\infty}
dt_{1}^{i}\,
\hat{\cal H}'_{\rm I}(t_{1}^{i})\right]
\nonumber \\
&&\hspace{-3mm}
={\cal T}_{C}\exp\left[-\frac{i}{2\hbar}\int d1\int d2
\sum_{i=1}^{2}(-1)^{i-1}\bar{V}(1-2)
\hat{\psi}^{\dagger}(1^{i})\hat{\psi}^{\dagger}(2^{i})
\hat{\psi}(2^{i})\hat{\psi}(1^{i})\right] ,
\label{S_C2}
\end{eqnarray}
where $\bar{V}$ is defined by
\begin{equation}
\bar{V}(1-2)\equiv\delta(t_{1}-t_{2})V({\bm r}_{1}-{\bm r}_{2}) .
\end{equation}
This interaction can be expressed graphically as Fig.\ 3,
where the wavy line denotes $(-1)^{i-1}\bar{V}$ and the two incoming (outgoing) arrows signify
$\hat{\psi}$ ($\hat{\psi}^{\dagger}$).

Let us compare Eq.\ (\ref{S_C2}) with the corresponding S-matrix of the equilibrium Matsubara formalism: \cite{KB62}
\begin{equation}
\hat{\cal S}_{\rm M}(\beta,0)\equiv {\cal T}_{\tau}\exp\left[-\frac{i}{2\hbar}\int d1\int d2\,
\bar{V}(1-2)
\hat{\psi}^{\dagger}(1)\hat{\psi}^{\dagger}(2)
\hat{\psi}(2)\hat{\psi}(1)\right],
\label{S-M}
\end{equation}
where $\beta\equiv 1/k_{\rm B}T$ with $T$ the temperature, every time integration extends over $t\in [0,-i\hbar\beta]$  with the periodic (antiperiodic) boundary condition for bosons (fermions), and ${\cal T}_{\tau}$ denotes the ``time''-ordering operator on the imaginary contour; see also Eq.\ (\ref{S(b)}) of Appendix \ref{sec:SCPE} for the derivation of Eq.\ (\ref{S-M}).
We then notice that, besides the difference in the integration contours, Eq.\ (\ref{S_C2}) contains an additional summation over the paths $i=1,2$ with the potential $(-1)^{i-1}\bar{V}(1-2)$.

This observation enables us to obtain the Feynman rules for the $n$th-order contribution $(n=1,2,\cdots)$ to $\check{G}$ straightforwardly from those of the Matsubara formalism.\cite{AGD63,FW71,KB62,LW60} They are summarized as follows:
\begin{itemize}
\item[(a)] Draw all possible $n$th-order connected diagrams.

\item[(b)] With each such diagram, associate a factor
\begin{equation}
\frac{(i\hbar)^{n}(\pm 1)^{n_{\ell}}}{2^{n}n! } ,
\label{Fyn2G}
\end{equation}
where $n_{\ell}$ denotes the number of closed particle loops, 
and the factor $(i\hbar)^{n}$ results from multiplying   
$(-i/\hbar)^{n}$ from the expansion of Eq.\ (\ref{S_C2}) by $(i\hbar)^{2n}$; the latter cancels
each $-i/\hbar$ in definition (\ref{G-def}) of Green's function.

\item[(c)] For an interaction line on the contour $C_{i}$ ($i=1,2$), associate the potential $(-1)^{i-1}\bar{V}$.

\item[(d)] With a particle line arriving at $1^{i}$ from $2^{j}$, associate $G^{ij(0)}(1,2)$.
If the line enters and leaves the same interaction line with $i=j$,
associate $G^{11(0)}(1,1_{+})$ and $G^{22(0)}(1_{+},1)$ for $i=1,2$, respectively,
where the subscript $_{+}$ denotes the presence of an extra infinitesimal positive constant 
in the relevant time argument.
This rule works to recover the operator arrangement of Eq.\ (\ref{H_int})
that $\hat{\psi}^\dagger$ lies to the left of $\hat{\psi}$.
Note that $G^{11(0)}(1,1_{+})=G^{22(0)}(1_{+},1)=G^{12(0)}(1,1)$, as seen from Eq.\ (\ref{Gcomp}).

\item[(e)] Integrate or sum over all the internal arguments.

\item[(f)] Spin degrees of freedom may be incorporated by multiplying 
the contribution of each closed particle line by $2S\!+\! 1$, where $S$ denotes the magnitude of spin.

\end{itemize}
It is worth pointing out that the factor $(2^{n}n!)^{-1}$ in Eq.\ (\ref{Fyn2G}) is canceled eventually by the number of topologically identical diagrams in the perturbation expansion for $\check{G}$.

Thus, one of the features inherent in the real-time nonequilibrium formalism lies in the additional
summation over the paths $C_{1}$ and $C_{2}$. Several alternative versions of the Feynman rules have been presented for carrying out this extra summation. The Langreth theorem\cite{Langreth76,HJ98} is among such alternatives.
However, this rule with the path decomposition $C\rightarrow C_{1}+C_{2}$ may be the most natural, straightforward, and simple.

\begin{figure}[t]
\begin{center}
  \includegraphics[width=0.28\linewidth]{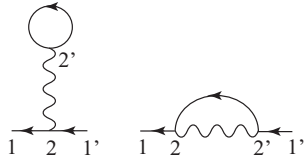}
\end{center}
\caption{Topologically distinct first-order diagrams for $\check{G}$. Path indices are omitted. Those obtained from above with $2\leftrightarrow 2'$ yield the same contribution.\label{fig:G1st}}
\end{figure}

As an example of these rules, let us consider the first-order contributions, where 
topologically distinct Feynman diagrams are given in Fig.\ 4.
Those obtained with the replacement $2\leftrightarrow 2'$ yield the same contribution to cancel the factor $1/(2^{1}1!)$ in rule (b).
Performing the summation over the paths $C_{i}$ ($i=1,2$),
we obtain an analytic expression for the first-order Green's function as
\begin{eqnarray}
&&\hspace{-1mm}
{G}^{ii'(1)}(1,1')
= i\hbar\sum_{j=1}^{2}\int d2 \int d2'\,
(-1)^{j-1}\bar{V}(2\!-\!2')\bigl[\pm G^{ij(0)}(1,2)G^{jj(0)}(2',2')
\nonumber \\
&&\hspace{23.4mm}
\times G^{ji'(0)}(2,1')
+G^{ij(0)}(1,2)G^{jj(0)}(2,2')
G^{ji'(0)}(2',1')\bigr]
\nonumber \\
&&\hspace{19.5mm}
=\sum_{jj'}\!\int d2 \!\int d2'\, G^{ij(0)}(1,2)\delta_{jj'}(-1)^{j-1}i\hbar \biggl[\pm \delta(2,2')\! \int \!d3\,
\bar{V}(2\!-\!3)
\nonumber \\
&&\hspace{23.4mm}
\times  G^{jj(0)}(3,3)
+\bar{V}(2\!-\!2')
G^{jj(0)}(2,2')\biggr] G^{j'i'(0)}(2',1'),
\label{G^(1)}
\end{eqnarray}
where, in the second expression, we have factored out the two Green's functions connected with the external lines.
A comparison of  Eq.\ (\ref{G^(1)}) with Eq.\ (\ref{Dyson-check2}) leads to the expression of the first-order self-energy:
\begin{subequations}
\label{Sigma^(1)-bare}
\begin{equation}
{\Sigma}^{ij({\rm s}1)}(1,2)
=\delta_{ij}(-1)^{i-1}i\hbar \biggl[\pm \delta(1,2)\! \int \!d1'\,
\bar{V}(1\!-\!1')
G^{ii(0)}(1',1')
+\bar{V}(1\!-\!2)
G^{ii(0)}(1,2)\biggr]  ,
\label{Sigma^(1)a}
\end{equation}
where the superscript $^{{\rm s}1}$ signifies that the expression belongs to the ``simple'' first-order perturbation expansion.
Thus, the first-order self-energy is diagonal in the Keldysh space.
We further consider the Feynman rule (d) above to make the replacements:
$G^{11(0)}(1,1')\!\rightarrow\! G^{11(0)}(1,1_{+}')\!=\! G^{12(0)}(1,1')$
and $G^{22(0)}(1,1')\!\rightarrow\! G^{22(0)}(1_{+},1')\!=\! G^{12(0)}(1,1')$.
We thereby obtain an expression for the first-order self-energy matrix $\check{\Sigma}^{({\rm s}1)}$ as
\begin{equation}
\check{\Sigma}^{({\rm s}1)}(1,2)=\check{\tau}_{3}\Sigma^{{\rm sHF}}(1,2),
\label{Sigma^(1)b}
\end{equation}
\end{subequations}
with
\begin{equation}
\Sigma^{{\rm sHF}}(1,2)\equiv
\pm \delta(1,2)\int d1'\,
\bar{V}(1\!-\!1')
i\hbar   G^{12(0)}(1',1')
+\bar{V}(1\!-\!2)i\hbar
G^{12(0)}(1,2) .
\label{Sigma-HF(1)}
\end{equation}
We present further examples of the perturbation expansion with the Feynman rules in the next subsection.

\subsection{Self-consistent perturbation expansion for $\check{G}$\label{subsec:Phi}}

The procedure described in \S 3.3 enables us to carry out the simple (bare) perturbation expansion concisely up to a desired order or to sum up some partial series. There is another category in the perturbation expansion, i.e., the self-consistent perturbation expansion, which is characterized by a unique property for describing nonequilibrium dynamical phenomena, i.e., various conservation laws are satisfied automatically. It was pioneered for a normal Fermi system by Luttinger and Ward \cite{LW60} in their effort to write down its thermodynamic potential rigorously as a functional of the exact self-energy $\Sigma$. Luttinger\cite{Luttinger60} subsequently pointed out that the expansion reproduces the Hartree-Fock approximation as the lowest-order approximation and may be used as a general approximation scheme. Baym\cite{Baym62} later presented a sufficient condition for various conservation laws to be obeyed automatically and showed that the self-consistent perturbation expansion based on the Luttinger-Ward functional meets the condition. The practical approximation scheme Baym presented there is the same as the one suggested by Luttinger\cite{Luttinger60} but superior to the latter in that it is given explicitly in terms of the functional $\Phi$.
Indeed, it is with $\Phi$ that the conservation laws are proved; see \S 8 on this point. Moreover, the functional enables us to calculate two-particle and higher-order correlations systematically, as shown in \S 7. 

It is more convenient to express the Luttinger-Ward thermodynamic potential as a functional of Green's function $G$.\cite{Baym62} Its key ingredient is the functional $\Phi=\Phi[G]$ given as a power series in terms of the interaction potential, i.e., the so-called ``skeleton diagram expansion''. Using $\Phi$, we can also construct self-consistent approximations systematically up to a desired order. We now explain this approximation scheme in detail with its advantages together with further examples of the perturbation expansion with the Feynman rules.
See Appendices \ref{sec:SCPE} and \ref{LW} for the self-consistent perturbation expansion and the Luttinger-Ward functional, respectively, in the equilibrium Matsubara formalism. Although we limit our consideration below to normal phases, it is worth pointing out that the Luttinger-Ward functional was extended also for superconductors,\cite{dDM64,Kita96} Bose-Einstein condensates,\cite{Kita09} and relativistic nonequilibrium phenomena in high-energy physics.\cite{CJT74,Berges04}

Let us introduce a nonequilibrium extension of the functional, i.e., $\Phi=\Phi[G^{ij}]$, in terms of S-matrix (\ref{S_C2}) by
\begin{equation}
\Phi \equiv \bigl[\ln \langle \hat{\cal S}_{C}\rangle\bigr]_{{\rm skeleton},
{G}^{ij(0)}\rightarrow {G}^{ij}}= \bigl[\langle \hat{\cal S}_{C}\rangle_{{\rm c}}-1\bigr]_{{\rm skeleton},
{G}^{ij(0)}\rightarrow {G}^{ij}} .
\label{Phi-def}
\end{equation}
Here, the subscript ``skeleton'' denotes retaining only skeleton diagrams without the self-energy insertion in the perturbation expansion for $\langle \hat{\cal S}_{C}\rangle_{{\rm c}}-1$, and ${G}^{ij(0)}\rightarrow {G}^{ij}$ signifies replacing every ${G}^{ij(0)}$ by the renormalized one ${G}^{ij}$. See, e.g., Ref.\ \citen{AGD63} for the proof of the second equality that taking the logarithm of $\langle \hat{\cal S}_{C}\rangle$ amounts to retaining only connected diagrams of $\langle \hat{\cal S}_{C}\rangle-1$.  Note $\Phi^{*}=\Phi$.

The Feynman rules for $\Phi$ are identical with those of Green's function above. 
It turns out that the exact self-energy is obtained from this $\Phi$ by\cite{LW60}
\begin{equation}
{\Sigma}^{ij}(1,2)=\pm  (-1)^{i+j}
\frac{\delta \Phi}{\delta {G}^{ji}(2,1)} .
\label{Sigma-Phi}
\end{equation}
This functional differentiation graphically amounts to removing a Green's function line from every Feynman diagram for $\Phi$  in all possible ways.
The factor $\pm $ originates from a reduction of a closed particle, and the factor $(-1)^{i+j}$ 
reflects two $\check{\tau}_{3}$'s in Eq.\ (\ref{Dyson-check2}).
It follows from Eq.\ (\ref{checkG-symm}), $\Phi^{*}=\Phi$, and Eq.\ (\ref{Sigma-Phi}) that
$\check{\Sigma}$ obeys
\begin{equation}
\check{\Sigma}(1,2)=-\check{\tau}_{1}\check{\Sigma}^{\dagger}(2,1)\check{\tau}_{1}.
\label{checkSigma-symm}
\end{equation}

Now, our self-consistent perturbation expansion denotes: (i) carrying out the expansion of Eq.\ (\ref{Phi-def}) 
up to some order or retain some partial infinite series, and (ii) determining $\check{G}$ and $\check{\Sigma}$ self-consistently using Dyson's equation (\ref{Dyson-check}) and relation (\ref{Sigma-Phi}). In this procedure, $\check{G}$ is determined with $\check{\Sigma}$ using Eq.\ (\ref{Dyson-check}), whereas $\check{\Sigma}$ is given in terms of $\check{G}$ as Eq.\ (\ref{Sigma-Phi}); hence, the word ``self-consistent''. The advantages of the approach are summarized as follows.

\begin{itemize}
\item[(a)] It becomes exact when all the Feynman diagrams are retained, thus having the same structure as the exact theory.
\item[(b)] It satisfies various conservation laws, as shown by Baym.\cite{Baym62} Indeed, it is clearly in the category of $\Phi$-derivable approximation.
\item[(c)] Vertex corrections are taken into account automatically.
\item[(d)] Two-particle and higher-order correlations can be calculated from any approximate $\Phi$, i.e.,\ providing some $\Phi$ amounts to determining the BBGKY hierarchy\cite{Cercignani88} completely.

\end{itemize}
Point (b) makes the approximation quite suitable for describing dynamical phenomena. Indeed, no other systematic approximation scheme with the property seems to have been known to date within quantum field theory. It is worth pointing out that the Boltzmann equation
can be derived by the approach with further approximations such as the quasiparticle approximation for the spectral function;\cite{KB62} see \S \ref{QPA} below on this point.

\begin{figure}[b]
\begin{center}
  \includegraphics[width=0.3\linewidth]{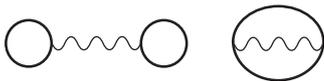}
\end{center}
\caption{Topologically distinct first-order diagrams for $\Phi$.\label{fig:Phi1New}}
\end{figure}

We now consider the first few series for $\Phi$ and write down the corresponding $\check{\Sigma}$ explicitly. As already mentioned, Feynman rules for $\Phi$ are identical with those of Green's function given around Eq.\ (\ref{Fyn2G}). 

To begin with, we consider the first-order contribution to $\Phi$. Topologically distinct diagrams are given in Fig.\ 5, where thick lines denote $\check{G}$; arrows, space-time arguments, and path indices have all been omitted. The corresponding analytic expression is given by
\begin{equation}
\Phi^{(1)}=\frac{i\hbar}{2} \sum_{i=1}^{2} \!\int \!d1 \!\int\!d1'\, (-1)^{i-1}\bar{V}(1\!-\!1')
\left[G^{ii}(1,1)G^{ii}(1',1')
\pm G^{ii}(1,1')G^{ii}(1',1) \right] .
\label{Phi^(1)}
\end{equation}
It should be noted that there is an extra factor $1/2$ in Eq.\ (\ref{Phi^(1)}) compared with Eq.\ (\ref{G^(1)}).
In general, the factor $1/2n$ remains uncanceled in the $n$th-order contribution to $\Phi$
due to the fact that diagrams of $\Phi$ are closed so that the $2n$ Green's functions are equivalent there.\cite{LW60}

The first-order self-energy is obtained from $\Phi^{(1)}$ using Eq.\ (\ref{Sigma-Phi}). Noting the rule (d) below Eq.\ (\ref{Fyn2G}), we can put the result into the concise form:
\begin{equation}
\check{\Sigma}^{(1)}(1,2)=\check{\tau}_{3}\Sigma^{{\rm HF}}(1,2),
\label{Sigma^(1)}
\end{equation}
with
\begin{equation}
\Sigma^{{\rm HF}}(1,2)\equiv
\pm \delta(1,2)\int d1'\,
\bar{V}(1\!-\!1')
i\hbar   G^{12}(1',1')
+\bar{V}(1\!-\!2)i\hbar
G^{12}(1,2) .
\label{Sigma-HF}
\end{equation}
Except for the difference between $G^{12(0)}$ and $G^{12}$, Eq.\ (\ref{Sigma^(1)}) coincides with Eq.\ (\ref{Sigma^(1)-bare}), as it should.

\begin{figure}[t]
\begin{center}
  \includegraphics[width=0.3\linewidth]{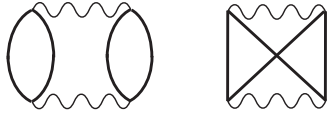}
\end{center}
\caption{Topologically distinct second-order diagrams for $\Phi$.\label{fig:Phi2New}}
\end{figure}

We next consider the second-order contribution to $\Phi$. Topologically distinct diagrams are enumerated in Fig.\ 6.
The corresponding analytic expression is given by
\begin{eqnarray}
&&\hspace{0mm}
\Phi^{(2)}=
\frac{(i\hbar)^{2}}{4}\sum_{i,j=1}^{2}(-1)^{i+j}\int d1 \int d1' 
\int  d2\int d2'\,
\bar{V}(1\!-\!1')\bar{V}(2\!-\!2')\bigl[G^{ij}(1,2)G^{ji}(2,1)
\nonumber \\
&&\hspace{12mm}
\times G^{ij}(1',2')G^{ji}(2',1')
\pm G^{ij}(1,2)G^{ji}(2,1')G^{ij}(1',2')G^{ji}(2',1)\bigr].
\label{Phi^(2)}
\end{eqnarray}
The second-order self-energy is obtained from $\Phi^{(2)}$ using Eq.\ (\ref{Sigma-Phi}) as
\begin{eqnarray}
&&\hspace{0mm}
\Sigma^{ij(2)}(1,2)=
(i\hbar)^{2} \int d1' \int d2'\,
\bar{V}(1\!-\!1')\bar{V}(2\!-\!2')
\bigl[\pm G^{ij}(1,2)G^{ij}(1',2')G^{ji}(2',1')
\nonumber \\
&&\hspace{23mm}
+ G^{ij}(1,2')G^{ji}(2',1')G^{ij}(1',2)\bigr] .
\label{Sigma^(2)}
\end{eqnarray}
Let us substitute Eqs.\ (\ref{Gcomp11}) and (\ref{Gcomp22}) into Eq.\ (\ref{Sigma^(2)}) with $i=j=1$ and
compare the result with the expressions of 
${\Sigma}^{12(2)}$ and ${\Sigma}^{21(2)}$.
We thereby obtain
\begin{eqnarray}
&&\hspace{0mm}
\Sigma^{11(2)}(1,2)=
(i\hbar)^{2} \int d1' \int d2'\,
\bar{V}(1\!-\!1')\bar{V}(2\!-\!2')\bigl \{ \theta(t_{1}-t_{2})\bigl[\pm G^{21}(1,2)G^{21}(1',2')
\nonumber \\
&&\hspace{23.5mm}
\times G^{12}(2',1')
+ G^{21}(1,2')G^{12}(2',1')G^{21}(1',2)\bigr] 
+\theta(t_{2}-t_{1})\bigl[\pm G^{12}(1,2)
\nonumber \\
&&\hspace{23.5mm}
\times 
G^{12}(1',2')
G^{21}(2',1')
+ G^{12}(1,2')G^{21}(2',1')G^{12}(1',2)\bigr] 
\nonumber \\
&&\hspace{19.7mm}
=\theta(t_{1}-t_{2})\Sigma^{21(2)}(1,2)+\theta(t_{2}-t_{1})\Sigma^{12(2)}(1,2) .
\label{Sigma^(2)11}
\end{eqnarray}
Similarly, we have
\begin{equation}
\Sigma^{22(2)}(1,2)=\theta(t_{2}-t_{1})\Sigma^{21(2)}(1,2)+\theta(t_{1}-t_{2})\Sigma^{12(2)}(1,2) .
\label{Sigma^(2)22}
\end{equation}
Thus, we realize that the relations (\ref{Gcomp11}) and (\ref{Gcomp22}) between the diagonal and off-diagonal elements
of $\check{G}$ also hold among the elements of $\check{\Sigma}^{(2)}$.
They are also satisfied generally among the elements of $\check{\Sigma}^{(n)}$ ($n=3,4,\cdots$), as may be checked by inspecting higher-order contributions.
Combining this fact with the first-order result (\ref{Sigma^(1)}), we conclude that the diagonal elements of $\check{\Sigma}$ can be expressed generally as
\begin{subequations}
\label{Sigma-prop}
\begin{eqnarray}
&&\hspace{-10mm}
\Sigma^{11}(1,2)=\Sigma^{{\rm HF}}(1,2)
+\theta(t_{2}\!-\!t_{1})\Sigma^{12}(1,2)
+\theta(t_{1}\!-\!t_{2})\Sigma^{21}(1,2) ,
\label{Sigma-prop2}
\\
&&\hspace{-10mm}
\Sigma^{22}(1,2)=-\Sigma^{{\rm HF}}(1,2)+\theta(t_{1}\!-\!t_{2})\Sigma^{12}(1,2)
+\theta(t_{2}\!-\!t_{1})\Sigma^{21}(1,2) .
\label{Sigma-prop3}
\end{eqnarray}
\end{subequations}
Hence, it follows that, besides Eq.\ (\ref{checkSigma-symm}),
$\check{\Sigma}$ also satisfies
\begin{equation}
{\rm Tr}\check{\Sigma}(1,2)={\rm Tr}\,\check{\tau}_{1} \check{\Sigma}(1,2).
\label{checkSigma-symm2}
\end{equation}
Thus, $\check{\Sigma}$ carries the same symmetry as Eq.\ (\ref{checkG-symm}) of Green's function.
Equation (\ref{Sigma-prop}) tells us that we only need to calculate $\Sigma^{12}$ and $\Sigma^{21}$ of $\check{\Sigma}$ for $n\geq 2$.

\subsection{Keldysh rotation}

To see the structures of nonequilibrium Green's function (\ref{G-check-def}) and Dyson's equation (\ref{Dyson-check}) more clearly, it is sometimes convenient to perform the following ``Keldysh rotation'' to $\check{\tau}_{3}\check{G}$:\cite{Keldysh64,RS86,LO75}
\begin{eqnarray}
&&\hspace{-10mm}
\check{G}^{\rm K}\equiv \check{L}\check{\tau}_{3}
\check{G}\check{L}^{\dagger} 
= 
\left[
\begin{array}{cc}
G^{\rm R} & G^{\rm K} \\
0 & G^{\rm A}
\end{array}
\right] ,\hspace{10mm}
\check{L}\equiv \frac{1}{\sqrt{2}}\left[
\begin{array}{cc}
1 & -1 \\
1 & 1
\end{array}
\right].
\label{GK-check-def}
\end{eqnarray}
By using Eqs.\ (\ref{Gcomp}) and (\ref{Gcomp-symm2}), the new elements $G^{\rm R,A,K}$ can be expressed explicitly 
in terms of $G^{21}$ and $G^{12}$ as
\begin{subequations}
\label{G-RAK-def}
\begin{eqnarray}
&&\hspace{-5mm}
G^{\rm R}(1,2) 
= 
\theta (t_{1}- t_{2})
\bigl[G^{21}(1,2)-G^{12}(1,2)\bigr],
\label{G-R-def}
\\
&&\hspace{-5mm}
G^{\rm A}(1,2) 
= -
\theta (t_{2}- t_{1})
\bigl[G^{21}(1,2)-G^{12}(1,2)\bigr],
\label{G-A-def}
\\
&&\hspace{-5mm}
G^{\rm K}(1,2)=
G^{21}(1,2)+G^{12}(1,2) 
=-\frac{i}{\hbar}\langle \hat{\psi}_{\rm H}(1)\hat{\psi}_{\rm H}^{\dagger}(2)
\pm \hat{\psi}_{\rm H}^{\dagger}(2)\hat{\psi}_{\rm H}(1)\rangle .
\label{G-K-def}
\end{eqnarray}
\end{subequations}
The functions
$G^{\rm R}$ and $G^{\rm A}$ thus introduced are the retarded and advanced Green's functions, respectively.\cite{AGD63}
Using Eq.\ (\ref{Gcomp-symm1}), one may check that $G^{\rm R,A,K}$ satisfies
\begin{equation}
[ G^{\rm R}(1,2)]^{*}
=G^{\rm A}(2,1), \hspace{10mm}
[ G^{\rm K}(1,2)]^{*}
=-G^{\rm K}(2,1) ,
\label{G-RAK-symm}
\end{equation}
which is expressed alternatively in terms of $\check{G}^{\rm K}$ as
\begin{equation}
\check{G}^{{\rm K}}(1,2)=
\check{\tau}_{2}[\check{G}^{{\rm K}}(2,1)]^{\dagger}\check{\tau}_{2} .
\label{checkG-K-symm}
\end{equation}

It is convenient to carry out the same transformation for $\check{\Sigma}$ as
\begin{equation}
\check{\Sigma}^{\rm K}(1,2)\equiv \check{L}\check{\tau}_{3}
\check{\Sigma}(1,2)\check{L}^{\dagger}
=
\left[
\begin{array}{cc}
\Sigma^{\rm R}(1,2) & \Sigma^{\rm K}(1,2) \\
0 & \Sigma^{\rm A}(1,2)
\end{array}
\right] .
\label{SigmaK-check-def}
\end{equation}
Equation (\ref{Sigma-prop}) enables us to write down the elements of $\check{\Sigma}^{\rm K}$ as
\begin{subequations}
\label{Sigma-RAK-def}
\begin{eqnarray}
&&\hspace{-10mm}
\Sigma^{\rm R}(1,2)
= 
\Sigma^{\rm HF}(1,2)+\theta (t_{1}- t_{2})
\bigl[\Sigma^{21}(1,2)-\Sigma^{12}(1,2)\bigr],
\label{Sigma-R-def}
\\
&&\hspace{-10mm}
\Sigma^{\rm A}(1,2)
= \Sigma^{\rm HF}(1,2)-
\theta (t_{2}- t_{1})
\bigl[\Sigma^{21}(1,2)-\Sigma^{12}(1,2)\bigr],
\label{Sigma-A-def}
\\
&&\hspace{-10mm}
\Sigma^{\rm K}(1,2)
=\Sigma^{12}(1,2)+\Sigma^{21}(1,2) .
\label{Sigma-K-def}
\end{eqnarray}
\end{subequations}
It follows from Eq.\ (\ref{checkSigma-symm}) that
$\check{\Sigma}^{\rm K}$ also satisfies
\begin{equation}
\check{\Sigma}^{\rm K}(1,2)=\check{\tau}_{2}[\check{\Sigma}^{{\rm K}}(2,1)]^{\dagger}
\check{\tau}_{2}.
\label{checkSigma-K-symm}
\end{equation}
With these preliminaries, we perform the Keldysh rotation to Dyson's equation (\ref{Dyson-check}).
Let us multiply Eq.\ (\ref{Dyson-check}) by
$\check{L}\check{\tau}_{3}$ and $\check{L}^{\dagger}$ from the left and right, respectively,
and use $\check{L}\check{L}^{\dagger}\!=\!
\check{1}$ with $\check{1}$ the unit matrix. 
We thereby obtain Dyson's equation for 
$\check{G}^{\rm K}$ as
\begin{equation}
\left(i\hbar \frac{\partial }{\partial t_{1}}-\hat{K}_{1}\right)
\check{G}^{\rm K}(1,2)-
\int \check{\Sigma}^{\rm K}(1,3)
\check{G}^{\rm K}(3,2)\,d3 =\check{1}
\delta(1,2) .
\label{Dyson-K-check}
\end{equation}
It follows from Eqs.\ (\ref{GK-check-def}) and (\ref{SigmaK-check-def}) that: (i) the $(2,1)$ element of this equation
vanishes identically, and (ii) the equation satisfies a symmetry relation corresponding to Eqs.\ (\ref{checkG-K-symm}) and (\ref{checkSigma-K-symm}). Thus, the Keldysh rotation makes Dyson's equation more transparent. It should be noted at the same time that the perturbation expansion for $\check{G}$ or $\check{\Sigma}$ is performed more conveniently in terms of $G^{ij}$.

The transformation (\ref{GK-check-def}) adopted here is due to Larkin and Ovchinnikov\cite{LO75} and different in form from the original one by Keldysh.\cite{Keldysh64}. However, the physical contents are the same between the two transformations; moreover, the one by Larkin and Ovchinnikov has been used more frequently in condensed matter physics, especially in the field of superconductivity.\cite{RS86}

\section{Quantum transport equations and nonequilibrium entropy\label{sec:QTE}}

Dyson's equation (\ref{Dyson-check}) and Eq.\ (\ref{Sigma-Phi}) for the self-energy form a closed set of self-consistent equations for Green's  function $\check{G}$. Solving them, we can trace the nonequilibrium time evolution of the system. Over the last decade, active investigations have been performed in the field of high-energy physics to solve those equations numerically for several approximate $\Phi$'s without any further approximations.\cite{BC01,AB01,AABBS02,Berges02,Berges03,CDM03,JCG04,AST05,Berges06}

Here, we reduce those equations further to obtain quantum transport equations in the phase space based on the standard prescriptions of the Wigner transformation and a subsequent gradient expansion.
 We thereby remove some variables of Dyson's equations, which are irrelevant in many cases. The approximation was numerically checked to hold excellently over a wide range except for an initial time interval much shorter than the time scale for thermalization.\cite{Berges06} The resultant transport equation will be  shown to contain information of entropy flow, i.e., it enables us to define nonequilibrium entropy that evolves with time so as to be compatible with an equilibrium expression. Thus, we will be able to clarify on what conditions the concept of entropy holds. The derivation below will be based on Ref.\ \citen{Ivanov00} by Ivanov et al.\ and Ref.\ \citen{Kita06a}.

\subsection{Wigner representation\label{subsec:Wigner}}

The Wigner representation was introduced by Wigner in 1932 to consider quantum corrections to classical statistical mechanics.\cite{Wigner32,HOSW84}
The representation enables us to extend the concept of ``phase space' in classical statistical mechanics to quantum statistical mechanics. It also has a close connection with the Weyl quantization\cite{Weyl31} and was used by Moyal in 1949 in his statistical formulation of quantum mechanics in phase space.\cite{Moyal49}
Thus, the Wigner representation has had quite an impact in clarifying the foundation of quantum mechanics as well as its connection to classical mechanics. Moreover, it provides us with an indispensable tool for deriving quantum transport equations.

To be specific, the Wigner transformation to $\check{G}(1,2)$ is defined by
\begin{subequations}
\label{checkG-Wig}
\begin{equation}
\check{G}({\bm p}\varepsilon,{\bm r}_{12}t_{12})\equiv \int d^{3}\bar{r}_{12}\,d\bar{t}_{12}\,\check{G}(1,2)
\,e^{-i({\bm p}\,\cdot\bar{\bm r}_{12}-\varepsilon \bar{t}_{12})/\hbar} ,
\label{checkG-Wig1}
\end{equation}
where ${\bm r}_{12}\equiv({\bm r}_{1}+{\bm r}_{2})/2$, 
$t_{12}\equiv(t_{1}+t_{2})/2$, 
$\bar{\bm r}_{12}\equiv{\bm r}_{1}-{\bm r}_{2}$, and 
$\bar{t}_{12}\equiv t_{1}-t_{2}$. 
Equation (\ref{checkG-Wig1}) is nothing but the Fourier transform of $\check{G}(1,2)$ in terms of the ``relative'' coordinates $(\bar{\bm r}_{12},\bar{t}_{12})$. The quantity 
$\check{G}({\bm p}\varepsilon,{\bm r}_{12}t_{12})$ thus introduced is called the Wigner representation.
The inverse transformation of Eq.\ (\ref{checkG-Wig1}) is given by
\begin{equation}
\check{G}(1,2)=  \int\frac{d^{3}p\,d\varepsilon}{(2\pi\hbar)^{4}}\,
\check{G}({\bm p}\varepsilon,{\bm r}_{12}t_{12})
\,e^{i({\bm p}\cdot\bar{\bm r}_{12}-\varepsilon \bar{t}_{12})/\hbar} .
\label{checkG-Wig2}
\end{equation}
\end{subequations}
The function Wigner introduced in 1932 corresponds to the case with no time dependence. \cite{Wigner32} It is sometimes called the Wigner quasi-probability distribution with only $({\bm p},{\bm r})$ as arguments.
Note that $\check{G}({\bm p}\varepsilon,{\bm r}t)$ is different in form from $\check{G}(1,2)$; using the same symbol may not cause any confusion.
Below every $\check{G}$ given without arguments will denote $\check{G}({\bm p}\varepsilon,{\bm r}t)$.

Equation (\ref{Gcomp}) tells us that the independent components of 
$\check{G}$ are $G^{12}$ and $G^{21}$.
It is useful for later purposes to express their Wigner representations in terms of 
two alternative functions. Let us first define the spectral function $A$ by
\begin{equation}
A(1,2)\equiv i\left[G^{21}(1,2)-G^{12}(1,2)\right]=
\frac{1}{\hbar}
\langle \hat{\psi}_{\rm H}(1)\hat{\psi}_{\rm H}^{\dagger}(2)
\!\mp\! \hat{\psi}_{\rm H}^{\dagger}(2)\hat{\psi}_{\rm H}(1)\rangle ,
\label{A-def}
\end{equation}
which satisfies $A^{*}(1,2)\!=\!A(2,1)$, as shown with Eq.\ (\ref{Gcomp-symm1}). It follows from this symmetry and Eq.\ (\ref{commun1}) at equal times that the Wigner representation of $A(1,2)$ has the following properties:
\begin{equation}
A^{*}({\bm p}\varepsilon,{\bm r}t)=A({\bm p}\varepsilon,{\bm r}t),\hspace{10mm}
\int_{-\infty}^{\infty}\frac{d\varepsilon}{2\pi}\,A({\bm p}\varepsilon,{\bm r}t)
= 1 .
\label{A-sum}
\end{equation}
Equation (\ref{A-def}) extends the equilibrium spectral function in the Matsubara formalism\cite{AGD63,FW71} to nonequilibrium cases.
We next introduce the distribution $\phi$
 directly in the Wigner representation as
\begin{equation}
\phi({\bm p}\varepsilon,{\bm r}_{12}t_{12})
\equiv\frac{\displaystyle \int_{-\infty}^{\infty}\! d\bar{t}_{12} \!
\int \!d^{3}\bar{r}_{12} \,
\frac{1}{\hbar}\langle \hat{\psi}^{\dagger}_{\rm H}(2)\hat{\psi}_{\rm H}(1)\rangle
e^{-i({\bm p}\cdot\bar{\bm r}_{12}-\varepsilon \bar{t}_{12})/\hbar} 
}
{A({\bm p}\varepsilon,{\bm r}_{12}t_{12})} 
=\phi^{*}({\bm p}\varepsilon,{\bm r}_{12}t_{12}).
\label{phi-def}
\end{equation}
As will be seen later,  $\phi({\bm p}\varepsilon,{\bm r}t)$ in equilibrium reduces to the Bose or Fermi distribution function 
$f(\varepsilon)=(e^{\varepsilon/k_{\rm B}T}\mp 1 )^{-1}$.
The two functions $A$ and $\phi$ form an independent pair of functions in $\check{G}$ alternative to 
$G^{12}$ and $G^{21}$.  Thus, in nonequilibrium cases, we also need to determine the distribution function besides the spectral function.

In the conventional study of transport equations,\cite{KB62} one almost immediately integrates $A$ out completely with some approximations such as the quasiparticle and quasiclassical approximations. One thereby obtains a simplified equation for $\phi$ alone. However, there may be cases where the approximations are not appropriate, such as those where the energy-level spacing is large or the density of states changes substantially. 
We proceed here by retaining $A$ as it is, which also helps us to clarify the structure of the equations to be solved 
and its connection with the equilibrium formalism.

The two quantities $A$ and $\phi$ completely determine $\check{G}$. Indeed, it follows from Eqs.\ (\ref{Gcomp}), (\ref{A-def}), and (\ref{phi-def}) that $G^{12}$ and $G^{21}$ are expressed as
\begin{subequations}
\label{G-Wig}
\begin{eqnarray}
G^{12}({\bm p}\varepsilon,{\bm r}t)
&=& \mp i
A({\bm p}\varepsilon,{\bm r}t)\phi({\bm p}\varepsilon,{\bm r}t)  ,
\label{G_12-Wig}
\\
G^{21}({\bm p}\varepsilon,{\bm r}t)
&=&-iA({\bm p}\varepsilon,{\bm r}t)[1\pm \phi({\bm p}\varepsilon,{\bm r}t)]  .
\label{G_21-Wig}
\end{eqnarray}
\end{subequations}
Moreover, the Wigner representation of Eq.\ (\ref{G-RAK-def}) can also be written in terms of $A$ and $\phi$ as
\begin{subequations}
\label{G^RK-Wig}
\begin{eqnarray}
G^{{\rm R}}({\bm p}\varepsilon,{\bm r}t)&=&[G^{{\rm A}}({\bm p}\varepsilon,{\bm r}t)]^{*}
=
\int_{-\infty}^{\infty}\frac{d\varepsilon'}{2\pi}\,
\frac{A({\bm p}\varepsilon',{\bm r}t)}{\varepsilon_{+}-\varepsilon'} ,
\label{G^R-Wig}
\\
G^{{\rm K}}({\bm p}\varepsilon,{\bm r}t)
&=& -iA({\bm p}\varepsilon,{\bm r}t)[1\pm 2 \phi({\bm p}\varepsilon,{\bm r}t)]  ,
\label{G^K-Wig}
\end{eqnarray}
\end{subequations}
with $\varepsilon_{+}\equiv\varepsilon+i0_{+}$. 
Note that Eq.\ (\ref{G^R-Wig}) is a convolution
of $A$ and $\varepsilon_{+}^{-1}$, where $\varepsilon_{+}^{-1}$ is the Fourier transform of step function (\ref{step-fn}).
Thus, ${G}^{\rm K}$ contains information on the distribution function, whereas $G^{\rm R,A}$ can be expressed solely in terms of the spectral function.

The same transformation is also possible for the self-energy matrix $\check{\Sigma}$,
whose diagonal elements can be expressed in terms of the off-diagonal elements as Eq.\ (\ref{Sigma-prop}).
Let us write $\Sigma^{12}$ and $\Sigma^{21}$ as
\begin{subequations}
\label{Sigma-Wig}
\begin{eqnarray}
\Sigma^{12}({\bm p}\varepsilon,{\bm r}t)
&=&\mp i
A_{\Sigma}({\bm p}\varepsilon,{\bm r}t)\phi_{\Sigma}({\bm p}\varepsilon,{\bm r}t) ,
\label{Sigma_12-Wig}
\\
\Sigma^{21}({\bm p}\varepsilon,{\bm r}t)
&=&-iA_{\Sigma}({\bm p}\varepsilon,{\bm r}t)[1\pm \phi_{\Sigma}({\bm p}\varepsilon,{\bm r}t)] ,
\label{Sigma_21-Wig}
\end{eqnarray}
\end{subequations}
where $A_{\Sigma}$ and $\phi_{\Sigma}$ are two alternative independent functions.
It then follows from Eqs.\ (\ref{Sigma-HF}) and  (\ref{Sigma-RAK-def}) that the elements of $\check{\Sigma}^{\rm K}$ are expressed as
\begin{subequations}
\label{Sigma^RK-Wig}
\begin{eqnarray}
\Sigma^{{\rm R}}({\bm p}\varepsilon,{\bm r}t)
&=&[\Sigma^{{\rm A}}({\bm p}\varepsilon,{\bm r}t)]^{*}
=\Sigma^{{\rm HF}}({\bm p},{\bm r}t)
+\int_{-\infty}^{\infty}\frac{d\varepsilon'}{2\pi}\,
\frac{A_{\Sigma}({\bm p}\varepsilon',{\bm r}t)}{\varepsilon_{+}-\varepsilon'} ,
\label{Sigma^R-Wig}
\\
\Sigma^{{\rm K}}({\bm p}\varepsilon,{\bm r}t)
&=&-iA_{\Sigma}({\bm p}\varepsilon,{\bm r}t)[1\pm 2 \phi_{\Sigma}({\bm p}\varepsilon,{\bm r}t)] .
\label{Sigma^K-Wig}
\end{eqnarray}
\end{subequations}

Finally, it is worth pointing out that the operator:
\begin{equation}
G_{0}^{-1}(1,3)\equiv \left(i\hbar\frac{\partial}{\partial t_{1}}-\hat{K}_{1}\right)\delta(1,3),
\label{G_0^-1-real}
\end{equation}
where $\hat{K}_{1}$ is defined by Eq.\ (\ref{K_1}), 
can also be transformed into the Wigner representation as
\begin{equation}
G_{0}^{-1}({\bm p}\varepsilon,{\bm r}t)
\equiv\varepsilon-\frac{p^{2}}{2m}-U({\bm r}t)+\mu .
\label{G_0^-1}
\end{equation}
Equation (\ref{G_0^-1}) is useful when converting Eq.\ (\ref{Dyson-K-check}) into the Wigner representation.

\subsection{Groenewold-Moyal product and gradient expansion}

We have introduced the Wigner representation as Eq.\ (\ref{checkG-Wig1}), where $\check{G}(1,2)$ can be regarded as a matrix with the arguments $1$ and $2$. With this viewpoint, the self-energy term in Dyson's equation (\ref{Dyson-K-check}) is a matrix product. Hence, we  need to clarify how a matrix product changes through the Wigner transformation before expressing Eq.\ (\ref{Dyson-K-check}) in the Wigner representation.
The answer to this question is given in terms of two arbitrary matrices $C(1,2)$ and $D(1,2)$ by
\begin{equation}
\int C(1,3)D(3,2) \,d3
=\int\frac{d^{3}p\,d\varepsilon}{(2\pi\hbar)^{4}} 
C({\bm p}\varepsilon,{\bm r}_{12}t_{12})
\ast
D({\bm p}\varepsilon,{\bm r}_{12}t_{12}) 
\,e^{i({\bm p}\cdot
\bar{\bm r}_{12}-\varepsilon\bar{t}_{12})/\hbar}.
\label{Identity-otimes}
\end{equation}
Here, the operator $\ast$ is defined in terms of ${\bm \partial}_{{\bm r}}\!\equiv\!\partial/\partial{\bm r}$
 and $\partial_{t}\!\equiv\!\partial/\partial t$ by
\begin{eqnarray}
C({\bm p}\varepsilon,{\bm r}t)\ast
D({\bm p}\varepsilon,{\bm r}t)
&\equiv& C({\bm p}\varepsilon,{\bm r}t)
\exp\!\left[\,\frac{i\hbar}{2}\!
\left(\overleftarrow{\bm \partial}_{{\bm r}}\!\cdot\!\overrightarrow{\bm \partial}_{{\bm p}}
-\overleftarrow\partial_{t}\overrightarrow{\partial}_{\varepsilon}
-\overleftarrow{\bm \partial}_{{\bm p}}\!\cdot\!\overrightarrow{\bm \partial}_{{\bm r}}
+\overleftarrow{\partial}_{\varepsilon}\overrightarrow{\partial}_{t}\right)
\right]
\nonumber \\
& & \times D({\bm p}\varepsilon,{\bm r}t) ,
\label{Moyal}
\end{eqnarray}
where the left (right) arrow on each differential operator denotes that it works on the left (right) function.
Equation (\ref{Identity-otimes}) was derived independently by 
Groenewold in 1946\cite{Groenewold46} and by Moyal in 1949.\cite{Moyal49}
We call it the Groenewold-Moyal product.

By considering only the space coordinates for simplicity, Eq.\ (\ref{Identity-otimes}) is proved as follows. Let us express
$C({\bm r}_{1},{\bm r}_{3})$ and $D({\bm r}_{3},{\bm r}_{2})$ on the left-hand side as Eq.\ (\ref{checkG-Wig2}).
We next expand the ``center-of-mass'' coordinates ${\bm r}_{13}={\bm r}_{12}+ \frac{1}{2}\bar{\bm r}_{32}$ and
${\bm r}_{32}={\bm r}_{12}- \frac{1}{2}\bar{\bm r}_{13}$ in Taylor series from ${\bm r}_{12}$ as 
$C({\bm p},{\bm r}_{12}+ \frac{1}{2}\bar{\bm r}_{32})
=\exp\bigl[\frac{1}{2}\bar{\bm r}_{32}\cdot{\bm \partial}_{{\bm r}_{12}}\bigr]C({\bm p},{\bm r}_{12})$ and $D({\bm p}',{\bm r}_{12}- \frac{1}{2}\bar{\bm r}_{13})
=\exp\bigl[-\frac{1}{2}\bar{\bm r}_{13}\cdot{\bm \partial}_{{\bm r}_{12}}\bigr]D({\bm p}',{\bm r}_{12})$.
We then remove the relative coordinates $\bar{\bm r}_{32}$ and
$\bar{\bm r}_{13}$ in the expansions as
$e^{i{\bm p}\cdot\bar{\bm r}_{13}/\hbar}
\exp\bigl[-\frac{1}{2}\bar{\bm r}_{13}\cdot{\bm \partial}_{{\bm r}_{12}}\bigr]D({\bm p}',{\bm r}_{12})
=
\exp\bigl[\frac{i\hbar}{2}\partial_{{\bm p}}\cdot{\bm \partial}_{{\bm r}_{12}}\bigr]e^{i{\bm p}\cdot\bar{\bm r}_{13}/\hbar}D({\bm p}',{\bm r}_{12})$, and subsequently perform partial integrations over 
${\bm p}$ and ${\bm p}'$ as $C({\bm p},{\bm r}_{12})\exp\bigl[\frac{i\hbar}{2}\partial_{{\bm p}}\cdot{\bm \partial}_{{\bm r}_{12}}\bigr]e^{i{\bm p}\cdot\bar{\bm r}_{13}/\hbar}\longrightarrow 
C({\bm p},{\bm r}_{12})\exp\bigl[-\frac{i\hbar}{2}\overleftarrow{\partial}_{{\bm p}}\cdot{\bm \partial}_{{\bm r}_{12}}\bigr]e^{i{\bm p}\cdot\bar{\bm r}_{13}/\hbar}$. Finally, we carry out the integration over ${\bm r}_{3}$
with only plane waves in the integrand to obtain the factor $\delta_{{\bm p}{\bm p}'}e^{i{\bm p}\cdot\bar{\bm r}_{12}}$.

Retaining up to the first order in the gradient expansion of Eq.\ (\ref{Moyal}) yields
\begin{equation}
C\!\ast\! D
\approx CD+\frac{i\hbar}{2}\{C,D\} ,
\label{otimes1}
\end{equation}
where $\{C,D\}$ denotes the generalized Poisson bracket:
\begin{equation}
\{C,D\}\equiv\frac{\partial C}{\partial{\bm r}}\!\cdot\!
\frac{\partial D}{\partial{\bm p}}
-\frac{\partial C}{\partial t}\,
\frac{\partial D}{\partial\varepsilon}
-\frac{\partial C}{\partial{\bm p}}\!\cdot\!
\frac{\partial D}{\partial{\bm r}}
+\frac{\partial C}{\partial \varepsilon}\,
\frac{\partial D}{\partial t}.
\label{Poisson}
\end{equation}
This approximation will be called the first-order gradient expansion.
It will hold excellently when the microscopic length scale $l_{\rm m}$ and time scale $t_{\rm m}$
of the system are much smaller than the macroscopic length $l_{\rm M}$ and time $t_{\rm M}$
characterizing the inhomogeneity of the system.
For a system of low-temperature fermions with the Fermi wave length $k_{\rm F}$ and Fermi energy $\varepsilon_{\rm F}$, for example, the microscopic length and time are given by
$l_{m}\sim 1/k_{\rm F}$ and $t_{m}\sim \hbar/\varepsilon_{\rm F}$, and those in a classical gas
are the mean spacing between two particles and 
$\hbar/k_{\rm B}T$, respectively.
When the space-time variations of the system occur over scales much longer than them, 
Eq.\ (\ref{otimes1}) will hold excellently.
This condition is certainly met over a wide range of nonequilibrium systems.
Indeed, the approximation was numerically checked to hold over a wide range except for an initial time interval much shorter than the time scale for thermalization.\cite{Berges06}

\subsection{Quantum transport equations\label{subsec:QTE}}

With these preliminaries, we now derive quantum transport equations
from Dyson's equation  (\ref{Dyson-K-check}).
It is useful for this purpose to write its first term
as a matrix product with Eq.\ (\ref{G_0^-1-real}).
Then, we can use Eq.\ (\ref{Identity-otimes}) to obtain the Wigner representation of Eq.\  
(\ref{Dyson-K-check}) as
\begin{equation}
({G}_{0}^{-1}\check{1}-\check{\Sigma}^{\rm K})\ast \check{G}^{\rm K}=\check{1},
\label{Dyson-K-Wig0}
\end{equation}
where $G_{0}^{-1}$ is defined by Eq.\ (\ref{G_0^-1}).

We next adopt approximation (\ref{otimes1}) for Eq.\ (\ref{Dyson-K-Wig0}).
By noting Eqs.\ (\ref{GK-check-def}) and (\ref{SigmaK-check-def}), the (1,1) element of the resultant equation can be written down explicitly as
\begin{subequations}
\label{Dyson-R-Wig}
\begin{equation}
(G_{0}^{-1}-\Sigma^{{\rm R}})G^{{\rm R}}
+\frac{i\hbar}{2}\{G_{0}^{-1}-\Sigma^{{\rm R}},G^{{\rm R}}\}
=1  .
\label{Dyson-Rl-Wig}
\end{equation}
The (2,2) element is obtained from the above with the  replacement R$\rightarrow$A
in the superscripts. 
Taking its complex conjugate subsequently with $G^{{\rm A}*}=G^{{\rm R}}$ in mind,
we have
\begin{equation}
(G_{0}^{-1}-\Sigma^{{\rm R}})G^{{\rm R}}
-\frac{i\hbar}{2}\{G_{0}^{-1}-\Sigma^{{\rm R}},G^{{\rm R}}\}
=1  .
\label{Dyson-Rr-Wig}
\end{equation}
\end{subequations}
This equation can also be derived directly from the (1,1) element of the right-hand Dyson's equation:
$\check{G}^{\rm K}\ast
({G}_{0}^{-1}\check{1}-\check{\Sigma}^{\rm K})=\check{1}$.
Let us add the above two equations. 
We then obtain
\begin{equation}
G^{{\rm R}}=(G_{0}^{-1}-\Sigma^{{\rm R}})^{-1}  .
\label{G^R}
\end{equation}
Noting Eq.\ (\ref{G^R-Wig}), we realize that this equation determines the 
spectral function $A$ for a given $\Sigma^{{\rm R}}$.

Equation (\ref{G^R}) has been derived by adding Eqs.\ (\ref{Dyson-Rl-Wig}) and (\ref{Dyson-Rr-Wig}).
However, the (1,1) and (2,2) elements of original
Eq.\ (\ref{Dyson-K-check}) are complex conjugates to each other so that they are equivalent mathematically.
Hence, one may ask whether this equivalence is still kept between Eqs.\ (\ref{Dyson-Rl-Wig}) and (\ref{Dyson-Rr-Wig}) obtained with the first-order gradient expansion.
To answer this question raised by Ivanov et al.,\cite{Ivanov00}
we alternatively subtract Eq.\ (\ref{Dyson-Rr-Wig}) from Eq.\ (\ref{Dyson-Rl-Wig}) and subsequently substitute
Eq.\ (\ref{G^R}). It yields
$$
0=\{(G^{{\rm R}})^{-1},G^{{\rm R}}\}=
-(G^{{\rm R}})^{-2}\{G^{{\rm R}},G^{{\rm R}}\},$$
which is just the identity $\{G^{{\rm R}},G^{{\rm R}}\}=0$.
Thus, Eqs.\ (\ref{Dyson-Rl-Wig}) and (\ref{Dyson-Rr-Wig}) have been checked to be identical also within the 
first-order gradient expansion.
This consideration shows that 
$G^{\rm R}$ from Eq.\ (\ref{G^R}) is correct up to the first order.

We next focus on the (1,2) element of Eq.\ (\ref{Dyson-K-Wig0}).
Noting Eqs.\ (\ref{GK-check-def}) and (\ref{SigmaK-check-def}) and adopting first-order gradient expansion (\ref{otimes1}), we obtain
\begin{subequations}
\label{Dyson-K-Wig}
\begin{equation}
(G_{0}^{-1}-\Sigma^{{\rm R}})G^{{\rm K}}-\Sigma^{{\rm K}}G^{{\rm A}}
+\frac{i\hbar}{2}\{G_{0}^{-1}-\Sigma^{{\rm R}},G^{{\rm K}}\}
-\frac{i\hbar}{2}\{\Sigma^{{\rm K}},G^{{\rm A}}\}
=0 .
\label{Dyson-Kl-Wig}
\end{equation}
Its complex conjugate can be expressed with Eqs.\ (\ref{G^RK-Wig}) and (\ref{Sigma^RK-Wig})
as
\begin{equation}
-(G_{0}^{-1}-\Sigma^{{\rm A}})G^{{\rm K}}+\Sigma^{{\rm K}}G^{{\rm R}}
+\frac{i\hbar}{2}\{G_{0}^{-1}-\Sigma^{{\rm A}},G^{{\rm K}}\}
-\frac{i\hbar}{2}\{\Sigma^{{\rm K}},G^{{\rm R}}\}
=0  .
\label{Dyson-Kr-Wig}
\end{equation}
\end{subequations}
Equation (\ref{Dyson-Kr-Wig}) may also be derived directly from the (1,2) element of the right-hand Dyson's equation $\check{G}^{\rm K}\ast
({G}_{0}^{-1}\check{1}-\check{\Sigma}^{\rm K})=\check{1}$.

Let us add Eqs.\ (\ref{Dyson-Kl-Wig}) and (\ref{Dyson-Kr-Wig}) and make use of
${\rm Im}\{G_{0}^{-1}-\Sigma^{{\rm R}},G^{{\rm R}}\}=0$ as well as 
Eqs.\ (\ref{G^RK-Wig}) and (\ref{Sigma^RK-Wig}).
We thereby obtain
$$
\{G_{0}^{-1}-{\rm Re}\Sigma^{{\rm R}},A \phi\}-\{A_{\Sigma}\phi_{\Sigma},{\rm Re}G^{{\rm R}}\}
=\frac{AA_{\Sigma}(\phi_{\Sigma}-\phi)}{\hbar}.
$$
The left-hand side of this equation consists only of space-time derivatives, which are first-order in the gradient expansion.
Hence, it follows that the term $\phi_{\Sigma}-\phi$ on the right-hand side
 is also of first-order;
 as seen below in \S\ref{$H$-theorem}, the right-hand side may be identified as the collision integral in the transport theory, which vanishes in equilibrium.
Hence, it follows that we may replace $\phi_{\Sigma}$ by
$\phi$ on the left-hand side as adopted by Botermans and Malfliet.\cite{BM90,Ivanov00}
The approximation yields 
\begin{equation}
\{G_{0}^{-1}-\!{\rm Re}\Sigma^{{\rm R}},A \phi\}-\{A_{\Sigma} \phi,{\rm Re}G^{{\rm R}}\}
={\cal C},
\label{transport}
\end{equation}
where ${\cal C}$ is the collision integral defined by
\begin{equation}
{\cal C}\equiv \frac{AA_{\Sigma}(\phi_{\Sigma}-\phi)}{\hbar}=
\mp\frac{G^{21}\Sigma^{12}-G^{12}\Sigma^{21}}{\hbar} .
\label{collision}
\end{equation}
Equation (\ref{transport}) determines the distribution function $\phi$ for given 
$A$ and $\check{\Sigma}$.

Alternatively, one may subtract Eq.\ (\ref{Dyson-Kr-Wig}) from Eq.\ (\ref{Dyson-Kl-Wig}).
Then, a procedure similar to that for deriving Eq.\ (\ref{transport}) yields
\begin{equation}
\frac{\{A_{\Sigma} \phi,A\}-\{A_{\Sigma},A \phi\}}{4}
=\frac{A_{\Sigma} \phi_{\Sigma}{\rm Re}G^{{\rm R}}
-(G_{0}^{-1}\!-\!{\rm Re}\Sigma^{{\rm R}})A \phi}{\hbar}.
\label{transport-d}
\end{equation}
This equation is nothing but Eq.\ (\ref{transport}) in disguise. Indeed, 
multiplying Eq.\ (\ref{transport-d}) by
$A_{\Sigma}/(G_{0}^{-1}\!-\!{\rm Re}\Sigma^{{\rm R}})$ reproduces Eq.\ (\ref{transport}).
This can be checked with Eq.\ (\ref{G^R}) by expressing ${\rm Re}G^{{\rm R}}$ and $A=-2{\rm Im}G^{{\rm R}}$ in terms of 
$M\!\equiv\!G_{0}^{-1}\!-\!{\rm Re}\Sigma^{{\rm R}}$ and 
$A_{\Sigma}\!\equiv\! -2{\rm Im}\Sigma^{\rm R}$
as
$$
{\rm Re}G^{{\rm R}}=\frac{M}{M^{2}+(A_{\Sigma}/2)^{2}},\hspace{10mm}
A=\frac{A_{\Sigma}}{M^{2}+(A_{\Sigma}/2)^{2}},
$$
and substituting them into the two apparently different equations.
Thus, we have seen that the equivalence between Eqs.\ (\ref{Dyson-Kl-Wig}) and (\ref{Dyson-Kr-Wig})
is recovered appropriately with the replacement $\phi_{\Sigma}\!\rightarrow\! \phi$ in the space-time derivatives.

Equations (\ref{G^R}) and (\ref{transport}) form a closed set of equations for the two unknown functions 
$A$ and $\phi$.
Moreover, the replacement $\phi\rightarrow f(\varepsilon)\equiv
(e^{\varepsilon/k_{\rm B}T}\mp 1)^{-1}$
in those equations transforms: (i) Eq.\ (\ref{G^R}) into Dyson's equation for the retarded Green's function in  equilibrium;
and (ii) Eq.\ (\ref{transport}) into the trivial relation $0=0$, i.e., 
both the space-time derivatives and collision integral become null with the replacement (see \S \ref{$H$-theorem}
below).
Thus, Eqs.\ (\ref{G^R}) and (\ref{transport}) form a natural extension of the equilibrium Dyson's equation into nonequilibrium cases,
where we also have to determine the distribution function $\phi$ with Eq.\ (\ref{transport}).

Finally, it is worth pointing out that a Wigner representation may not be positive definite in general.  
However, the following argument suggests that the spectral function $A$ and distribution function $\phi$ obtained using Eqs.\ (\ref{G^R}) and (\ref{transport}) are both positive.
First, they are real, as Eqs.\ (\ref{A-sum}) and (\ref{phi-def}) show. Second, it follows from the retarded nature of $G^{{\rm R}}(1,2)$ in Eq.\ (\ref{G-R-def})
that all the singularities of $G^{{\rm R}}({\mib p}\varepsilon,{\mib r}t)$
in Eq.\ (\ref{G^R}) lie on the lower half of the complex $\varepsilon$ plane.
This implies that ${\rm Im}\Sigma^{{\rm R}}({\mib p}\varepsilon,{\mib r}t)
\leq 0$ and, hence, 
${\rm Im}G^{{\rm R}}({\mib p}\varepsilon,{\mib r}t)\leq 0$. 
By using Eq.\ (\ref{G^R-Wig}), the latter condition can be written 
alternatively as
\begin{equation}
A({\mib p}\varepsilon,{\mib r}t)\geq 0 .
\end{equation}
As for $\phi$, it reduces in equilibrium to the Bose/Fermi distribution function, which is definitely positive.
Noting that it obeys Eq.\ (\ref{transport}) of the first-order gradient expansion
so that its deviation from the equilibrium form will not be substantial, $\phi$ is also expected to be positive.

\subsection{Approximation for the self-energies\label{subsec:local}}

We now discuss how to calculate the self-energies within the first-order gradient expansion
of the self-consistent $\Phi$-derivable approximation.
In brief, they should be estimated with the local approximation
of neglecting the space-time derivatives.
This may be realized by noting: (i) Eq.\ (\ref{G^R})
is correct up to the first order in the gradient expansion
even with $\Sigma^{\rm R}$ of the local approximation, as pointed out in the paragraph just below Eq.\ 
(\ref{G^R}); (ii) Eq.\ (\ref{transport}), which is composed of space-time derivatives and the collision integral, is first order by itself so that we should make use of the zeroth-order self-energies in the equation.

We now write down the self-energies in the local approximation
up to the second order in the perturbation expansion. 
Using Eq.\ (\ref{V-Fourier}), we can transform Eq.\ (\ref{Sigma-HF})
in the Hartree-Fock approximation 
into the Wigner representation as 
\begin{equation}
\Sigma^{{\rm HF}}({\bm p},{\bm r}t)=\pm i\hbar
\int \frac{d^{3}p'd\varepsilon'}{(2\pi \hbar)^{4}}
(V_{\bm 0}\pm V_{{\bm p}-{\bm p}'}) 
G^{12}({\bm p}'\varepsilon',{\bm r}t) .
\label{Sigma-HF-Wig}
\end{equation}
Similarly, second-order self-energy (\ref{Sigma^(2)}) may be expressed with Eq.\ (\ref{Identity-otimes}) in the Wigner representation. Adopting the local approximation subsequently, we obtain
\begin{eqnarray}
\Sigma^{ij(2)}({\bm p}\varepsilon,{\bm r}t)
&=& \mp (\hbar)^{2} \! 
\int\left[\prod_{k=2}^{4}\frac{d^{3}p_{k}d\varepsilon_{k}}{(2\pi\hbar)^{4}}\right]
\frac{1}{2}|V_{{\bm p}-{\bm p}_{3}}\pm V_{{\bm p}-{\bm p}_{4}}|^{2}
(2\pi\hbar)^{4}
\delta({\bm p}\!+\!{\bm p}_{2}\!-\!{\bm p}_{3}\!-\!{\bm p}_{4})
\nonumber \\
& &
\times \delta(\varepsilon\!+\!\varepsilon_{2}\!-\!\varepsilon_{3}\!-\!\varepsilon_{4})
G^{ji}({\bm p}_{2}\varepsilon_{2},{\bm r}t)G^{ij}({\bm p}_{3}\varepsilon_{3},{\bm r}t)
G^{ij}({\bm p}_{4}\varepsilon_{4},{\bm r}t).
\label{Sigma^(2)-Wig}
\end{eqnarray}
By writing $G^{12}\!=\!\mp  i G^{<}$, $G^{21}\!=\!- iG^{>}$,
$\Sigma^{12}\!=\!\mp i \Sigma^{<}$, and $\Sigma^{21}\!=\!- i\Sigma^{>}$, 
this expression acquires exactly the same expression as Eq.\ (4-16) of Kadanoff and Baym.\cite{KB62}

Having given the self-energies as a functional of $G^{ij}$ in the Wigner representation, we can solve Eqs.\ (\ref{G^R}) and (\ref{transport}) self-consistently to trace their time evolutions. The exact procedure is summarized as follows: 
(i) With $\phi({\bm p}\varepsilon,{\bm r}t)$ given at a certain time $t$, 
we calculate the spectral function $A({\bm p}\varepsilon,{\bm r}t)$ self-consistently using Eq.\
(\ref{G^R}) and $\Sigma^{\rm R}=\Sigma^{\rm R}({\bm p}\varepsilon,{\bm r}t;A,\phi)$ of the local $\Phi$-derivable approximation. (ii) The distribution function $\phi$ at $t+dt$ is 
determined subsequently with Eq.\ (\ref{transport}).
Hence, it follows that the explicit time dependence in the procedure originates from $\phi$.

\subsection{Expression of entropy density\label{subsec:entropy-density}}

Equation (\ref{transport}), which describes the time evolution of the distribution function, also
contains information on the heat flow. Indeed, it enables us to derive an expression of nonequilibrium entropy density and its continuity equation. To see this, let us multiply Eq.\ (\ref{transport}) by $\hbar k_{\rm B}(2\pi\hbar)^{-4}
\ln [(1\pm \phi)/\phi]$, carry out an integration over 
${\bm p}\varepsilon$, and make use of ${\rm Im}\{G_{0}^{-1}\!-\!\Sigma^{{\rm R}},G^{{\rm R}}\}\!=\!0$
and $
\ln[(1\pm \phi)/\phi]d\phi\!=\! d\sigma$ with
\begin{equation}
\sigma[\phi] \equiv -\phi\ln\phi\pm (1\pm \phi)\ln (1\pm \phi) .
\label{sigma-def}
\end{equation}
We thereby obtain
\begin{equation}
\frac{\partial s}{\partial t}+{\bm\nabla}\cdot{\bm j}_{s}=\frac{\partial s_{\rm coll}}{\partial t},
\label{continuity-s}
\end{equation}
where $s\!=\!s({\bm r}t)$, ${\bm j}_{s}\!=\!{\bm j}_{s}({\bm r}t)$, and
${\partial s_{\rm coll}}/{\partial t}$ are defined by
\begin{eqnarray}
s&\equiv& \hbar k_{\rm B} \!\int \!\frac{d^{3}p\,d\varepsilon}{(2\pi\hbar)^{4}}
\sigma
\!\left[A\frac{\partial (G_{0}^{-1}\!-\!{\rm Re}\Sigma^{{\rm R}})}{\partial \varepsilon}
+A_{\Sigma} \frac{\partial {\rm Re}G^{{\rm R}}}{\partial \varepsilon}\right] \! ,
\label{entropy-density}
\\
{\bm j}_{s}&\equiv& \hbar k_{\rm B} \!\int \!\frac{d^{3}p\,d\varepsilon}{(2\pi\hbar)^{4}}
\sigma
\!\left[-A\frac{\partial (G_{0}^{-1}\!-\!{\rm Re}\Sigma^{{\rm R}})}{\partial {\bm p}}
-A_{\Sigma} \frac{\partial {\rm Re}G^{{\rm R}}}{\partial {\bm p}}\right] \! ,
\label{entropy-current}
\\
\frac{\partial s_{\rm coll}}{\partial t}&\equiv& \hbar k_{\rm B}
\int\frac{d^{3}p\,d\varepsilon}{(2\pi\hbar)^{4}}\,
{\cal C}\ln\frac{1\pm \phi}{\phi} .
\label{ds_col}
\end{eqnarray}
Equations (\ref{entropy-density}) and (\ref{entropy-current}) may be identified as entropy density and entropy flux density.
Indeed, if we put $\phi\rightarrow f(\varepsilon)\equiv
(e^{\varepsilon/k_{\rm B}T}\mp 1)^{-1}$ in Eq.\ (\ref{entropy-density}) and perform an
integration over ${\bm r}$, 
we reproduce an expression of equilibrium entropy derived in Ref.\ \citen{Kita99};
see also Appendix \ref{sec:S_eq} for the equilibrium expression.
On the other hand, Eq.\ (\ref{ds_col}) denotes entropy production per unit time and unit volume due to collisions; see the next subsection on this point.
Thus, we have been able to define nonequilibrium entropy density so as to be compatible with equilibrium statistical mechanics.
As will be shown in \S \ref{QPA}, it also embraces Boltzmann's nonequilibrium entropy density as the dilute high-temperature limit.

Expression (\ref{entropy-density}) is different from the one obtained earlier by Ivanov et al.\cite{Ivanov00} which contains extra terms stemming from space-time derivatives in the self-energies, i.e.,  terms due to memory effects.
The corresponding expression of nonequilibrium entropy density does not reduce adequately to the entropy density in equilibrium\cite{Kita99,Kita06a} as $\phi\rightarrow f(\varepsilon)\equiv
(e^{\varepsilon/k_{\rm B}T}\mp 1)^{-1}$.
It was derived so as to be compatible with the equilibrium entropy by Carneiro and Pethick.\cite{CP75}  However, the derivation of Carneiro and Pethick was based on the zero-temperature perturbation expansion technique of Goldstone\cite{FW71} and suffers from an inappropriate treatment of energy denominators, as pointed out in Ref.\ \citen{Kita06a}.
Thus, it is Eq.\ (\ref{entropy-density}) that is compatible with equilibrium statistical mechanics.

\subsection{$H$-theorem and thermodynamic equilibrium\label{$H$-theorem}}

Here, we show that Eq.\ (\ref{continuity-s}) satisfies the $H$-theorem, i.e., the law of increase in entropy, 
within the second-order perturbation expansion.

Let us substitute Eq.\ (\ref{Sigma^(2)-Wig}) into Eq.\ (\ref{collision})
and the corresponding expression of ${\cal C}$ subsequently into Eq.\ (\ref{ds_col}).
Using Eq.\  (\ref{G-Wig}),  we further express the result in terms of $A$ and $\phi$
and symmetrize the formula with respect to integration variables. 
The second-order quantity ${\partial s_{\rm coll}^{(2)}}/{\partial t}$ is thereby transformed into
\begin{eqnarray}
&&\hspace{0mm}
\frac{\partial s_{\rm coll}^{(2)}}{\partial t}
=\frac{k_{\rm B}\hbar^{2}}{8}  \prod_{j=1}^{4} \int\!
d^{3}p_{j}\,d\varepsilon_{j}
|V_{{\bm p}_{1}-{\bm p}_{3}}\!\pm \! V_{{\bm p}_{1}-{\bm p}_{4}}|^{2}
\delta({\bm p}_{1}\!+\!{\bm p}_{2}\!-\!{\bm p}_{3}\!-\!{\bm p}_{4})
\delta(\varepsilon_{1}\!+\!\varepsilon_{2}\!-\!\varepsilon_{3}\!-\!\varepsilon_{4})
\nonumber \\
&&\hspace{14mm}
\times
 A_1A_2
A_3A_4
\bigl[(1\!\pm \!\phi_1)(1\pm\phi_2)\phi_{3}\phi_4
\!-\! \phi_1\phi_2(1\pm\phi_3)(1\pm\phi_4)
\bigr]
\nonumber \\
&&\hspace{14mm}
\times\ln\frac{(1\pm\phi_1)(1\pm\phi_2)\phi_{3}\phi_4}
{\phi_1\phi_2(1\pm\phi_3)(1\pm\phi_4)},
\label{ds_col/dt}
\end{eqnarray}
with $A_{j}=A({\bm p}_{j}\varepsilon_{j},{\bm r}t)$, etc.
Noting $(x-y)\ln(x/y)\geq 0$, which holds for any positive $x$ and $y$, 
we conclude ${\partial s_{\rm coll}^{(2)}}/{\partial t}\geq 0$.
This is nothing but a quantum-mechanical extension of Boltzmann's $H$-theorem.\cite{Boltzmann72,Cercignani88}
The equality in Eq.\ (\ref{ds_col/dt}) holds when
\begin{eqnarray*}
\ln \frac{1\pm\phi_1}{\phi_1}+\ln \frac{1\pm\phi_2}{\phi_2}
-\ln \frac{1\pm\phi_3}{\phi_3}-\ln \frac{1\pm\phi_4}{\phi_4}=0
\end{eqnarray*}
is satisfied.
Noting the delta functions in Eq.\  (\ref{ds_col/dt}),
we realize that $\ln [({1\!\pm\! \phi_1})/{\phi_1}]$ must have linear ${\bm p}_{1}$ and $\varepsilon_{1}$ dependences as
\begin{equation}
\ln \frac{1\pm\phi_1}{\phi_1}=\alpha + \beta (\varepsilon_{1}-{\bm v}\cdot {\bm p}_{1}),
\end{equation}
where $\alpha$, $\beta$ and ${\bm v}$ are arbitrary functions of ${\bm r}t$.
We thereby obtain, for $\phi=\phi^{({\rm le})}$ that makes 
the collision integral vanish, the expression:
\begin{equation}
\phi^{({\rm le})}({\bm p}\varepsilon,{\bm r}t)=\frac{1}{e^{\beta({\bm r}t) 
[\,\varepsilon\,-\,{\bm v}({\bm r}t)\,\cdot\, {\bm p}\,]+\alpha({\bm r}t)}\mp 1},
\label{phi^le}
\end{equation}
which is exactly the local equilibrium distribution.

Thus, we have proved the $H$-theorem within the second-order perturbation
and also provided an expression for the distribution function $\phi^{({\rm le})}$ that makes
entropy production vanish.
It can also be shown easily within the second-order perturbation that this $\phi^{({\rm le})}$ makes the collision integral (\ref{collision}) vanish.

It still remains to be clarified whether ${\partial s_{\rm coll}}/{\partial t}\geq 0$ does hold
up to the infinite order. However, it seems quite natural physically to expect it.

\section{Quasiparticle and quasiclassical approximations}

In many cases, quantum transport equations (\ref{G^R}) and (\ref{transport}) are
further simplified by integrating out the spectral function $A$ completely.
There are a couple of standard approximations to carry it out, i.e.,
the quasiparticle and quasiclassical approximations.
We explain them here in detail using second-order self-energy (\ref{Sigma^(2)-Wig}).
It is thereby shown that quantum transport equations embrace the Boltzmann equation as a limit.
It will also become clear under what conditions the concept of ``distribution function in the phase space'',
which is frequently assumed to exist from the very beginning in the classical transport theory,\cite{Cercignani88}
is justified.

\subsection{Quasiparticle approximation\label{QPA}}

This approximation becomes excellent whenever there is a sharp $\delta$-function-like peak 
in the spectral function $A$. 
We first consider the weak-coupling limit as a typical example.
Here, we may drop the self-energy in Eq.\ (\ref{G^R}). Hence,
the spectral function $A=-2{\rm Im}G^{\rm R}$ is obtained from Eq.\
(\ref{G_0^-1}) as
\begin{equation}
A({\bm p}\varepsilon,{\bm r}t)\approx 2\pi 
\delta\left(\varepsilon-\xi_{\bm p}\right),\hspace{10mm}
\xi_{\bm p}({\bm r}t)\equiv \frac{{\bm p}^{2}}{2m}+U({\bm r}t)-\mu .
\label{A-QPA}
\end{equation}
As for Eq.\ (\ref{transport}), we can neglect 
terms with $\Sigma^{\rm R}$ on its left-hand side to the same order of approximation.
By noting Eqs.\ (\ref{G_0^-1}) and (\ref{Poisson}),
the resultant equation can be written down explicitly as
\begin{equation}
\frac{\partial (A\phi)}{\partial t}+\frac{{\bm p}}{m} \cdot
\frac{\partial (A\phi)}{\partial {\bm r}}
+\frac{\partial U}{\partial t}
\frac{\partial (A\phi)}{\partial \varepsilon}
-\frac{\partial U}{\partial {\bm r}}\cdot
\frac{\partial (A\phi)}{\partial {\bm p}}={\cal C}.
\label{transport-QPA0}
\end{equation}
We next divide Eq.\ (\ref{transport-QPA0}) by $2\pi$ and perform an integration over 
$-\infty\leq\varepsilon\leq \infty$.
The corresponding equation can be expressed concisely in terms of the distribution function
in the phase space:
\begin{equation}
f({\bm p},{\bm r},t)\equiv \int_{-\infty}^{\infty} \frac{d\varepsilon}{2\pi}
A({\bm p}\varepsilon,{\bm r}t)\phi({\bm p}\varepsilon,{\bm r}t)=\phi({\bm p}\xi_{\bm p},{\bm r}t),
\label{f-def-QP}
\end{equation}
as
\begin{subequations}
\label{Boltzmann}
\begin{equation}
\frac{\partial f}{\partial t}+\frac{{\bm p}}{m}\cdot\frac{\partial f}{\partial {\bm r}}
-\frac{\partial U}{\partial {\bm r}}\cdot\frac{\partial f}{\partial {\bm p}}=I_{\bm p}[f].
\label{Boltzmann-Q}
\end{equation}
The collision integral $I_{\bm p}[f]$ can be written down by
substituting Eqs.\ (\ref{G-Wig}), (\ref{Sigma^(2)-Wig}), and (\ref{A-QPA}) into Eq.\ 
(\ref{collision}) and carrying out the integration over
$\varepsilon$ as
\begin{eqnarray}
&&\hspace{-10mm}
I_{\bm p}[f]\equiv
\frac{\hbar^{2}}{2}  \prod_{j=2}^{4} \int
\frac{d^{3}p_{j}}{(2\pi\hbar)^{3}}
|V_{{\bm p}-{\bm p}_{3}}\pm V_{{\bm p}-{\bm p}_{4}}|^{2}
(2\pi)^{4}
\delta({\bm p}+{\bm p}_{2}-{\bm p}_{3}-{\bm p}_{4})
\nonumber \\
&&\hspace{3mm}
\times 
\delta(\xi_{\bm p}+\xi_{{\bm p}_{2}}-\xi_{{\bm p}_{3}}-\xi_{{\bm p}_{4}})
\bigl[(1\pm f)(1\pm f_2)f_{3}f_4
- f f_2(1\pm f_3)(1\pm f_4) 
\bigr],
\label{I_p}
\end{eqnarray}
\end{subequations}
with $f_{j}=f({\bm p}_{j},{\bm r},t)$.
Equation (\ref{Boltzmann}) may be regarded as the Boltzmann equation with
quantum effects.
Indeed, if we neglect quantum effects in Eq.\ (\ref{I_p}) as
$\pm V_{{\bm p}-{\bm p}_{3}}V_{{\bm p}-{\bm p}_{4}}^{*}\rightarrow  0$ and 
$(1\pm f)\rightarrow  1$,
we obtain the standard Boltzmann equation in the dilute classical limit.\cite{Cercignani88}
Note that the function $f$ here is dimensionless and normalized so as to reproduce $(e^{\beta \xi_{\bm p}}\mp 1)^{-1}$ in equilibrium.

The corresponding expressions of entropy density, entropy flux density, and entropy production 
are obtained by adopting the same approximation in Eqs.\ 
(\ref{entropy-density}), (\ref{entropy-current}), and (\ref{ds_col/dt}), respectively.
They are given in terms of Eq.\ (\ref{sigma-def}) by
\begin{equation}
s\equiv k_{\rm B} \int \frac{d^{3}p}{(2\pi\hbar)^{3}}\,
\sigma[f] ,
\label{entropy-density-QPA}
\end{equation}
\begin{equation}
{\bm j}_{s}\equiv k_{\rm B} \int \frac{d^{3}p}{(2\pi\hbar)^{3}}\,
\frac{{\bm p}}{m}  \sigma[f] ,
\label{entropy-current-QPA}
\end{equation}
\begin{eqnarray}
&&\hspace{-2mm}
\frac{\partial s_{\rm coll}^{(2)}}{\partial t}
=\frac{k_{\rm B}\hbar^{2}}{8}  \prod_{j=1}^{4} \int
\frac{d^{3}p_{j}}{(2\pi\hbar)^{3}}
|V_{{\bm p}_{1}-{\bm p}_{3}}\pm V_{{\bm p}_{1}-{\bm p}_{4}}|^{2}
(2\pi)^{4}\delta({\bm p}_{1}+{\bm p}_{2}-{\bm p}_{3}-{\bm p}_{4})
\nonumber \\
&&\hspace{12mm}
\times
\delta(\xi_{\bm p}+\xi_{{\bm p}_{2}}-\xi_{{\bm p}_{3}}-\xi_{{\bm p}_{4}})
\bigl[(1\pm f_1)(1\pm f_2)f_{3}f_4
- f_1f_2(1\pm f_3)(1\pm f_4)
\bigr]
\nonumber \\
&&\hspace{12mm}
\times
\ln\frac{(1\pm f_1)(1\pm f_2)f_{3}f_4}
{f_1f_2(1\pm f_3)(1\pm f_4)}.
\label{ds_col/dt-QPA}
\end{eqnarray}
Equation (\ref{entropy-density-QPA}) is the familiar
entropy density for noninteracting systems.\cite{LL80}
These expressions can be obtained alternatively by multiplying Eq.\ 
(\ref{Boltzmann}) by $k_{\rm B}\ln [({1\pm f})/{f}]$ and performing an integration over
${\bm p}$.
Boltzmann's entropy and $H$-theorem are reproduced from the above with
$\pm V_{{\bm p}-{\bm p}_{3}}V_{{\bm p}-{\bm p}_{4}}^{*}\rightarrow  0$ and
$1\pm f\rightarrow 1$.

Thus, we have seen that a transport equation for the distribution function $f$
in the phase space results naturally as the weak-coupling limit of quantum transport equations
(\ref{G^R}) and (\ref{transport}).
It should also be noted that the weakness of the interaction is not essential for the quasiparticle approximation. To be specific, it is sufficient for the approximation to hold that the imaginary part of
$\Sigma^{\rm R}$, i.e., the lifetime of quasiparticles responsible for the excitation, is sufficiently small in the relevant energy range.
Thus, the quasiparticle approximation can also be applied excellently for normal Fermi liquids at low temperatures.\cite{AGD63}
We will discuss it in detail below.

First, let us neglect terms with $A_{\Sigma}=-2{\rm Im}\Sigma^{\rm R}$ on the left-hand side of Eq.\ (\ref{transport}).
We also expand $G_0^{-1}-{\rm Re}\Sigma^{\rm R}$ up to the first order 
from $\varepsilon=0$ and ${\bm p}={\bm p}_{\rm F}$ (${\bm p}_{\rm F}$: Fermi momentum) as
\begin{equation}
G_0^{-1}-{\rm Re}\Sigma^{\rm R}\approx 
\frac{\varepsilon -\xi_{\bm p}}{a}.
\label{ReG^R-1}
\end{equation}
Here, $a$ and $\xi_{\bm p}$ are the renormalization constant and quasiparticle energy:
\begin{subequations}
\begin{eqnarray}
\frac{1}{a}&\equiv& 1-\left.\frac{\partial {\rm Re}\Sigma^{\rm R}}{\partial \varepsilon}
\right|_{\varepsilon=0,{\bm p}={\bm p}_{\rm F}},
\\
\hspace{10mm}
\xi_{\bm p}&\equiv& a\left[\frac{{\bm p}_{\rm F}^{2}}{2m}+U+
{\rm Re}\Sigma^{\rm R}({\bm 0},0)-\mu\right]
+ {\bm v}_{\rm F}\cdot({\bm p}-{\bm p}_{\rm F}),
\label{xi_p}
\end{eqnarray}
respectively, with the Fermi velocity ${\bm v}_{\rm F}$ defined by
\begin{equation}
{\bm v}_{\rm F}\equiv a\!\left(\frac{{\bm p}_{\rm F}}{m}+\left.
\frac{\partial {\rm Re}\Sigma^{\rm R}}{\partial {\bm p}}
\right|_{\varepsilon=0,{\bm p}={\bm p}_{\rm F}}\right).
\end{equation}
\end{subequations}
The spectral function $A=-2{\rm Im}G^{\rm R}$ in this approximation is obtained from
Eq.\ (\ref{G^R}) as $A({\bm p}\varepsilon,{\bm r}t)\approx 2\pi a
\delta\left(\varepsilon-\xi_{\bm p}\right)$.
Let us introduce the distribution function $f$ as
\begin{equation}
f({\bm p},{\bm r},t)\equiv \int_{-\infty}^{\infty} \frac{d\varepsilon}{2\pi a}
A({\bm p}\varepsilon,{\bm r}t)\phi({\bm p}\varepsilon,{\bm r}t).
\label{f-qp}
\end{equation}
We then obtain Eq.\ (\ref{Boltzmann}) from Eq.\ (\ref{transport}) using exactly the same procedure as above for the weak interaction.
However,  the factor
$V_{{\bm p}-{\bm p}_{3}} \pm V_{{\bm p}-{\bm p}_{4}}$ in Eq.\ (\ref{I_p}) is to be replaced by
the effective interaction between quasiparticles, as described in detail in Ref.\ \citen{AGD63}, for example.

\subsection{Quasiclassical approximation\label{QCA}}

For Fermi systems at low temperatures, there is an alternative approximation called ``quasiclassical approximation''. It holds excellently even when the lifetime of quasiparticles is substantial, as for electrons in metals
with strong impurity potentials or electron-phonon interactions.
The approximation was introduced by Prange and Kadanoff for the case of electron-phonon interactions,\cite{PK64}
and extended subsequently to describe superconductors and superfluids in equilibrium\cite{Eilenberger68,Rainer83} and in nonequilibrium.\cite{LO75,Rainer83,LO86,Kita01}
In the case of conventional superconductors, for example, the impurity scattering plays a crucial role 
in changing their properties in magnetic fields from type I to type II,\cite{Parks69}
and the lifetime may not be negligible in the latter cases.
It was Eilenberger\cite{Eilenberger68} who first adopted the approximation to integrate out an irrelevant variable from the Gor'kov equations in equilibrium. He thereby obtained quasiclassical equations of superconductivity, i.e., Eilenberger equations, which extend the Ginzburg-Landau equations\cite{Parks69} over whole temperatures and magnetic fields.
Those equations have been used extensively to clarify and understand the properties of inhomogeneous superconductors and superfluids.\cite{Rainer83,LO86}

In the quasiclassical approximation, one carries out an integration over $\xi_{\bm p}$
defined by Eq.\ (\ref{xi_p}) 
instead of an integration over $\varepsilon$ in the quasiparticle approximation.
To be specific, proceed here as follows. Let us write down Eq.\ (\ref{transport}) explicitly
with Eqs.\ (\ref{Poisson}), (\ref{ReG^R-1}), and (\ref{xi_p}) as
\begin{eqnarray}
&&
\frac{1}{a}\frac{\partial (A\phi)}{\partial t}+\frac{{\bm v}_{\rm F}}{a}\cdot
\frac{\partial (A\phi)}{\partial {\bm r}}
+\frac{\partial (U\!+\! {\rm Re}\Sigma^{\rm R})}{\partial t}
\frac{\partial (A\phi)}{\partial \varepsilon}
-\frac{\partial (U\!+\! {\rm Re}\Sigma^{\rm R})}{\partial {\bm r}}\cdot
\frac{\partial (A\phi)}{\partial {\bm p}}
\nonumber \\
&&+ i \{\Sigma^{12},{\rm Re}G^{\rm R}\}
 ={\cal C},
\label{transport-QC0}
\end{eqnarray}
where we have approximated $A_{\Sigma}\phi\approx -i \Sigma^{12}$ on the left-hand side.
The term $\partial \check{\Sigma}({\bm p}\varepsilon,{\bm r}t)/\partial {\bm p}$
is expected to be of the order of
$|\check{\Sigma}({\bm p}\varepsilon,{\bm r}t)|/p_{\rm F}$ so that we may neglect
the $|{\bm p}|$ dependence in $\check{\Sigma}$ as
\begin{equation}
\check{\Sigma}({\bm p}\varepsilon,{\bm r}t)\approx 
\check{\Sigma}({\bm p}_{\rm F}\varepsilon,{\bm r}t).
\end{equation}
By noting
${\rm Im}\Sigma^{\rm R}=-\frac{1}{2} A_{\Sigma}$ and Eq.\ (\ref{ReG^R-1}), the corresponding spectral function $A=-2{\rm Im}G^{\rm R}$ is obtained as
\begin{equation}
A({\bm p}\varepsilon,{\bm r}t)\approx
\frac{a^{2}A_{\Sigma}({\bm p}_{\rm F}\varepsilon,{\bm r}t)}
{(\varepsilon-\xi_{\bm p})^{2}+\bigl[\frac{1}{2}a
A_{\Sigma}({\bm p}_{\rm F}\varepsilon,{\bm r}t)\bigr]^{2}}.
\label{A-QC}
\end{equation}
The $|{\bm p}|$ dependence in Eq.\ (\ref{A-QC}) lies only in $\xi_{\bm p}$
and may be approximated excellently using a Lorentzian in terms of $\xi_{\bm p}$ with the width $\frac{1}{2}aA_{\Sigma}$ and the area $2\pi a$.
With this observation, we introduce the distribution function in the quasiclassical approximation as
\begin{equation}
{f}^{\rm qc}({\bm p}_{\rm F}\varepsilon,{\bm r}t)\equiv \lim_{\xi_{\rm c}\rightarrow\infty}
\int_{-\xi_{\rm c}}^{\xi_{\rm c}}
\frac{d\xi_{\bm p}}{2\pi a}A({\bm p}\varepsilon,{\bm r}t)\phi({\bm p}\varepsilon,{\bm r}t).
\label{f-def-QC}
\end{equation}
Thus, the distribution function here has $\varepsilon$ as an argument
instead of $|{\bm p}|$ in the quasiparticle approximation, and the momentum
has only the angular dependence on the Fermi surface.
We next perform an integration over $\xi_{\bm p}$ in Eq.\ (\ref{transport-QC0})
in the same way as in Eq.\ (\ref{f-def-QC}).
It then follows that the fourth term on the left-hand side vanishes upon the integration.
Moreover, the $\xi_{\bm p}$ integration of ${\rm Re}G^{\rm R}$ may be estimated with Eqs.\
(\ref{G^R-Wig}) and (\ref{A-QC}) as
\begin{equation}
\lim_{\xi_{\rm c}\rightarrow\infty}
\int_{-\xi_{\rm c}}^{\xi_{\rm c}}
\frac{d\xi_{\bm p}}{2\pi a}{\rm Re}G^{\rm R}({\bm p}\varepsilon,{\bm r}t)
={\rm P}\int_{-\infty}^{\infty}\frac{d\varepsilon'}{2\pi}\frac{1}{\varepsilon-\varepsilon'} .
\end{equation}
Hence, it follows that the fifth term with derivatives of ${\rm Re}G^{\rm R}$ vanishes after the 
$\xi_{\bm p}$ integration.
As for the collision integral ${\cal C}$
obtained by substituting Eqs.\ (\ref{G-Wig}) and (\ref{Sigma^(2)-Wig}) into
Eq.\ (\ref{collision}), let us introduce the density of states by $N(\xi)\equiv \int \frac{d^{3}p}{(2\pi\hbar)^{3}}\delta(\xi-\xi_{\bm p})$ and subsequently approximate it as
$N(\xi)\approx N(0)$ to carry out the $\xi$ integration.
We thereby obtain the quasiclassical transport equation as
\begin{equation}
\frac{\partial {f}^{\rm qc}}{\partial t}+{\bm v}_{\rm F}\cdot
\frac{\partial {f}^{\rm qc}}{\partial {\bm r}}
+a \frac{\partial (U\!+\! {\rm Re}\Sigma^{\rm R})}{\partial t}
\frac{\partial {f}^{\rm qc}}{\partial \varepsilon}={I}_{\bm p}^{\rm qc}[{f}^{\rm qc}] ,
\label{transport-QC}
\end{equation}
with
\begin{eqnarray}
&&\hspace{0mm}
I_{\bm p}^{\rm qc}[f^{\rm qc}]\equiv
\frac{\hbar^{2}}{2}[N(0)]^{3} (2\pi a)^{4} \prod_{j=2}^{4} \int
\frac{d\varepsilon_{j}d\Omega_{j}}{4\pi}
|V_{{\bm p}_{\rm F}-{\bm p}_{{\rm F}3}}\!-\! V_{{\bm p}_{\rm F}-{\bm p}_{{\rm F}4}}|^{2}
\nonumber \\
&&\hspace{17.5mm}
\times 
\delta({\bm p}_{{\rm F}}\!+\!{\bm p}_{{\rm F}2}
\!-\!{\bm p}_{{\rm F}3}\!-\!{\bm p}_{{\rm F}4})
\delta(\varepsilon\!+\!\varepsilon_{2}\!-\!\varepsilon_{3}\!-\!\varepsilon_{4})
\nonumber \\
&&\hspace{17.5mm}
\times 
\bigl[(1\!-\!\ f^{\rm qc})(1\!-\! f^{\rm qc}_2)f^{\rm qc}_{3}f^{\rm qc}_4
\!-\! f^{\rm qc} f^{\rm qc}_2(1\!-\! f^{\rm qc}_3)(1\!-\! f^{\rm qc}_4) 
\bigr],
\label{I_p-QC}
\end{eqnarray}
where $d\Omega_{j}$ denotes an infinitesimal solid angle along ${\bm p}_{{\rm F}j}$.

Thus, we have obtained a transport equation in the quasiclassical approximation.
As already pointed out at the beginning of this subsection, it has an advantage of being applicable even for cases with short lifetimes.
The derivation here has been carried out with second-order self-energy (\ref{Sigma^(2)-Wig})
of a weak interaction. However, the approximation is justified even for low temperature 
Fermi liquids with considerable interactions, and also applicable to superconductors and superfluids.
See Ref.\  \citen{Rainer83} for details.

\section{Transport equations for electrons in electromagnetic fields\label{QTE-charge}}

We next focus on electrons in electromagnetic fields, where special consideration of the gauge invariance is necessary for
deriving quantum transport equations appropriately.
To be specific, we should modify Eq.\ (\ref{checkG-Wig}) to introduce a gauge-covariant Wigner transformation in terms of the ``center-of-mass'' coordinate 
and make use of the corresponding Groenewold-Moyal product.
We will discuss it below following Ref.\ \citen{Kita01}.

We consider the following one-particle operator in place of Eq.\ (\ref{K_1}):
\begin{equation}
\hat{K}_{1}\equiv \frac{1}{2m}\left[-i\hbar{\bm\nabla}_{1}-\frac{e}{c}{\bm A}(1)\right]^{2}+eA_{4}(1)-\mu,
\label{K_1-gauge}
\end{equation}
where ${\bm A}(1)$ and $A_{4}(1)$ are the vector and scalar potentials of electromagnetism, respectively.
The corresponding Dyson's equation is still given by Eq.\ (\ref{Dyson-K-check}); it is invariant through the gauge transformation:
\begin{subequations}
\label{gauge-transformation}
\begin{equation}
{\bm A}(1)\rightarrow {\bm A}(1)+{\bm \nabla}_{1}\chi(1),\hspace{10mm}
A_{4}(1)\rightarrow A_{4}(1)-\frac{1}{c}\frac{\partial\chi(1)}{\partial t_{1}},
\label{A-gauge}
\end{equation}
\begin{equation}
\check{G}(1,2)\rightarrow \exp\!\left\{ i\frac{e}{\hbar c}[\chi(1)-\chi(2)]\right\}\check{G}(1,2),
\label{G-gauge}
\end{equation}
\end{subequations}
with $e<0$ and $\chi(1)$ denoting an arbitrary scalar function.

To discuss transport phenomena of electrons in solids under electromagnetic fields, we also need to incorporate interactions with phonons, impurities, etc., to obtain a finite conductivity. However, Dyson's equation for the electronic part is still given by Eq.\ (\ref{Dyson-K-check}) with modified self-energies. We will proceed by assuming that such effects have already been taken into account into the self-energy matrix of Eq.\ (\ref{Dyson-K-check}). See also Ref.\ \citen{KY08} on how to incorporate the periodic potential.

Following the standard prescription to derive quantum transport equations,
we first transform Eq.\ (\ref{Dyson-K-check}) into a Wigner representation.
If we adopt Eq.\ (\ref{checkG-Wig}) for this purpose, however, 
the resultant equation suffers from the flaws that: 
(i) the gauge invariance of the equation is not retained adequately;
(ii) the magnetic Lorentz force of deflecting electrons is absent from the equation.
This is because the transformation (\ref{checkG-Wig}) breaks the gauge covariance with respect to the ``center-of-mass'' coordinate.

To remove these drawbacks, we adopt the ``gauge-invariant Wigner transformation'' introduced by Stratonovich
in 1956.\cite{Stratonovich56}
It is defined in place of Eq.\ (\ref{checkG-Wig}) by
\begin{subequations}
\label{checkG-Wig-gauge}
\begin{eqnarray}
\check{G}({\bm p}\varepsilon,{\bm r}_{12}t_{12})&\equiv& \int d^{3}\bar{r}_{12}\,d\bar{t}_{12}\,\check{G}(1,2)\,e^{-iI(1,2)}
\,e^{-i({\bm p}\,\cdot\bar{\bm r}_{12}-\varepsilon \bar{t}_{12})/\hbar} ,
\label{checkG-Wig1-gauge}
\\
\check{G}(1,2)&=& e^{iI(1,2)}\int\frac{d^{3}p\,d\varepsilon}{(2\pi\hbar)^{4}}\,
\check{G}({\bm p}\varepsilon,{\bm r}_{12}t_{12})\,
e^{i({\bm p}\cdot\bar{\bm r}_{12}-\varepsilon \bar{t}_{12})/\hbar} ,
\label{checkG-Wig2-gauge}
\end{eqnarray}
\end{subequations}
where $I(1,2)$ is given in terms of the four vectors
$\vec{A}\equiv({\bm A},-cA_{4})$ and $\vec{r}_{1}=({\bm r}_{1},t_{1})$
by
\begin{equation}
I(1,2)
\equiv \frac{e}{\hbar c}\int_{\vec{r}_{2}}^{\vec{r}_{1}}\vec{A}(\vec{s}) \cdot d\vec{s},
\label{Idef}
\end{equation}
with $\vec{s}$ denoting the straight-line path from $\vec{r}_{2}$ to $\vec{r}_{1}$.
Equation (\ref{checkG-Wig-gauge}) was also derived by Fujita elegantly on the basis of the 
Weyl transformation.\cite{Fujita66}

The role of the extra phase factor in Eq.\ (\ref{G-gauge}) may be realized from its change under the gauge transformation (\ref{A-gauge}):
\begin{equation}
e^{iI(1,2)}\longrightarrow
\exp\!\left\{ i\frac{e}{\hbar c}[\chi(1)-\chi(2)]\right\}e^{iI(1,2)} .
\end{equation}
Thus, it transforms in exactly the same way as $\check{G}$ in Eq.\ (\ref{G-gauge}),
thereby making $\check{G}(1,2)e^{-iI(1,2)}$ in Eq.\ (\ref{checkG-Wig1-gauge})  gauge invariant.
It is worth looking at Eq.\ (\ref{checkG-Wig-gauge}) from a more general viewpoint for its extension to superconductors.
Setting $1=2$ in Eq.\ (\ref{G-gauge}) tells us that $\check{G}(1,2)$ is gauge invariant with respect to the ``center-of-mass'' coordinate. Thus, the gauge dependence of $\check{G}(1,2)$ lies solely in the relative coordinate, which should be removed before the Wigner transformation. The factor $e^{-iI(1,2)}$ of $\check{G}(1,2)e^{-iI(1,2)}$ exactly performs this task.
It has been shown\cite{Kita01} that this idea can be applied to superconductors so as to describe appropriately that the pair potential behaves like an effective wave function of charge $2e$ with the gauge covariance in terms of the center-of-mass coordinate; see the last paragraph of this section for more details.

\begin{figure}[t]
\begin{center}
  \includegraphics[width=0.25\linewidth]{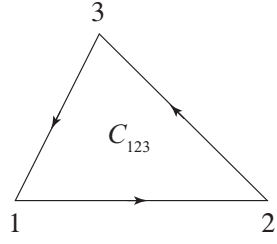}
\end{center}
\caption{Closed contour $C_{123}$.\label{fig:C123}}
\end{figure}
We now modify the Groenewold-Moyal product (\ref{Identity-otimes})
so as to be compatible with Eq.\ (\ref{checkG-Wig-gauge}).
We carry it out here for the simplest case of static electromagnetic fields;
see Ref.\ \citen{Kita01} for a more general treatment.
We first express $C(1,3)$ and $D(3,2)$ on the left-hand side of 
Eq.\ (\ref{Identity-otimes}) as Eq.\ (\ref{checkG-Wig2-gauge}). We then have the extra phase factor $e^{iI(1,3)+iI(3,2)}$ in the integrand on its right-hand side.
Let us express it as a product of $e^{iI(1,2)}$ and the other factor as
\begin{equation}
e^{iI(1,3)+iI(3,2)}=e^{iI(1,2)}e^{i\phi_{123}},\hspace{10mm} 
\phi_{123}\equiv\frac{e}{\hbar c}\oint_{C_{123}}\vec{A}(\vec{s}\,)\cdot d\vec{s},
\label{I+I}
\end{equation}
where the contour $C_{123}$ is given in Fig.\ 7.
Using Stokes' theorem, we can express $\phi_{123}$ in terms of the electric field ${\bm E}\equiv -{\bm \nabla}A_{4}-\frac{1}{c}{\partial}_{t}{\bm A}$ and the magnetic field ${\bm B}\equiv {\bm \nabla}\times {\bm A}$ as
\begin{equation}
\phi_{123}=
\frac{e}{2\hbar c}\bigl[
-c {\bm E}\cdot(\bar{\bm r}_{13}\bar{t}_{32}-\bar{t}_{13}\bar{\bm r}_{32})
-{\bm B}\cdot(\bar{\bm r}_{13}\times\bar{\bm r}_{32}) \bigr].
\end{equation}
The extra phase factor in Eq.\ (\ref{I+I}) modifies Eq.\ (\ref{Identity-otimes}) into
\begin{equation}
\int C(1,3)D(3,2) \,d3
=e^{iI(1,2)}\int\frac{d^{3}pd\varepsilon}{(2\pi\hbar)^{4}} \,
C({\bm p}\varepsilon,{\bm r}_{12}t_{12})
\ast
D({\bm p}\varepsilon,{\bm r}_{12}t_{12}) 
\,e^{i({\bm p}\cdot
\bar{\bm r}_{12}-\varepsilon\bar{t}_{12})/\hbar},
\label{Identity-otimes-EB}
\end{equation}
where the Groenewold-Moyal product is now defined by
\begin{eqnarray}
C({\bm p}\varepsilon,{\bm r}t)\ast
D({\bm p}\varepsilon,{\bm r}t) 
&\equiv& 
C({\bm p}\varepsilon,{\bm r}t)
\exp\!\left[\frac{i\hbar}{2}\left(\overleftarrow{\bm\partial}_{{\bm r}}\!\cdot\!
\overrightarrow{\bm\partial}_{{\bm p}}
\!-\!\overleftarrow{\partial}_{t}\overrightarrow{\partial}_{\varepsilon}
\!-\!\overleftarrow{\bm\partial}_{{\bm p}}\!\cdot\!\overrightarrow{\bm\partial}_{{\bm r}}
\!+\!\overleftarrow{\partial}_{\varepsilon}\overrightarrow{\partial}_{t}\right)\right]
\nonumber \\
& &
\times\exp\!\left\{\frac{i\hbar}{2}\biggl[
-e{\bm E}\cdot\!\left(\overleftarrow{\bm\partial}_{{\bm p}}\overrightarrow{\partial}_{{\varepsilon}}
\!-\!\overleftarrow{\partial}_{\varepsilon}\overrightarrow{\bm\partial}_{{\bm p}}\right)
+\frac{e}{c}{\bm B}\cdot\!\left(\overleftarrow{\bm\partial}_{{\bm p}}
\!\times\!\overrightarrow{\bm\partial}_{{\bm p}}\right)
\biggr]\!\right\}
\nonumber \\
& &
\times D({\bm p}\varepsilon,{\bm r}t) .
\label{Moyal-gauge}
\end{eqnarray}
This modified Groenewold-Moyal product was derived independently
in Refs.\ \citen{Kita01} and \citen{LF01}.
Its expansion up to the first order in the differential operators
can also be expressed as Eq.\ (\ref{otimes1})
with the generalized Poisson bracket:
\begin{eqnarray}
&&\hspace{0mm}
\{C,D\}
\equiv \frac{\partial C}{\partial {\bm r}}\cdot\frac{\partial D}{\partial {\bm p}}
-\frac{\partial C}{\partial t}\frac{\partial D}{\partial \varepsilon}
-\frac{\partial C}{\partial {\bm p}}\cdot\frac{\partial D}{\partial {\bm r}}
+\frac{\partial C}{\partial \varepsilon}\frac{\partial D}{\partial t}
\nonumber \\
&&\hspace{17mm}
-e{\bm E}\cdot\!\left(\frac{\partial C}{\partial {\bm p}}\frac{\partial D}{\partial \varepsilon}
-\frac{\partial C}{\partial \varepsilon}\frac{\partial D}{\partial {\bm p}}\right)
\!+\frac{e}{c}{\bm B}\cdot\!\left(\frac{\partial C}{\partial {\bm p}}
\!\times\!\frac{\partial D}{\partial {\bm p}}\right) .
\label{Poisson-gauge}
\end{eqnarray}

With these preliminaries, we now express Dyson's equation (\ref{Dyson-K-check})
with Eq.\ (\ref{K_1-gauge}) in the Wigner representation.
Let us substitute Wigner representation (\ref{checkG-Wig2-gauge}) into Eq.\ (\ref{Dyson-K-check}).
We then have space-time derivatives of $I(1,2)$.
To estimate them, we express the straight-line path $\vec{s}$
in Eq.\ (\ref{Idef}) in terms of $\vec{R}\equiv\frac{1}{2}(\vec{r}_{1}+\vec{r}_{2})$
 and $\vec{r}\equiv \vec{r}_{1}-\vec{r}_{2}$ as
$\vec{s}=\vec{R}+\eta\vec{r}$ $(-\frac{1}{2}\leq \eta\leq \frac{1}{2})$.
We also expand $\vec{A}(\vec{s})$ in a Taylor series from $\vec{s}=\vec{R}$ and subsequently perform the integration
over $\eta$ order by order.
We thereby obtain an alternative expression of $I(1,2)$ as
\begin{equation}
I(1,2)=\frac{e}{\hbar c}g\bigl(\vec{r}\cdot\vec{\partial}_{\vec{R}}\bigr)\,
\vec{r}\cdot\vec{A}(\vec{R}),\hspace{10mm}
g(x)\equiv \frac{\sinh(x/2)}{x/2}.
\end{equation}
We next operate $\partial_{\vec{r}_{1}}=\partial_{\vec{r}}+\frac{1}{2}\partial_{\vec{R}}$ on $I(1,2)$
and make use of the identities $r_{\nu}\partial_{\mu} A_{\nu}=r_{\nu}\partial_{\nu} A_{\mu}+r_{\nu}
(\partial_{\mu} A_{\nu}-\partial_{\nu} A_{\mu})$ $(\mu,\nu=1,2,3,4)$ and 
$g(x)+[g'(x)+\frac{1}{2}g(x)]x=e^{x/2}$.
The derivative ${\partial I(1,2)}/{\partial r_{1\mu}}$ is thereby transformed as
\begin{eqnarray}
&&\hspace{0mm}
\frac{\partial I(1,2)}{\partial r_{1\mu}}=\frac{e}{\hbar c} \left\{A_{\mu}(1)+
\left[g'(\vec{r}\cdot\vec{\partial}_{\vec{R}})+\frac{1}{2}g(\vec{r}\cdot\vec{\partial}_{\vec{R}})\right]
\sum_{\nu}r_{\nu}\left[\frac{\partial A_{\nu}(\vec{R})}{\partial R_{\mu}}
-\frac{\partial A_{\mu}(\vec{R})}{\partial R_{\nu}}\right]\right\}
\nonumber \\
&& \hspace{15mm}
=\left\{ \begin{array}{ll}
\displaystyle
\frac{e}{\hbar c} \left[{\bm A}(1)+\frac{1}{2} (c\bar{t}_{12}{\bm E}
+\bar{\bm r}_{12}\times {\bm B})\right]_{\mu} & :\mu =1,2,3 \\
\displaystyle
-\frac{e}{\hbar } \left[A_{4}(1)+\frac{1}{2} \bar{\bm r}_{12}\cdot {\bm E}
\right] & :\mu=4 
\end{array}
\right. .
\end{eqnarray}
Here, the first equality is an identity, whereas the second one holds only for static electromagnetic fields.

We now substitute Eqs.\ (\ref{K_1-gauge}) and (\ref{checkG-Wig2-gauge}) into the first term in Eq.\ (\ref{Dyson-K-check}) and follow the procedure of deriving Eq.\ (\ref{Identity-otimes}).
Comparing the resultant expressions with the definition of the $\ast$ product in Eq.\ (\ref{Moyal-gauge}),
we obtain the simple expressions:
\begin{subequations}
\label{diff-gauge}
\begin{equation}
\left[-i\hbar{\bm\nabla}_{1} -\frac{e}{c}{\bm A}(1)\right]
\check{G}^{\rm K}(1,2)
=e^{iI(1,2)}\int\frac{d^3 p\,d\varepsilon}{(2\pi\hbar)^{4}}\,
{\bm p}\ast \check{G}^{\rm K}({\bm p}\varepsilon,{\bm r}_{12}t_{12})\,
e^{i({\bm p}\cdot\bar{\bm r}_{12}-\varepsilon\bar{t}_{12})/\hbar} ,
\end{equation}
\begin{equation}
\left[i\hbar \frac{\partial }{\partial t_{1}} -eA_{4}(1)\right]\!
\check{G}^{\rm K}(1,2)
=e^{iI(1,2)}\int\frac{d^3 p\,d\varepsilon}{(2\pi\hbar)^{4}}\,
\varepsilon\ast \check{G}^{\rm K}({\bm p}\varepsilon,{\bm r}_{12}t_{12})
\,e^{i({\bm p}\cdot\bar{\bm r}_{12}-\varepsilon\bar{t}_{12})/\hbar} .
\end{equation}
\end{subequations}
They show clearly that the gauge-invariant Wigner representations of $-i\hbar{\bm\nabla}_{1} -\frac{e}{c}{\bm A}(1)$
and $i\hbar \frac{\partial }{\partial t_{1}} -eA_{4}(1)$ are given respectively by
${\bm p}$ and $\varepsilon$.

Using Eqs.\ (\ref{Identity-otimes-EB}) and (\ref{diff-gauge}), we can express
the Wigner representation of Eq.\ (\ref{Dyson-K-check}) in a concise form as
\begin{equation}
\left[\left(\varepsilon-\frac{p^{2}}{2m}+\mu\right)\!\check{1}
-{\check{\Sigma}}^{\rm K}({\bm p}\varepsilon,{\bm r}t)\right]
\ast {\check{G}}^{\rm K}({\bm p}\varepsilon,{\bm r}t)
=\check{1} ,
\label{Dyson-Wig}
\end{equation}
where the Groenewold-Moyal product is defined now by Eq.\ (\ref{Moyal-gauge}).

From this point, we can directly follow the procedure described in \S \ref{subsec:QTE}.
Particularly for the weak interaction case, we can adopt the transformation shown in \S \ref{QPA}
to an excellent approximation.
We thereby obtain the Boltzmann equation in static electromagnetic fields as
\begin{equation}
\frac{\partial f}{\partial t}+{\bm v}\cdot\frac{\partial f}{\partial {\bm r}}+
e\left({\bm E}+\frac{1}{c}{\bm v}\times{\bm B}\right)\cdot
\frac{\partial f}{\partial {\bm p}}=I_{\bm p}[f],
\label{Boltzmann-EB}
\end{equation}
where ${\bm v}\equiv{\bm p}/m$, and $f$ is defined by Eq.\ (\ref{f-def-QP}).
It is worth noting once again that, for discussing the electronic conductivity, 
we must incorporate the impurity scattering or electron-phonon interactions
besides the two-body interaction in the collision integral $I_{\bm p}[f]$.

We finally comment on an alternative method
that has been used  frequently for deriving gauge-invariant transport equations.\cite{Keldysh64,Kubo64,Onoda06}
It consists of adopting the ordinary Wigner transformation (\ref{checkG-Wig1}) followed by
the change of variables:
\begin{equation}
{\bm p}\longrightarrow {\bm \pi}\equiv {\bm p}-\frac{e}{c}{\bm A},
\hspace{10mm}
\varepsilon \longrightarrow \pi_{4}\equiv \varepsilon -eA_{4},
\label{Ch-V}
\end{equation}
in the integrand.
In the case of static electromagnetic fields, the corresponding Groenewold-Moyal product has been shown to be identical with Eq.\ (\ref{Moyal-gauge}).\cite{Onoda06}
There seem at least two difficulties in this change-of-variables method.
First, it does not retain gauge invariance for electromagnetic fields with space-time variations.
This may be realized by looking at the corresponding Wigner representation:
\begin{equation}
\check{G}({\bm p}\varepsilon,{\bm r}_{12}t_{12})=
\int d^{3}\bar{r}_{12}d\bar{t}_{12} \,
\check{G}(1,2)\, e^{i\{[{\bm p}-(e/c){\bm A}]\cdot\bar{\bm r}_{12}-
(\varepsilon-e A_{4}) \bar{t}_{12}\}/\hbar} .
\label{checkG-Wig-gauge2}
\end{equation}
If we compare this expression with Eq.\ (\ref{checkG-Wig1-gauge}),
we realize that the phase $I(1,2)$ in the latter is replaced here by $(e/c){\bm A}\cdot\bar{\bm r}_{12}-
e A_{4} \bar{t}_{12}$.
However, this transformation breaks gauge invariance through Eq.\ (\ref{gauge-transformation}),
except those cases where $\chi(1)$ has linear dependence in  $({\bm r}_{1},t_{1})$.
Moreover, it is unclear which space-time point we should adopt as the argument 
of electromagnetic potentials in Eq.\ (\ref{checkG-Wig-gauge2}).
An exceptional case is when electromagnetic fields are static, where Eq.\ (\ref{checkG-Wig-gauge2})  yields the same result as Eq.\ (\ref{checkG-Wig-gauge}).
This difficulty was already pointed out by Stratonovich\cite{Stratonovich56}
and also discussed in detail by Serimaa et al.\cite{SJV86}
Second, the change-of-variables method is apparently not applicable to 
superconductors where we have new types of functions,  such as the pair potential $\Delta(1,2)$ and the anomalous Green's function $F(1,2)$.\cite{AGD63}
The pair potential $\Delta (1,2)$, for example, transforms as
$\Delta (1,2)\rightarrow \exp\!\left\{ i\frac{e}{\hbar c}[\chi(1)+\chi(2)]\right\}\Delta(1,2)$
for Eq.\ (\ref{gauge-transformation}), thus behaving like a wave function of charge $2e$ for $2=1$.
If one handles the product $\Delta(1,3)F^{*}(3,2)$
using the change-of-variables method, one will end up having the operators ${\partial}/{\partial {\bm r}}$ and
${\partial}/{\partial t}$ in front of 
$\Delta({\bm p}\varepsilon,{\bm r}t)$, thus failing to describe the fact that $\Delta({\bm p}\varepsilon,{\bm r}t)$
behaves like an effective wave function of charge $2e$.

In contrast, the transformation of Eq.\ (\ref{checkG-Wig-gauge}) is free from such difficulties and can be extended straightforwardly to superconductors.\cite{Kita01} For the pair potential, for example, we only need to make use of the phase factor $\exp[-iI(\vec{r}_{1},\vec{R})-iI(\vec{r}_{2},\vec{R})]$  [$\vec{R}\equiv\frac{1}{2}(\vec{r}_{1}+\vec{r}_{2})$]
 instead of $e^{-iI(1,2)}$ in Eq.\ (\ref{checkG-Wig2-gauge}).
The resultant Wigner representation $\Delta({\bm p}\varepsilon,{\bm r}t)$ can be shown to transform as 
$\Delta({\bm p}\varepsilon,{\bm r}t) \rightarrow \exp\!\left[ i\frac{2e}{\hbar c}\chi({\bm r}t)\right]\!\Delta
({\bm p}\varepsilon,{\bm r}t)$ for Eq.\ (\ref{gauge-transformation}). Moreover, the differential operators
${\partial}/{\partial {\bm r}}-i(2e/\hbar c){\bm A}({\bm r}t)$ and 
${\partial}/{\partial t}+i(2e/\hbar)A_{4}({\bm r}t)$ naturally appear in front of $\Delta({\bm p}\varepsilon,{\bm r}t)$.

\section{Two-body correlations\label{two-body}}

Here, we show that,
with the self-consistent approximation considered in \S \ref{subsec:Phi},
we can also calculate two-body and higher-order correlations for a given $\Phi$.
Thus, choosing a definite $\Phi$ amounts to determining the whole BBGKY hierarchy\cite{Cercignani88}
in the $\Phi$-derivable approximation.

To be specific, we consider the following two-body correlation:
\begin{equation}
{\cal K}^{ij,kl}(12,34)\equiv \!\left(\!-\frac{i}{\hbar}\right)^{\!\!2}\!
\langle {\cal T}_{C}
\hat{\psi}_{\rm H}(1^{i})\hat{\psi}_{\rm H}(3^{k})\hat{\psi}_{\rm H}^{\dagger}(4^{l})
\hat{\psi}_{\rm H}^{\dagger}(2^{j})\rangle 
 -G^{ij}(1,2)G^{kl}(3,4) .
\label{K}
\end{equation}
The equation for ${\cal K}$ is obtained by introducing an extra nonlocal potential
$W(1^{C},2^{C})$ on the Keldysh contour.\cite{MS59,BK61}
Let us regard $\hat{\cal H}(t)$ in Eq.\ (\ref{H}) as the unperturbed Hamiltonian.
Then, the argument of \S \ref{subsec:int-rep} tells us that all the effects of $W(1^{C},2^{C})$ can be incorporated into the S-matrix:
\begin{equation}
\hat{\cal S}_{C}'\equiv {\cal T}_{C}\exp\!\left[-\frac{i}{\hbar}\sum_{i,j=1}^{2}
(-1)^{i+j}\int d1\int d2\,\hat{\psi}_{\rm H}^{\dagger}(1^{i}){W}^{ij}(1,2)
\hat{\psi}_{\rm H}(2^{j})\right] ,
\label{S'}
\end{equation}
where the factor $(-1)^{i+j}$ stems from the transformation of Eq.\ (\ref{int_C}), and the subscript $_{\rm H}$
signifies the representation defined by Eq.\ (\ref{O_H-def}).
Looking back at the second expression of Eq.\ (\ref{G-def}), we also realize that
Green's function with the extra potential ${W}^{ij}(1,2)\equiv W(1^{i},2^{j})$ can be defined by
\begin{equation}
G^{ij}(1,2;\check{W})=-\frac{i}{\hbar}\frac{\langle {\cal T}_{C}\hat{\cal S}_{C}'
\hat{\psi}_{\rm H}(1^{i})\hat{\psi}_{\rm H}^{\,\dagger}(2^{j})\rangle}{
\langle \hat{\cal S}_{C}'\rangle} .
\label{checkG-W}
\end{equation}
Now, it is easy to see that correlation function (\ref{K}) is obtained from this Green's function as
\begin{equation}
{\cal K}^{ii',jj'}(11',22')= \left. \pm (-1)^{j+j'}
\frac{\delta G^{ii'}(1,1';\check{W})}{
\delta W^{j'j}(2',2)}\right|_{\check{W}=\check{0}} .
\label{K2}
\end{equation}
Here, the factor $\pm$ is due to the permutation between $\hat{\psi}_{\rm H}(2^{j})$ and $\hat{\psi}_{\rm H}^{\,\dagger}(2^{\prime j'})$, and $(-1)^{j+j'}$ cancels the factor $(-1)^{i+j}$ in  Eq.\ (\ref{S'}).
Equation (\ref{K2}) may also be written concisely as
${\check{\check{\cal K}}}=\pm \delta \check{G}/\delta (\check{\tau}_{3}\check{W}\check{\tau}_{3})$.

The connection (\ref{K2}) enables us to derive the equation for 
${\check{\check{\cal K}}}$, i.e., the Bethe-Salpeter equation,\cite{SB51} from Dyson's equation for $\check{G}$ as follows.
First, Dyson's equation is obtained from Eq.\
(\ref{Dyson-check}) by adding an extra contribution of $\check{W}$.
It may be written symbolically as
\begin{subequations}
\label{Dyson-sym}
\begin{equation}
\check{G}^{-1}\check{G}=\check{1},
\label{Dyson-sym1}
\end{equation}
\begin{equation}
\check{G}^{-1}\equiv \!\left(i\hbar\frac{\partial}{\partial t}-\hat{K}\right)\!\check{\tau}_{3}-\check{\tau}_{3}(\check{W}+\check{\Sigma})\check{\tau}_{3}.
\label{Dyson-sym2}
\end{equation}
\end{subequations}
Hence, it follows that the first-order change $\delta \check{G}$ satisfies $\check{G}^{-1}\delta \check{G}+\delta \check{G}^{-1}
\check{G}=\check{0}$, or equivalently, 
\begin{equation}
\delta \check{G}=\check{G}(-\delta \check{G}^{-1})\check{G} .
\label{dG-eq}
\end{equation}
We also obtain
$$-\delta \check{G}^{-1}=\check{\tau}_{3}(
\delta\check{W}+\delta\check{\Sigma})\check{\tau}_{3}=
\delta(\check{\tau}_{3}\check{W}\check{\tau}_{3})
+\frac{\delta(\check{\tau}_{3}\check{\Sigma}\check{\tau}_{3})}{\delta\check{G}}
\frac{\delta\check{G}}{\delta(\check{\tau}_{3}\check{W}\check{\tau}_{3})}
\delta(\check{\tau}_{3}\check{W}\check{\tau}_{3})
$$
from Eq.\ (\ref{Dyson-sym2}). Let us substitute this $-\delta \check{G}^{-1}$ into Eq.\ (\ref{dG-eq}), divide the resultant equation by $\delta(\check{\tau}_{3}\check{W}\check{\tau}_{3})$, 
and set $\check{W}=\check{0}$. 
We then express it in terms of Eq.\ (\ref{K2}) and the vertex:
\begin{equation}
\Gamma^{i'i,j'j}(1'1,2'2)\equiv 
\mp\frac{i}{\hbar}(-1)^{i+i'} \frac{\delta \Sigma^{i'i}(1',1)}{
\delta G^{jj'}(2,2')}
=-\frac{i}{\hbar}
\frac{\delta^{2}\Phi}{\delta G^{ii'}(1,1')\delta G^{jj'}(2,2')} ,
\label{Xi}
\end{equation}
where Eq.\ (\ref{Sigma-Phi}) has been used in the second equality.
We thereby obtain the Bethe-Salpeter equation as
\begin{eqnarray}
&&\hspace{0mm}
{\cal K}^{ii',jj'}(11',22')=\pm G^{ij'}(1,2')G^{ji'}(2,1')
\pm i\hbar  \sum_{kk'll'}
\int d3\int d3'\int d4\int d4' \, G^{ik'}(1,3')
\nonumber \\
&&\hspace{30mm}
\times G^{ki'}(3,1')\Gamma^{k'k,l'l}(3'3,4'4) {\cal K}^{ll',jj'}(44',22') ,
\label{BS}
\end{eqnarray}
which determines ${\check{\check{\cal K}}}$ for given $\check{G}$ and ${\check{\check{\Gamma}}}$.

We now realize from Eqs.\ (\ref{Dyson-check}), (\ref{Sigma-Phi}), (\ref{Xi}), and (\ref{BS}) that the functional $\Phi=\Phi[\check{G}]$ enables us to calculate one-particle, two-particle, and also higher-order correlations in a unified way. To put it another way, $\Phi$ specifies the approximation definitely from one-particle to higher-order levels. This simple structure of the approximation is among the advantages of the self-consistent $\Phi$-derivable approximation.

The integral equation (\ref{BS}) can be solved formally as
\begin{equation}
\check{\check{\mbox{$\cal K$}}}= \pm \bigl(\,\check{\check{1}}\mp i\hbar
\check{G}\check{G}\,\check{\check{\Gamma}}\,\bigr)^{-1}\check{G}\check{G} ,
\label{K-2}
\end{equation}
where $\check{\check{1}}$ and $\check{G}\check{G}$ are defined by
\begin{subequations}
\label{checkcheck}
\begin{eqnarray}
(\check{\check{1}})^{ii',jj'}(11',22')&\equiv& \delta_{ij}\delta_{i'j'}\delta(1,2)
\delta(1',2') ,
\\
(\check{G}\check{G})^{ii',jj'}(11',22')&\equiv& G^{ij'}(1,2')G^{ji'}(2,1')  .
\label{GG}
\end{eqnarray}
\end{subequations}
It follows from Eq.\ (\ref{Xi}) that the vertex $\check{\check{\Gamma}}$ has the symmetry:
\begin{subequations}
\label{Lambda-symm}
\begin{equation}
\Gamma^{ii',jj'}(11',22')=\Gamma^{jj',ii'}(22',11') .
\end{equation}
It also satisfies 
\begin{eqnarray}
&&\hspace{-8mm}
\Gamma^{ii',jj'}(11',22')
=-
\sum_{kk'll'}(\check{\tau}_{1})_{ik}(\check{\tau}_{1})_{k'i'}
(\check{\tau}_{1})_{jl}(\check{\tau}_{1})_{l'j'}
[\Gamma^{k'k,l'l}(1'1,2'2)]^{*},
\end{eqnarray}
\end{subequations}
which results from Eq.\  (\ref{checkG-symm}).
These symmetries are useful in solving the Bethe-Salpeter equation.

In \S \ref{subsec:local}, we have adopted the local approximation for the self-energy.
This local approximation may also be useful when calculating two-body correlations.
Hence, we write down the Bethe-Salpeter equation in the local approximation.
First, the vertex function is expanded as
\begin{eqnarray}
&&\hspace{0mm}
\Gamma^{i'i,j'j}(1'1,2'2)
=\int\frac{d^{3}p_{1}d\varepsilon_{1}}{(2\pi\hbar)^{4}}
\int\frac{d^{3}p_{2}d\varepsilon_{2}}{(2\pi\hbar)^{4}}
\int\frac{d^{3}q\,d\omega}{(2\pi\hbar)^{4}}\,
\Gamma^{i'i,j'j}({\bm p}_{1}\varepsilon_{1},{\bm p}_{2}\varepsilon_{2};
{\bm q}\omega,{\bm r}t)
\nonumber \\
&&\hspace{30mm} 
\times  e^{i[({\bm p}_{1-}\cdot{\bm r}_{1}'-\varepsilon_{1-} t_{1}')-({\bm p}_{1+}\cdot{\bm r}_{1}-\varepsilon_{1+} t_{1})
+({\bm p}_{2+}\cdot{\bm r}_{2}'-\varepsilon_{2+} t_{2}')-({\bm p}_{2-}\cdot{\bm r}_{2}-\varepsilon_{2-} t_{2})]/\hbar} ,
\nonumber \\
\label{Xi-Wig}
\end{eqnarray}
where ${\bm p}_{j\pm}\!\equiv\!{\bm p}_{j}\pm{\bm q}/2$,
$\varepsilon_{j\pm}\!\equiv\!\varepsilon_{j}\pm\omega/2$, and
${\bm r}t$ is some space-time point around $1,1',2,2'$.
Let us substitute Eqs.\ (\ref{checkG-Wig2}) and (\ref{Xi-Wig}) into Eq.\ (\ref{BS}).
It then turns out that ${\check{\check{\cal K}}}$ in the local approximation can also be expanded as
\begin{eqnarray}
&&\hspace{0mm}
{\cal K}^{ii',jj'}(11',22')
=\int\frac{d^{3}p_{1}d\varepsilon_{1}}{(2\pi\hbar)^{4}}
\int\frac{d^{3}p_{2}d\varepsilon_{2}}{(2\pi\hbar)^{4}}
\int\frac{d^{3}q\,d\omega}{(2\pi\hbar)^{4}}\,
{\cal K}^{ii',jj'}({\bm p}_{1}\varepsilon_{1},{\bm p}_{2}\varepsilon_{2};
{\bm q}\omega,{\bm r}t)
\nonumber \\
&&\hspace{30mm} 
\times e^{i[({\bm p}_{1-}\cdot{\bm r}_{1}-\varepsilon_{1-} t_{1})
-({\bm p}_{1+}\cdot{\bm r}_{1}'-\varepsilon_{1+} t_{1}')
+({\bm p}_{2+}\cdot{\bm r}_{2}-\varepsilon_{2+} t_{2})
-({\bm p}_{2-}\cdot{\bm r}_{2}'-\varepsilon_{2-} t_{2}')]/\hbar} .
\nonumber \\
\label{K-Wig}
\end{eqnarray}
The equation for the Fourier coefficient ${\cal K}^{ii',jj'}({\bm p}_{1}\varepsilon_{1},{\bm p}_{2}\varepsilon_{2};
{\bm q}\omega,{\bm r}t)$ is formally the same as Eq.\ (\ref{K-2}) with the modifications:
(i) Every matrix such as $\check{\check{\cal K}}$ has the indices
$({\bm p}_{1}\varepsilon_{1}$,${\bm p}_{2}\varepsilon_{2})$ to distinguish rows and columns.
(ii) Integrations over internal variables are now given by $\int d^{3}p\,d\varepsilon/(2\pi\hbar)^{4}$.
(iii) The quantities $\check{\check{1}}$ and $\check{G}\check{G}$ are defined as
\begin{subequations}
\label{checkcheck-p}
\begin{eqnarray}
(\check{\check{1}})^{ii',jj'}({\bm p}\varepsilon,{\bm p}'\varepsilon')
&\equiv&
(2\pi\hbar)^{4}\delta({\bm p}-{\bm p}')\delta(\varepsilon-\varepsilon') 
\delta_{ij}\delta_{i'j'},
\\
(\check{G}\check{G})^{ii',jj'}({\bm p}\varepsilon,{\bm p}'\varepsilon';
{\bm q}\omega,{\bm r}t)
&\equiv&
(2\pi\hbar)^{4}\delta({\bm p}-{\bm p}')\delta(\varepsilon-\varepsilon') 
G^{ij'}({\bm p}_{-}\varepsilon_{-},{\bm r}t)G^{ji'}({\bm p}_{+}\varepsilon_{+},{\bm r}t).
\nonumber \\
\label{GG-p}
\end{eqnarray}
\end{subequations}

\section{$\Phi$-derivable approximation and conservation laws\label{sec:conserv}}

Careful consideration of conservation laws is necessary when tackling dynamical problems.
For example, if an approximation you have adopted for a closed system does not conserve particle number, you will end up obtaining results that are completely nonsense.
Thus, of crucial importance here is to find a general criterion with which conservation laws are satisfied. It is Baym \cite{Baym62} who presented a sufficient condition on this fundamental issue. To be more specific, Baym proved that conservation laws are obeyed automatically in the $\Phi$-derivable approximation described in \S \ref{subsec:Phi}, which still seems to be the only systematic approximation scheme with the essential property.
In this section, we will provide a detailed proof of it, modifying Baym's original one with the equilibrium Matsubara formalism\cite{Baym62} onto the Keldysh contour.\cite{Kita06a}

\subsection{Identities\label{subsec:identities}}

As preliminaries, we first derive various identities that are obeyed within the $\Phi$-derivable approximation. Since this approximation becomes exact when all the terms in Eq.\ (\ref{Phi-def}) are incorporated, those identities are also satisfied in the exact theory.

First, we consider the following gauge transformation:
\begin{equation}
\check{G}(2,1)\longrightarrow 
e^{i\check{\chi}(2)} 
\check{G}(2,1)
e^{-i\check{\chi} (1)} ,\hspace{10mm}
\check{\chi}(1)\equiv\left[
\begin{array}{cc}
\chi(1) & 0 \\
0 & 0
\end{array}
\right] .
\label{gauge}
\end{equation}
Since it is composed of closed particle lines, the functional $\Phi$ of Eq.\ (\ref{Phi-def}) is invariant through Eq.\ (\ref{gauge}), including any approximate one with a partial summation.
Thus, there is no change in $\Phi$ also at first order in $\chi$ as $\delta \Phi = 0$. By noting that $\Phi$ is a functional of $\check{G}$ with the property (\ref{Sigma-Phi}),
the condition $\delta \Phi = 0$ reads
\begin{equation}
\int d1\int d2 \, {\rm Tr}\,
\check{\tau}_{3}\check{\Sigma}(1,2)\check{\tau}_{3} 
\delta \check{G}(2,1) = 0 .
\label{Identity}
\end{equation}
On the other hand, $\delta\check{G}$ can be written down explicitly from Eq.\ (\ref{gauge}) as
\begin{equation}
\delta\check{G}(2,1)= i\left[\check{\chi}(2) 
\check{G}(2,1)-
\check{G}(2,1)\check{\chi} (1)\right] .
\label{deltaG-gauge}
\end{equation}
Substituting Eq.\ (\ref{deltaG-gauge}) into Eq.\ (\ref{Identity}) and noting that $\chi(1)$ is arbitrary, we obtain
\begin{subequations}
\label{gauge-identity}
\begin{equation}
\int d2 \,{\rm Tr}\, \frac{\check{1}\!+\!\check{\tau}_{3}}{2}
\bigl[ \check{\tau}_{3}\check{\Sigma}(1,2)\check{\tau}_{3} 
\check{G}(2,1)-
\check{G}(1,2)\check{\tau}_{3}
\check{\Sigma}(2,1)\check{\tau}_{3} 
\bigr] = 0  .
\label{gauge-identity1}
\end{equation}
It is further transformed into
\begin{eqnarray}
&&
\int d2\,
\bigl[ \Sigma^{{\rm R}}(1,2)G^{12}(2,1)
+ \Sigma^{12}(1,2) G^{{\rm A}}(2,1)
\nonumber \\
&&\hspace{10mm}
-G^{{\rm R}}(1,2)\Sigma^{12}(2,1)
-G^{12}(1,2) \Sigma^{{\rm A}}(2,1)
 \bigr] =0,
\label{gauge-identity2}
\end{eqnarray}
\end{subequations}
where we have used Eqs.\ (\ref{G-check-def}), (\ref{Gcomp}), and (\ref{G-RAK-def})
to express $\Sigma^{11}G^{11}-\Sigma^{12}G^{21}=\Sigma^{{\rm R}}G^{12}+\Sigma^{12}G^{{\rm A}}$.
Equation (\ref{gauge-identity2}) is what results from the gauge degree of freedom.

Second, we consider the following Galilean transformation:
\begin{equation}
\check{G}(2,1)\longrightarrow 
\exp\!\left[\check{\bm R}(t_{2})\!\cdot\!
\overrightarrow{\bm\nabla}_{2}\right]
\check{G}(2,1)
\exp\!\left[
\overleftarrow{\bm\nabla}_{1}\!\cdot\!\check{\bm R}(t_{1})\right]  ,
\hspace{5mm}
\check{\bm R}(t)\equiv \left[
\begin{array}{cc}
{\bm R}(t) & {\bm 0} \\
{\bm 0} & {\bm 0}
\end{array}
\right] .
\label{Galilean}
\end{equation}
Since the transformation only shifts the boundary of the system by ${\bm R}(t)$, $\Phi$ is invariant through Eq.\ (\ref{Galilean}).
Hence, Eq.\ (\ref{Identity}) also holds here.
The first-order change of $\check{G}$ is now given by
\begin{equation}
\delta\check{G}(2,1)= 
\check{\bm R}(t_{2})\!\cdot\!{\bm\nabla}_{2}\check{G}(2,1) + 
\check{\bm R}(t_{1})\!\cdot\!{\bm\nabla}_{1}\check{G}(2,1) . 
\label{deltaG-Galilean}
\end{equation}
Substituting Eq.\ (\ref{deltaG-Galilean}) into Eq.\ (\ref{Identity}) and noting that ${\bm R}(t)$ is arbitrary,
we obtain
\begin{equation}
\int{\bm Q}(1)\, d^{3}r_{1}={\bm 0} ,
\label{Galilean-identity}
\end{equation}
with
\begin{eqnarray}
&&
{\bm Q}(1)
\equiv \mp i\hbar\frac{\mbox{\boldmath$\nabla$}_{\! 1}\!-\!\mbox{\boldmath$\nabla$}_{\! 1'}}{2}
\int d2 \,
\bigl[
\Sigma^{\rm R}(1,2) G^{12}(2,1')
+\Sigma^{12}(1,2) G^{\rm A}(2,1')
\nonumber \\
&&\hspace{46mm}
- G^{\rm R}(1,2) \Sigma^{12}(2,1')
-G^{12}(1,2) \Sigma^{\rm A}(2,1')
\bigr]_{1'=1} ,
\label{vecS}
\end{eqnarray}
where terms with derivatives of self-energies are due to partial integrations.

Third, we consider a change of variables on $C_{1}$ given by
$t \rightarrow \theta(t)\equiv t+ \varphi(t)$.
Green's function is transformed accordingly as
\begin{equation}
\check{G}(2,1)\longrightarrow \!
\check{U}(t_{2})
\check{G}({\bm r}_{2}\theta_{2},{\bm r}_{1}\theta_{1})\check{U}(t_{1}),
\hspace{10mm}
\check{U}(t)\equiv \left[
\begin{array}{cc}
(d\theta/dt)^{1/4} & 0
\\
0 & 0
\end{array}
\right] ,
\label{rubber}
\end{equation}
where the factor $(d\theta/dt)^{1/4}$ is introduced to cancel $dt/d\theta$  in Eq.\ 
(\ref{S_C2}) upon the change of variables, thereby keeping $\Phi$ invariant through the transformation.
Hence, it follows that Eq.\ (\ref{Identity}) also holds in this case.
The corresponding $\delta\check{G}$ is given by
\begin{equation}
\delta G^{ji}(2,1)= 
\left\{\delta_{j1}\!\left[\frac{\varphi'(t_{2})}{4}
\!+\! \varphi(t_{2})\frac{\partial}{\partial t_{2}}\right]\right.
\left.+ \delta_{i1}\!\left[\frac{\varphi'(t_{1})}{4}
+ \varphi(t_{1})\frac{\partial}{\partial t_{1}}\right]\!\right\}
G^{ji}(2,1) .
\label{deltaG-rubber}
\end{equation}
Substituting Eq.\ (\ref{deltaG-rubber}) into Eq.\ (\ref{Identity}) and noting that
$\varphi(t)$ is arbitrary, we obtain
\begin{equation}
\frac{d\langle  \hat{\cal H}_{\rm int}(t_{1})\rangle}{d t_{1}}
=-\int  Q_{\varepsilon}(1)\,d^{3}r_{1},
\label{rubber-identity}
\end{equation}
where $\langle  \hat{\cal H}_{\rm int}(t_{1})\rangle$ and
$Q_{\varepsilon}(1)$ are defined by
\begin{eqnarray}
\langle \hat{\cal H}_{\rm int}(t_{1})\rangle&\equiv&
\pm \frac{i\hbar}{4}\int d^{3}r_{1}
\int d2 \, \bigl[
\Sigma^{\rm R}(1,2) G^{12}(2,1)
+\Sigma^{12}(1,2) G^{\rm A}(2,1)
\nonumber \\
& &
+ G^{\rm R}(1,2) \Sigma^{12}(2,1)
+G^{12}(1,2) \Sigma^{\rm A}(2,1)
\bigr],
\label{Hint}
\\
Q_{\varepsilon}(1)&\equiv&\mp i\hbar \frac{\partial}{\partial t_{1}}
\int d2 \, \bigl[
\Sigma^{\rm R}(1',2) G^{12}(2,1)
+\Sigma^{12}(1',2) G^{\rm A}(2,1)
\nonumber \\
& &
+ G^{\rm R}(1,2) \Sigma^{12}(2,1')+G^{12}(1,2) \Sigma^{\rm A}(2,1')
\bigr]_{1'=1} .
\label{Q}
\end{eqnarray}
Equation (\ref{rubber-identity}) is what results from Eq.\ (\ref{rubber}).

The quantity $\langle  \hat{\cal H}_{\rm int}(t_{1})\rangle$ of Eq.\ (\ref{Hint}) is the interaction energy
of the system. To see this, let us write down the (1,2) component of Eq.\ (\ref{Dyson-check}):
\begin{equation}
\left(i\hbar \frac{\partial }{\partial t_{1}}-\hat{K}_{1}
\right)\!
G^{12}(1,2)
-\int\left[\Sigma^{{\rm R}}(1,3)G^{12}(3,2)
+
\Sigma^{12}(1,3)G^{{\rm A}}(3,2)\right]d3 =0  ,
\label{Dyson-12-l}
\end{equation}
where we have used Eqs.\ (\ref{G-check-def}), (\ref{Gcomp}), and (\ref{G-RAK-def})
to express $\Sigma^{11}G^{12}-\Sigma^{12}G^{22}=\Sigma^{{\rm R}}G^{12}+\Sigma^{12}G^{{\rm A}}$.
There is an alternative method of deriving the equation for $G^{12}$.
It is based on
$$
i\hbar \frac{\partial }{\partial t_{1}}\hat{\psi}_{\rm H}(1)=
\hat{\cal U}^{\dagger}(t_{1},-\infty)[\hat{\psi}({\bm r}_{1}),\hat{\cal H}(t_{1})]\hat{\cal U}(t_{1},-\infty),
$$
which stems from $\hat{\psi}_{\rm H}(1)\equiv\hat{\cal U}^{\dagger}(t_{1},-\infty)
\hat{\psi}({\bm r}_{1})\hat{\cal U}(t_{1},-\infty)$ and Eq.\ (\ref{U-eq-motion}).
By calculating the commutation relation with Eq.\ (\ref{H}), this equation of motion is transformed into
\begin{equation}
\left( i\hbar \frac{\partial }{\partial t_{1}}-\hat{K}_{1}\right)\!
\hat{\psi}_{\rm H}(1)-
\int\bar{V}(1-1')\hat{\psi}_{\rm H}^{\dagger}(1')\hat{\psi}_{\rm H}(1')\hat{\psi}_{\rm H}(1) =0  .
\label{eq-motion}
\end{equation}
Let us multiply Eq.\ (\ref{eq-motion}) by
$\mp(i/\hbar)\hat{\psi}_{\rm H}^{\dagger}(2)$ from the left,
take the thermodynamic average with density matrix (\ref{rho_0}), and use Eq.\ (\ref{Gcomp12}).
Comparing the resultant equation with Eq.\ (\ref{Dyson-12-l}), we obtain the identity:
\begin{eqnarray}
\frac{1}{2}\int d1'\,
\bar{V}(1\!-\!1') \langle\hat{\psi}_{\rm H}^{\dagger}(2)\hat{\psi}_{\rm H}^{\dagger}(1')
\hat{\psi}_{\rm H}(1')
\hat{\psi}_{\rm H}(1)\rangle
&=& \pm \frac{i\hbar }{2}\int\bigl[\Sigma^{{\rm R}}(1,3)G^{12}(3,2)
\nonumber \\
& & + \Sigma^{12}(1,3)G^{{\rm A}}(3,2)\bigr] d3 .
\label{Hint-identity}
\end{eqnarray}
Setting $2\!=\! 1$ and performing an integration over ${\bm r}_{1}$ in Eq.\ (\ref{Hint-identity}), we obtain an expression of the interaction energy in terms of the self-energies. We further take its complex conjugate to average the resultant expression and the original one. 
Noting Eqs.\ (\ref{Gcomp-symm1}) and (\ref{G-RAK-symm}), we conclude that Eq.\ (\ref{Hint}) is indeed the interaction energy.

\subsection{Conservation laws}
\label{subsec:conserve}

We now show that conservation laws are automatically obeyed in the $\Phi$-derivable approximation.
First, taking complex conjugate of Eq.\ (\ref{Dyson-12-l}) and using Eqs.\ (\ref{Gcomp-symm1}) and (\ref{G-RAK-symm}), we obtain
\begin{equation}
\left(-i\hbar \frac{\partial }{\partial t_{2}}-\hat{K}_{2}\right)\!
G^{12}(1,2)-\int\left[G^{{\rm R}}(1,3)\Sigma^{12}(3,2)+G^{12}(1,3)\Sigma^{{\rm A}}(3,2)\right]d3 =0 .
\label{Dyson-12-r}
\end{equation}
We then subtract Eq.\ (\ref{Dyson-12-r}) from Eq.\ (\ref{Dyson-12-l}) with Eq.\ (\ref{K_1}) in mind, set
$2\!=\!1$, and make use of Eq.\ (\ref{gauge-identity2}).
We thereby arrive at the particle conservation law:
\begin{equation}
\frac{\partial n(1)}{\partial t_{1}}+{\bm\nabla}_{1}\!\cdot{\bm j}(1)=0 ,
\label{conserve-n}
\end{equation}
where $n(1)$ and ${\bm j}(1)$ are the particle density and the flux density, respectively,
defined by
\begin{subequations}
\label{nj}
\begin{eqnarray}
n(1)&\equiv& \pm  i\hbar  G^{12}(1,1) ,
\label{number}
\\
{\bm j}(1)&\equiv& \left.
\pm \hbar^{2}\frac{{\bm\nabla}_{1}\!-\!{\bm\nabla}_{2}}{2m} G^{12}(1,2) 
\right|_{2=1}.
\label{current}
\end{eqnarray}
\end{subequations}

Second, let us operate 
$\mp i\hbar({\bm\nabla}_{1}\!-\!{\bm\nabla}_{2})/2m$ on Eqs.\  (\ref{Dyson-12-l}) and (\ref{Dyson-12-r}),
subtract the latter from the former, and set $2\!=\! 1$.
It then turns out that the flux density ${\bm j}(1)$ satisfies
\begin{equation}
\frac{\partial}{\partial t_{1}}{\bm j}(1)
+\frac{1}{m}{\bm\nabla}_{1}\underline{\Theta}^{\rm K}(1)
+\frac{n(1)}{m}{\bm\nabla}_{1}U(1)
=\frac{1}{m}{\bm Q}(1) ,
\label{momentum-eq}
\end{equation}
where ${\bm Q}(1)$ is given by Eq.\ (\ref{vecS}), and the tensor $\underline{\Theta}^{\rm K}(1)$
is defined by
\begin{equation}
\Theta_{ij}^{\rm K}(1)=\mp\frac{i\hbar^{3}}{4m}
(\nabla_{\! 1i}\!-\!\nabla_{\! 2i})
(\nabla_{\! 1j}\!-\!\nabla_{\! 2j})G^{12}(1,2)\bigr|_{2=1} .
\label{Theta^K}
\end{equation}
We further integrate Eq.\ (\ref{momentum-eq}) over the whole space of the system and make use of Eq.\ (\ref{Galilean-identity}). We thereby obtain
\begin{equation}
\frac{\partial}{\partial t_{1}}\int {\bm j}(1)\,d^{3}r_{1}
=-\int \frac{n(1)}{m}{\bm\nabla}_{1}U(1)\,d^{3}r_{1}.
\label{conserve-p}
\end{equation}
Multiplying Eq.\ (\ref{conserve-p}) by $m$ yields the total momentum conservation law.

We finally focus on the energy conservation law.
Let us operate 
$\mp i\hbar\frac{\partial}{\partial t_{2}}$ and $\mp i\hbar\frac{\partial}{\partial t_{1}}$ on Eqs.\ 
(\ref{Dyson-12-l}) and (\ref{Dyson-12-r}), respectively, add the two equations subsequently, and set $2\!=\! 1$.
We then obtain
\begin{equation}
\frac{\partial {\cal E}^{\rm K}(1)}{\partial t_{1}} 
+{\bm\nabla}_{1}\cdot{\bm j}_{\varepsilon}'(1)
+U(1)\frac{\partial n(1)}{\partial t_{1}}
=Q_{\varepsilon}(1) ,
\label{energy-eq}
\end{equation}
where $Q_{\varepsilon}(1)$ is given by Eq.\ (\ref{Q}),
and ${\cal E}^{\rm K}(1)$ and ${\bm j}_{\varepsilon}'(1)$ are defined by
\begin{eqnarray}
{\cal E}^{\rm K}(1)&\equiv& \pm \frac{i\hbar^{3}}{2m}{\bm\nabla}_{1}\!\cdot\!{\bm\nabla}_{2}
G^{12}(1,2)\biggr|_{2=1}  ,
\label{Kinetic-e}
\\
{\bm j}_{\varepsilon}'(1)&\equiv& \mp \frac{i\hbar^{3}}{2m}
\!\left(\frac{\partial}{\partial t_{1}}{\bm\nabla}_{2}\!+\!
\frac{\partial}{\partial t_{2}}{\bm\nabla}_{1}\right)\! G^{12}(1,2)\biggr|_{2=1}.
\label{j_e'}
\end{eqnarray}
The quantity ${\cal E}^{\rm K}(1)$ above signifies the kinetic energy density.
We further integrate Eq.\ (\ref{energy-eq}) over the whole space of the system and make use of Eqs.\ 
(\ref{rubber-identity}) and (\ref{conserve-n}).
We thereby arrive at the total energy conservation law as
\begin{equation}
\frac{d}{d t_{1}}\!\left[\int {\cal E}^{\rm K}(1)\,d^{3}r_{1}
\!+\!\langle \hat{\cal H}_{\rm int}(t_{1})\rangle\right]=-
\int{\bm j}(1)\cdot\!{\bm\nabla}_{1}U(1)\,d^{3}r_{1}.
\label{conserve-e}
\end{equation}

Equations (\ref{conserve-n}), (\ref{conserve-p}), and 
(\ref{conserve-e}) are the main results of this subsection. They clearly show that conservation laws are obeyed in the $\Phi$-derivable approximation.

\subsection{Sum rule for the spectral function}

The spectral function $A$ defined by Eq.\ (\ref{A-def}) satisfies sum rule (\ref{A-sum}),
which stands for the equal-time commutation relation of $\hat{\psi}$ and $\hat{\psi}^{\dagger}$.
Although obvious intuitively, it may be worth showing that this property holds true in the dynamics of
the $\Phi$-derivable approximation.

Besides Eq.\ (\ref{Dyson-12-l}), we make use of the (2,1) component of Eq.\ (\ref{Dyson-check})
for this purpose, which reads as
\begin{equation}
\left(i\hbar \frac{\partial }{\partial t_{1}}-\hat{K}_{1} \right)\!
G^{21}(1,2)
-\int\left[\Sigma^{{\rm R}}(1,3)G^{21}(3,2)
+
\Sigma^{21}(1,3)G^{{\rm A}}(3,2)\right]d3 =0  ,
\label{Dyson-21-l}
\end{equation}
where we have used Eqs.\ (\ref{G-check-def}), (\ref{Gcomp}), and (\ref{G-RAK-def})
to express $\Sigma^{21}G^{11}-\Sigma^{22}G^{21}=\Sigma^{{\rm R}}G^{21}+\Sigma^{21}G^{{\rm A}}$.
Let us subtract Eq.\ (\ref{Dyson-12-l}) from Eq.\ (\ref{Dyson-21-l}) and set $t_{2}=t_{1}$ subsequently.
It then follows that the self-energy terms cancel in the resultant equation, which can be expressed
with Eqs.\ (\ref{K_1}) and (\ref{A-def}) as
\begin{equation}
\left[i\hbar \frac{\partial }{\partial t_{1}}+
\frac{\hbar^{2}\nabla_{1}^{2}}{2m}-U(1)+\mu
\right]\!
A(1,2)\biggl|_{t_{2}=t_{1}}=0 .
\label{A-l}
\end{equation}
Let us take complex conjugate of Eq.\ (\ref{A-l}) with $A^{*}(1,2)=A(2,1)$ in mind, subsequently exchange arguments as ${\bm r}_{1}\leftrightarrow {\bm r}_{2}$, and subtract the resultant equation from Eq.\ (\ref{A-l}). 
We thereby obtain an equation of motion for $A({\bm r}_{1}t_{1},{\bm r}_{2}t_{1})$ as
\begin{equation}
\frac{\partial A({\bm r}_{1}t_{1},{\bm r}_{2}t_{1})}{\partial t_{1}}
+\frac{\hbar(\nabla_{1}^{2}-\nabla_{2}^{2})}{2mi}A({\bm r}_{1}t_{1},{\bm r}_{2}t_{1})
-\frac{U({\bm r}_{1}t_{1})-U({\bm r}_{2}t_{1})}{i\hbar}
A({\bm r}_{1}t_{1},{\bm r}_{2}t_{1})=0 .
\label{A-eq}
\end{equation}
Now, suppose $A({\bm r}_{1}t_{1},{\bm r}_{2}t_{1})=\delta({\bm r}_{1}-{\bm r}_{2})$ at $t_{1}$.
Then, one may see easily that the second and third terms in Eq.\ (\ref{A-eq}) vanish, 
so that ${\partial A({\bm r}_{1}t_{1},{\bm r}_{2}t_{1})}/{\partial t_{1}}=0$ at $t_{1}$.
Hence, we conclude that sum rule (\ref{A-sum}) holds at any $t\geq t_{1}$.

\subsection{\label{app:conserve-W}Conservation laws in the local approximation}

Equations (\ref{conserve-n}), (\ref{momentum-eq}), and (\ref{energy-eq}), 
which describe time evolutions of the particle, momentum, and energy densities, respectively,
are of fundamental importance.
We express them here in the Wigner representation with the first-order gradient expansion
so as to be compatible with the transport Eqs.\ (\ref{G^R}) and (\ref{transport}).

Substituting Eqs.\ (\ref{checkG-Wig2}) and (\ref{G_12-Wig}) into Eq.\ (\ref{nj}), 
we can express $n({\bm r}t)$ and ${\bm j}({\bm r}t)$ in terms of $A$ and $\phi$.
Those expressions are free from approximations and satisfy Eq.\ (\ref{conserve-n}).
To be more specific, the particle density $n({\bm r}t)$ and the fluid velocity ${\bm v}({\bm r}t)\!\equiv\!{\bm j}({\bm r}t)/n({\bm r}t)$
are given by
\begin{subequations}
\label{nj-p}
\begin{eqnarray}
n({\bm r}t)&=& 
\hbar\int\frac{d^{3}p\,d\varepsilon}{(2\pi\hbar)^{4}}
A({\bm p}\varepsilon,{\bm r}t)\phi({\bm p}\varepsilon,{\bm r}t) ,
\label{rho}
\\
{\bm v}({\bm r}t)
&=&\frac{\hbar}{n({\bm r}t)}\int\frac{d^{3}p\,d\varepsilon}
{(2\pi\hbar)^{4}}\,
\frac{\bm p}{m}A({\bm p}\varepsilon,{\bm r}t)\phi({\bm p}\varepsilon,{\bm r}t) .
\label{v}
\end{eqnarray}
\end{subequations}
Equation (\ref{conserve-n}) is expressed in terms of these quantities as
\begin{equation}
\frac{\partial n}{\partial t}+{\bm\nabla}(n{\bm v})
=0 .
\label{continuity}
\end{equation}

We next consider Eq.\ (\ref{momentum-eq}).
By substituting Eqs.\ (\ref{checkG-Wig2}) and (\ref{G_12-Wig}) into it, Eq.\ (\ref{Theta^K}) is transformed into
\begin{equation}
\Theta_{ij}^{\rm K}({\bm r}t)= \hbar
\int \frac{d^{3}p\,d\varepsilon}{(2\pi\hbar)^{4}}\,
\frac{p_{i}p_{j}}{m}
A({\bm p}\varepsilon,{\bm r}t)
\phi({\bm p}\varepsilon,{\bm r}t) 
= mn({\bm r}t) v_{i}({\bm r}t)v_{j}({\bm r}t) +\Pi_{ij}^{\rm K}({\bm r}t) ,
\label{Theta^K-p}
\end{equation}
where $\Pi_{ij}^{\rm K}({\bm r}t)$ signifies the kinetic-energy part of the momentum flux density tensor in the coordinate system moving with the local fluid velocity, given explicitly by
\begin{equation}
\Pi_{ij}^{\rm K}({\bm r}t) \equiv \hbar
\int \frac{d^{3}p\,d\varepsilon}{(2\pi\hbar)^{4}}\,
\frac{\bar{p}_{i}\bar{p}_{j}}{m}
A({\bm p}\varepsilon,{\bm r}t)
\phi({\bm p}\varepsilon,{\bm r}t)  ,\hspace{10mm}\bar{\bm p}\equiv{\bm p}-m{\bm v}.
\label{Pi^K}
\end{equation}
We next consider ${\bm Q}(1)$ in Eq.\ (\ref{momentum-eq}).
Substituting Eq.\ (\ref{Hint-identity}) and its complex conjugate into Eq.\ 
(\ref{vecS}), we obtain
$$
{\bm Q}(1)\!
=-\int d^{3}r_{1}'
\frac{\partial V({\bm r}_{1}\!-\!{\bm r}_{1}')}{\partial {\bm r}_{1}}
\langle\hat{\psi}_{\rm H}^{\dagger}(1)\hat{\psi}_{\rm H}^{\dagger}(1')
\hat{\psi}_{\rm H}(1')
\hat{\psi}_{\rm H}(1)\rangle ,
$$
with $1'\!=\!{\bm r}_{1}'t_{1}$ in this expression.
We next express $\langle\hat{\psi}_{\rm H}^{\dagger}(1)\hat{\psi}_{\rm H}^{\dagger}(1')
\hat{\psi}_{\rm H}(1')
\hat{\psi}_{\rm H}(1)\rangle=\rho_{2}\bigl({\bm r}_{1}\!-\!{\bm r}_{1}',
\frac{{\bm r}_{1}+{\bm r}_{1}'}{2},t_{1}\bigr)$ and expand  
$\frac{{\bm r}_{1}+{\bm r}_{1}'}{2}$ from ${\bm r}_{1}$ up to the first order in the Taylor series.
We thereby obtain
\begin{equation}
{\bm Q}(1) = -{\bm\nabla}_{1} \underline{\Pi}^{{\rm P}}(1),
\label{vecQ-2}
\end{equation}
where $\Pi_{ij}^{{\rm P}}(1)$ is defined by
\begin{equation}
\Pi_{ij}^{{\rm P}}(1)\equiv-\frac{1}{2}
\int d^{3}\bar{r}\,
\bar{r}_{i} \frac{\partial V(\bar{\bm r})}{\partial \bar{r}_{j}}
\langle\hat{\psi}_{\rm H}^{\dagger}(1_{+})\hat{\psi}_{\rm H}^{\dagger}(1_{-})
\hat{\psi}_{\rm H}(1_{-})
\hat{\psi}_{\rm H}(1_{+})\rangle ,
\label{Pi^V}
\end{equation}
with $1_{\pm}\!\equiv\!({\bm r}_{1}\!\pm\!{\bar{\bm r}}/{2},t_{1})$.
Noting Eq.\ (\ref{vecQ-2}), we realize that $\Pi_{ij}^{{\rm P}}$ should be estimated with the local approximation
in the first-order gradient expansion scheme.
Hence, we express $\Pi_{ij}^{{\rm P}}$  with ${\check{\check{\cal K}}}$ in Eq.\ (\ref{K})
and substitute Eqs.\ (\ref{V-Fourier}), (\ref{checkG-Wig2}), and (\ref{K-Wig}) into the resultant expression. Carrying out a partial integration over $\bar{\bm r}$ subsequently, 
we obtain
\begin{eqnarray}
&&\hspace{-10mm}
\Pi_{ij}^{{\rm P}}({\bm r}t)=\frac{(i\hbar)^{2}}{2}\int 
\frac{d^{3}q\,d\omega}{(2\pi\hbar)^{4}}
\int \frac{d^{3}p\,d\varepsilon}{(2\pi\hbar)^{4}}
\int \frac{d^{3}p'd\varepsilon'}{(2\pi\hbar)^{4}}
\left(V_{q}\delta_{ij}\!+\!\frac{q_{i}q_{j}}{q}\frac{dV_{q}}{dq}\right)\!
\nonumber \\
&&\hspace{7mm}
\times\bigl[{\cal K}^{12,12}({\bm p}\varepsilon,{\bm p}'\varepsilon';{\bm q}\omega,{\bm r}t)
+(2\pi\hbar)^{4}
\delta({\bm q})\delta(\omega)G^{12}({\bm p}\varepsilon,{\bm r}t)
G^{12}({\bm p}'\varepsilon',{\bm r}t)
\bigr] .
\label{Pi^V2}
\end{eqnarray}
Now, let us substitute ${\bm j}=n{\bm v}$, Eq.\ (\ref{Theta^K-p}), and Eq.\ (\ref{vecQ-2}) into Eq.\ (\ref{momentum-eq}) and make use of Eq.\ (\ref{continuity}).
We thereby arrive at 
\begin{equation}
\frac{\partial {\bm v}}{\partial t}+{\bm v}\!\cdot\!{\bm\nabla}{\bm v}
+\frac{1}{mn}{\bm\nabla}\underline{\Pi}+\frac{{\bm\nabla}U}{m}=0 ,
\label{NS}
\end{equation}
where $\Pi_{ij}\equiv \Pi_{ij}^{\rm K}+\Pi_{ij}^{{\rm P}}$.
By using Eqs.\ (\ref{Pi^K}) and (\ref{Pi^V2}), the  tensor $\Pi_{ij}$ can be written explicitly as
\begin{eqnarray}
&&\hspace{-10mm}
\Pi_{ij}({\bm r}t)
\equiv \hbar
\int \frac{d^{3}p\,d\varepsilon}{(2\pi\hbar)^{4}}\,
\frac{\bar{p}_{i}\bar{p}_{j}}{m}
A({\bm p}\varepsilon,{\bm r}t)
\phi({\bm p}\varepsilon,{\bm r}t)
\nonumber \\
&&\hspace{6mm}
+\frac{(i\hbar)^{2}}{2}\int 
\frac{d^{3}q\,d\omega}{(2\pi\hbar)^{4}}
\int \frac{d^{3}p\,d\varepsilon}{(2\pi\hbar)^{4}}
\int \frac{d^{3}p'd\varepsilon'}{(2\pi\hbar)^{4}}
\left(V_{q}\delta_{ij}\!+\!\frac{q_{i}q_{j}}{q}\frac{dV_{q}}{dq}\right)
\nonumber \\
&&\hspace{6mm}
\times 
\bigl[{\cal K}^{12,12}({\bm p}\varepsilon,{\bm p}'\varepsilon';{\bm q}\omega,{\bm r}t)
+(2\pi\hbar)^{4}
\delta({\bm q})\delta(\omega)G^{12}({\bm p}\varepsilon,{\bm r}t)
G^{12}({\bm p}'\varepsilon',{\bm r}t)
\bigr] ,
\label{Pi}
\end{eqnarray}
where $\bar{\bm p}={\bm p}-m{\bm v}$,
$G^{12}$ is given by Eq.\ (\ref{G_12-Wig}), and 
${\check{\check{\cal K}}}$
is the solution of the Bethe-Salpeter equation (\ref{K-2}) in the local approximation.
Note $\underline{\Pi}$ is symmetric.

We finally derive the differential energy conservation law.
Instead of Eq.\ (\ref{energy-eq}), it is convenient for this purpose to start from the expression of the energy density:
\begin{equation}
{\cal E}(1)\equiv 
\frac{\hbar^{2}}{2m}{\bm\nabla}_{1}'\!\cdot\!{\bm\nabla}_{1}
\langle\hat{\psi}^{\dagger}_{\rm H}(1')\hat{\psi}_{\rm H}(1)\rangle\bigr|_{1'=1}
+\frac{1}{2}\!\int\!d^{3}r_{1}'\,
V({\bm r}_{1}\!-\!{\bm r}_{1}')
\langle\hat{\psi}_{\rm H}^{\dagger}(1)\hat{\psi}_{\rm H}^{\dagger}(1')
\hat{\psi}_{\rm H}(1')
\hat{\psi}_{\rm H}(1)\rangle ,
\label{E-density}
\end{equation}
where $t_{1}'\!=\!t_{1}$ is implied. Let us differentiate Eq.\ (\ref{E-density}) with respect to $t_{1}$ and remove the 
time derivatives of the field operators in the resultant expression with Eq.\ (\ref{eq-motion}). 
We also carry out the first-order gradient expansion for the interaction term by the procedure of deriving Eq.\ 
(\ref{vecQ-2}). We thereby obtain the differential energy conservation law:
\begin{equation}
\frac{\partial{\cal E}({\bm r}t)}{\partial t}+{\bm\nabla}\cdot 
{\bm j}_{\varepsilon}({\bm r}t)=
-{\bm j}({\bm r}t)\cdot{\bm\nabla}U({\bm r}t) .
\label{continuity-E}
\end{equation}
Here, ${\bm j}_{\varepsilon}$ denotes the energy flux density defined in terms of the operator
\begin{equation}
\hat{\!\bm j}({\bm r}t)\equiv \frac{\hbar}{2mi}({\bm\nabla}\!-\!{\bm\nabla}')
\hat{\psi}_{\rm H}^{\dagger}({\bm r}'t)\hat{\psi}_{\rm H}({\bm r}t)
\bigr|_{{\bm r}'={\bm r}} ,
\end{equation}
by
\begin{eqnarray}
{\bm j}_{\varepsilon}({\bm r}t)
&\equiv& \pm
\frac{\hbar^{4}}{4m^{2}}({\bm\nabla}\!-\!
{\bm\nabla}'){\bm\nabla}\!\cdot\!{\bm\nabla}'G^{12}({\bm r}t,{\bm r}'t)
\biggr|_{{\bm r}'={\bm r}} \!
+\frac{1}{2}\!\int d\bar{\bm r}\,V(\bar{\bm r})\langle \hat{\psi}_{\rm H}^{\dagger}({\bm r}_{-}t)\,
\hat{\!\bm j}({\bm r}_{+}t)
\nonumber \\
& &
\times
\hat{\psi}_{\rm H}({\bm r}_{-}t)\rangle
-\frac{1}{4}\int d^{3}\bar{r}\,
\bar{\bm r}\,\frac{\partial V(\bar{\bm r})}{\partial \bar{\bm r}}\cdot
\bigl[\langle \hat{\psi}_{\rm H}^{\dagger}({\bm r}_{-}t)\,
\hat{\!\bm j}({\bm r}_{+}t)\hat{\psi}_{\rm H}({\bm r}_{-}t)\rangle
+\langle \hat{\psi}_{\rm H}^{\dagger}({\bm r}_{+}t)\,
\nonumber \\
& &
\times \,\hat{\!\bm j}({\bm r}_{-}t)
\hat{\psi}_{\rm H}({\bm r}_{+}t)\rangle\bigr] .
\label{j_e}
\end{eqnarray}
Noting Eq.\ (\ref{continuity-E}), we realize that Eqs.\ 
(\ref{E-density}) and (\ref{j_e}) should be estimated with the local approximation in the first-order gradient expansion scheme.
It is carried out by the procedure of deriving Eqs.\ (\ref{Theta^K-p}) and (\ref{Pi^V2}).
We thereby obtain
\begin{subequations}
\label{Ej_e}
\begin{eqnarray}
{\cal E}&=&\frac{1}{2}mnv^{2}+{\cal E}_{\rm i} ,
\\
{\bm j}_{\varepsilon}&=&\frac{1}{2}mnv^{2}{\bm v}+{\cal E}_{\rm i}{\bm v}
+\underline{\Pi}{\bm v}+{\bm j}_{Q} .
\end{eqnarray}
\end{subequations}
Here, $\underline{\Pi}$ is given by Eq.\ (\ref{Pi}), and ${\cal E}_{\rm i}$ and 
${\bm j}_{Q}$ are defined in terms of $\bar{\bm p}\!=\!{\bm p}\!-\!m{\bm v}$
by
\begin{subequations}
\label{tildeEj_e}
\begin{eqnarray}
{\cal E}_{\rm i}({\bm r}t)&\equiv&
\hbar\int\frac{d^{3}p\,d\varepsilon}{(2\pi\hbar)^{4}}\frac{\bar{p}^{2}}{2m}
A({\bm p}\varepsilon,{\bm r}t)
\phi({\bm p}\varepsilon,{\bm r}t) 
\nonumber \\
& &
\pm \frac{i\hbar}{2} 
\int \frac{d^{3}p\,d\varepsilon}{(2\pi\hbar)^{4}}
[\Sigma^{\rm R}({\bm p}\varepsilon,{\bm r}t)
G^{12}({\bm p}\varepsilon,{\bm r}t)
+\Sigma^{12}({\bm p}\varepsilon,{\bm r}t)G^{\rm A}({\bm p}\varepsilon,{\bm r}t)],
\label{tildeE}
\\
{\bm j}_{Q}({\bm r}t)&\equiv&
\hbar\int\frac{d^{3}p\,d\varepsilon}
{(2\pi\hbar)^{4}}\frac{\bar{p}^{2}}{2m^{2}}\bar{\bm p}
A({\bm p}\varepsilon,{\bm r}t)
\phi({\bm p}\varepsilon,{\bm r}t)
\nonumber \\
& &
+\frac{(i\hbar)^{2}}{2}\int 
\frac{d^{3}q\,d\omega}{(2\pi\hbar)^{4}}
\int \frac{d^{3}p\,d\varepsilon}{(2\pi\hbar)^{4}}
\int \frac{d^{3}p'd\varepsilon'}{(2\pi\hbar)^{4}}
\frac{1}{m}\!\left(2\bar{\bm p} V_{q}
+{\bm q} \frac{{\bm q}\cdot\bar{\bm p}}{q}
\frac{d V_{q}}{d q}\right)
\nonumber \\
& &
\times\bigl[{\cal K}^{12,12}({\bm p}\varepsilon,{\bm p}'\varepsilon';{\bm q}\omega,{\bm r}t)
+(2\pi\hbar)^{4}
\delta({\bm q})\delta(\omega)G^{12}({\bm p}\varepsilon,{\bm r}t)
G^{12}({\bm p}'\varepsilon'\!,{\bm r}t)
\bigr] ,
\label{j_Q}
\end{eqnarray}
\end{subequations}
respectively.  In deriving Eq.\ (\ref{tildeE}), we have used Eq.\ (\ref{Hint-identity}).
The quantity ${\cal E}_{\rm i}$ signifies the internal-energy density in the coordinate system moving with the local fluid velocity ${\bm v}$, and ${\bm j}_{Q}$ denotes the heat flux density.

Let us substitute Eq.\ (\ref{Ej_e}) into Eq.\ (\ref{continuity-E}) and transform the resultant expression with Eqs.\ 
(\ref{continuity}) and (\ref{NS}). 
We thereby obtain the differential energy conservation law as
\begin{equation}
\frac{\partial{\cal E}_{\rm i}}{\partial t}+
{\bm\nabla}\cdot({\cal E}_{\rm i}{\bm v}+{\bm j}_{Q})
+\sum_{ij}\Pi_{ij}\frac{\partial v_{j}}{\partial r_{i}}=0 ,
\label{continuity-E2}
\end{equation}
which for ${\bm v}={\bm 0}$ reduces to the heat equation.

Equations (\ref{continuity}), (\ref{NS}), and (\ref{continuity-E2}) are the differential particle, momentum, and energy
conservation laws in the first-order gradient expansion, respectively. The quantities in those equations are defined by Eqs.\ (\ref{nj-p}), (\ref{Pi}), and (\ref{tildeEj_e}). They can be calculated microscopically with the solutions of Eqs.\ (\ref{G^R}), (\ref{transport}), and (\ref{K-2}) in the local approximation.

\section{Transport equation and fluid mechanics\label{sec:FM}}

In most cases, fluid mechanics is constructed phenomenologically on the basis of the local conservation laws of particle number, momentum, and energy\cite{LL-FM} without recourse to statistical mechanics. The Navier-Stokes equation thereby obtained is an archetype of nonlinear evolution equations embracing a wide range of nonequilibrium phenomena such as turbulence, chaos, and pattern formation. Since it consists of many particles, however, fluids should be described on a microscopic ground in a statistical mechanical manner, including its fluctuations.

Taking a dilute classical monatomic gas as an example, we show here that the basic equations of fluid mechanics can be derived microscopically from the transport equation. To be specific, we start from the Boltzmann equation and subsequently adopt Enskog's expansion from local equilibrium\cite{CC90,HCB54} to derive equations of fluid mechanics appropriate for dilute classical gases. The consideration here enables us (i) to provide a microscopic basis of those equations, (ii) to calculate various transport coefficients such as the viscosity coefficient from first principles, and also (iii) to determine the form of the nonequilibrium distribution function. The distribution function thus obtained may be used with the method of \S \ref{two-body} to calculate two-body correlations and fluctuations in nonequilibrium fluids. Thus, nonequilibrium systems are not mysterious at all, but can be handled in a statistical mechanical manner with the transport equation. One may also convince oneself after reading this section that typical nonequilibrium phenomena such as turbulence, chaos, and pattern formation in fluid mechanics lie fairly close to equilibrium.

The contents of this section are described in detail in the classic textbooks by Chapman and Cowling \cite{CC90} and by Hirschfelder et al.\ \cite{HCB54}
It may be useful, however, to include them here in the general framework of nonequilibrium  quantum field theory, which can describe quantum effects and strong correlations
far beyond the Boltzmann equation, to elucidate its great potential.
Moreover, we proceed here by regarding Enskog's expansion as an approximate method to determine the distribution function, i.e.,  the key quantity in statistical mechanics, rather than jumping directly onto calculations of transport coefficients. \cite{CC90}  The derivation here follows that in Ref.\ \citen{Kita06b}.

Our consideration starts from the Boltzmann Eq.\ (\ref{Boltzmann}), i.e., 
\begin{equation}
\frac{\partial f}{\partial t}+\frac{{\bm p}}{m}\cdot\frac{\partial f}{\partial {\bm r}}
-\frac{\partial U}{\partial {\bm r}}\cdot\frac{\partial f}{\partial {\bm p}}=I_{\bm p}[f].
\label{Boltzmann-C}
\end{equation}
The collision integral $I_{\bm p}[f]$ in the dilute classical limit is obtained from Eq.\ (\ref{I_p})
by neglecting the exchange effect as 
$\pm V_{{\bm p}-{\bm p}_{3}}V_{{\bm p}-{\bm p}_{4}}^{*}\rightarrow  0$ and taking the classical limit as $(1\pm f)\rightarrow  1$.
Performing a change of variables ${\bm p}_{2}={\bm p}+{\bm q}$,
${\bm p}_{3}={\bm P}-{\bm q}'/2$, and ${\bm p}_{4}={\bm P}+{\bm q}'/2$ subsequently, we obtain
$I_{\bm p}[f]$ as
\begin{equation}
I_{\bm p}[f]=\int\frac{d^{3}q}{(2\pi\hbar)^{3}}\,\frac{q}{m}\int
d\sigma\, [f_{{\bm p}+({\bm q}-{\bm q}')/2}
f_{{\bm p}+({\bm q}+{\bm q}')/2}-f_{{\bm p}}f_{{\bm p}+{\bm q}}] .
\label{C2}
\end{equation}
Here, $d\sigma\equiv d\Omega_{{\bm q}'}\int dq'
\delta(q'-q)\left[m|V_{({\bm q}'-{\bm q})/2}|/4\pi\hbar^{2}\right]^{2}$ is the differential cross section 
in the coordinate system moving with the mass center\cite{LL-M} and $d\Omega_{{\bm q}'}$
signifies the infinitesimal solid angle along ${\bm q}'$.
In the special case of the contact potential with  $V_{{\bm q}}=\mbox{const}$, the differential cross section can be expressed concisely in terms of the scattering length $a\!\equiv\! m|V|/4\pi\hbar^{2}$ by
\begin{equation}
d\sigma=a^{2}d\Omega_{{\bm q}'}\int dq'
\delta(q'-q) .
\label{dsigma}
\end{equation}
The expression corresponds to the scattering of two rigid particles with radius $a$. \cite{LL-M}
We will adopt Eq.\ (\ref{dsigma}) below for the differential cross section.

We next write down the conservation laws obtained from the Boltzmann equation.
Relevant quantities for this purpose are the particle density $n({\bm r},t)$, fluid velocity ${\bm v}({\bm r},t)$, local temperature
$T({\bm r},t)$, momentum flux density tensor $\underline{\Pi}({\bm r},t)$ in the coordinate system moving with ${\bm v}({\bm r},t)$,
and heat flux density ${\bm j}_{Q}({\bm r},t)$. They are defined by
\begin{subequations}
\label{thermoQ}
\begin{eqnarray}
n({\bm r},t)&\equiv& \int\frac{d^{3}p}{(2\pi\hbar)^{3}}f({\bm p},{\bm r},t) ,
\label{n-C}
\\
{\bm v}({\bm r},t)&\equiv& \frac{1}{n({\bm r},t)}
\int\frac{d^{3}p}{(2\pi\hbar)^{3}}\frac{{\bm p}}{m}
f({\bm p},{\bm r},t) ,
\label{v-C}
\\
T({\bm r},t)&\equiv& \frac{2}{3k_{\rm B}n({\bm r},t)}
\int\frac{d^{3}p}{(2\pi\hbar)^{3}}\frac{\bar{p}^{2}}{2m}
f({\bm p},{\bm r},t) ,
\label{T-C}
\\
\underline{\Pi}({\bm r},t)&\equiv& 
\int\frac{d^{3}p}{(2\pi\hbar)^{3}}\frac{\bar{\bm p}\bar{\bm p}}{m}
f({\bm p},{\bm r},t) ,
\label{Pi-C}
\\
{\bm j}_{Q}({\bm r},t)&\equiv& 
\int\frac{d^{3}p}{(2\pi\hbar)^{3}}\frac{\bar{p}^{2}}{2m}
\frac{\bar{\bm p}}{m}f({\bm p},{\bm r},t) ,
\label{j_Q-C}
\end{eqnarray}
\end{subequations}
with $\bar{\bm p}\!\equiv\!{\bm p}-m{\bm v}$.
These can also be obtained from general expressions
(\ref{rho}), (\ref{v}), (\ref{tildeE}), (\ref{Pi}), and (\ref{j_Q})
by neglecting the interaction, rewriting the resultant expressions with $f$ in Eq.\ (\ref{f-def-QP}),
and approximating the internal energy density by
${\cal E}_{\rm i}\!=\!\frac{3}{2}nk_{\rm B}T$ for ideal monatomic gases.

With these preliminaries, we now derive particle number, momentum, and energy conservation laws. 
Let us multiply Eq.\ (\ref{Boltzmann-C}) by $1$, ${\bm p}/m$, and
$\bar{p}^{2}/2m$, and carry out an integration over ${\bm p}$.
It then turns out that the contributions of collision integral (\ref{C2}) vanish in the resultant equations
due to particle number, momentum, and energy conservations in the two-particle collision.
Thus, the conservation laws are expressed with the quantities in Eq.\ (\ref{thermoQ}) as
\begin{subequations}
\label{continuity-C}
\begin{equation}
\frac{\partial n}{\partial t}+{\bm \nabla}(n{\bm v})
=0  ,
\label{continuity-C-n}
\end{equation}
\begin{equation}
\frac{\partial {\bm v}}{\partial t}+{\bm v}\!\cdot\!{\bm \nabla}{\bm v}
+\frac{1}{mn}{\bm \nabla}\underline{\Pi}+\frac{{\bm\nabla}U}{m} ={\bm 0} ,
\label{continuity-C-p}
\end{equation}
\begin{equation}
\frac{3}{2}nk_{\rm B}\!\left(\frac{\partial T}{\partial t}+
{\bm v}\cdot{\bm\nabla}T\right)\!+{\bm\nabla}\!\cdot{\bm j}_{Q}
+\underline{\Pi}:\!{\bm \nabla}{\bm v}=0 ,
\label{continuity-C-E}
\end{equation}
\end{subequations}
where $\underline{A}\!:\!\underline{B}$ denotes the tensor product:
\begin{equation}
\underline{A}\!:\!\underline{B}\!\equiv\!\sum_{ij}A_{ij}B_{ji}.
\end{equation}
Equations (\ref{continuity-C-n})--(\ref{continuity-C-E}) are the classical-gas versions of the general conservation laws
(\ref{continuity}), (\ref{NS}), and (\ref{continuity-E2}) in the local approximation, respectively. 
Indeed, Eqs.\ (\ref{continuity-C-n}) and (\ref{continuity-C-p}) are exactly identical with Eqs.\ 
(\ref{continuity}) and (\ref{NS}), respectively, and Eq.\ (\ref{continuity-C-E}) can be derived from Eq.\ (\ref{continuity-E2}) by setting  ${\cal E}_{\rm i}\!=\!\frac{3}{2}nk_{\rm B}T$.

On the basis of Enskog's expansion from the local equilibrium,\cite{CC90,HCB54} we now reduce the Boltzmann Eq.\ (\ref{Boltzmann-C}), which is nonlinear with the variables
$({\bm p},{\bm r},t)$, to the problem of solving Eq.\ (\ref{continuity-C}) with 
 the variables $({\bm r},t)$.
Let us expand $f$ as
\begin{equation}
f({\bm p},{\bm r},t)=f^{({\rm le})}({\bm p},{\bm r},t)
\left[1\!+\!\varphi^{(1)}({\bm p},{\bm r},t)
+\cdots\right] ,
\label{Enskog}
\end{equation}
where $f^{({\rm le})}$ denotes the local-equilibrium distribution:
\begin{equation}
f^{({\rm le})}=\frac{(2\pi\hbar)^{3}n}{(2\pi mk_{\rm B}T)^{3/2}}\exp\!\left(
-\frac{\bar{p}^{2}}{2mk_{\rm B}T}\right) ,
\label{f_eq}
\end{equation}
with $\bar{\bm p}\equiv{\bm p}-m{\bm v}$.
This form of $f^{({\rm le})}$ has been chosen to satisfy: \cite{CC90,HCB54} (i) the local-equilibrium condition that $I_{\bm p}[f]$ vanishes, and (ii) Eqs.\ (\ref{n-C})--(\ref{T-C}) by itself.
Hence, it follows that the higher-order corrections $\varphi^{(j)}$
$(j\!=\!1,2,\cdots)$ in Eq.\ (\ref{Enskog}) should satisfy 
\begin{equation}
\int\frac{d^{3}p}{(2\pi\hbar)^{3}}\bar{\bm p}^{n}f^{({\rm le})}\varphi^{(j)}=0  .
\hspace{10mm}(n=0,1,2)
\label{constraint}
\end{equation}
There are five unknown functions  in $f^{({\rm le})}$, i.e., $n({\bm r},t)$, ${\bm v}({\bm r},t)$, and $T({\bm r},t)$,
which are the same in number as Eq.\ (\ref{continuity-C}).
However, the latter still contains two unknown functions 
$\underline{\Pi}$ and ${\bm j}_{Q}$.

We now express $\underline{\Pi}$ and ${\bm j}_{Q}$ in terms of $(n,{\bm v},T)$ so that 
Eq.\ (\ref{continuity-C}) becomes a closed equation for $(n,{\bm v},T)$.
Let us substitute Eq.\ (\ref{Enskog}) into Eq.\ (\ref{Boltzmann-C}) and regard the differential operators
on the left-hand side as the first-order quantities in Enskog's expansion.
Also noting $I_{\bm p}[f^{({\rm le})}]=0$, we obtain the following first-order equation:
\begin{equation}
\frac{\partial f^{({\rm le})}}{\partial t}+
\frac{{\bm p}}{m}\cdot\frac{\partial f^{({\rm le})}}{\partial{\bm r}}
+{\bm \nabla}U \cdot \frac{\bar{\bm p}}{m k_{\rm B}T}f^{({\rm le})}=I_{\bm p}^{(1)}[\varphi^{(1)}] ,
\label{Boltzmann1}
\end{equation}
where $I_{\bm p}^{(1)}[\varphi^{(1)}]$ is given with Eqs.\ (\ref{C2}) and (\ref{Enskog}) by
\begin{equation}
I_{\bm p}^{(1)}[\varphi^{(1)}]=\int\!\frac{d^{3}q}{(2\pi\hbar)^{3}}\,\frac{q}{m}\int\!
d\sigma\, f^{({\rm le})}_{{\bm p}}f^{({\rm le})}_{{\bm p}+{\bm q}}
\left(\varphi^{(1)}_{{\bm p}+({\bm q}-{\bm q}')/2}+
\varphi^{(1)}_{{\bm p}+({\bm q}+{\bm q}')/2}-\varphi^{(1)}_{{\bm p}}
-\varphi^{(1)}_{{\bm p}+{\bm q}}\right).
\label{C^1}
\end{equation}
Using Eq.\ (\ref{f_eq}), we next express derivatives of $f^{({\rm le})}$ in Eq.\ (\ref{Boltzmann1})
with respect to those of $(n,{\bm v},T)$. We further remove resultant time derivatives 
with the conservation laws (\ref{continuity-C}) in the local approximation.
To this end, we make use of the following relations:
\begin{equation}
\underline{\Pi}^{({\rm le})}\!=\!P\underline{1},\hspace{10mm}
{\bm j}_{Q}^{({\rm le})}\!=\!{\bm 0},
\label{j_QPi^eq}
\end{equation}
with $P\!\equiv\!nk_{\rm B}T$ and 
$\underline{1}$ as the unit tensor, which are obtained by substituting $f=f^{({\rm le})}$ into Eq.\ (\ref{thermoQ}).
The left-hand side of Eq.\ (\ref{Boltzmann1}) is thereby transformed into
\begin{equation}
\frac{\partial f^{({\rm le})}}{\partial t}+
\frac{{\bm p}}{m}\cdot\frac{\partial f^{({\rm le})}}{\partial{\bm r}}
+\frac{\bar{\bm p}\cdot{\bm\nabla}U}{mk_{\rm B}T}f^{({\rm le})}
=\left[2
\!\left({\bm k}{\bm k}\!-\!\frac{k^{2}}{3}\underline{1}\right)
\!:\!{\bm \nabla}{\bm v}
+\!\left(k^{2}\!-\!\frac{5}{2}\right)\!
\frac{\bar{\bm p}}{m}\!\cdot\!{\bm \nabla}\ln T\right] \! f^{({\rm le})},
\label{Boltzmann1-l}
\end{equation}
where ${\bm k}$ is a dimensionless vector defined by
\begin{subequations}
\label{dimensionless}
\begin{equation}
{\bm k}\equiv\bar{\bm p}/\sqrt{2mk_{\rm B}T} .
\label{k-def}
\end{equation}
We now introduce a couple of additional dimensionless quantities:
\begin{equation}
\hat{\bm v}\equiv \sqrt{\frac{2k_{\rm B}T}{m}}{\bm v} ,
\hspace{10mm}\hat{\bm r}\equiv n^{-1/3}{\bm r}  ,
\label{v-rHat}
\end{equation}
\end{subequations}
and the mean-free path:
\begin{equation}
l\equiv \frac{1}{4\sqrt{2}\pi a^{2}n} .
\label{l}
\end{equation}
Let us rewrite Eqs.\  (\ref{dsigma}) and (\ref{f_eq}) in terms of these dimensionless quantities, 
put the resultant expressions into Eq.\ (\ref{C^1}), 
and substitute it together with Eq.\ (\ref{Boltzmann1-l}) into Eq.\ (\ref{Boltzmann1}).
We thereby obtain
\begin{eqnarray}
&&\hspace{0mm}
\frac{2\sqrt{2}}{\sqrt{\pi}ln^{1/3}}\!\int\!\frac{d^{3}q}{4\pi}
\!\int\!\frac{d^{3}q'}{4\pi}\frac{\delta(q'\!-\!q)}{q}e^{-k^{2}
-({\bm k}+{\bm q})^{2}} 
\bigl[\varphi^{(1)}_{{\bm k}+({\bm q}-{\bm q}')/2}+
\varphi^{(1)}_{{\bm k}+({\bm q}+{\bm q}')/2}\!-\!\varphi^{(1)}_{{\bm k}}
\!-\!\varphi^{(1)}_{{\bm k}+{\bm q}}\bigr] 
\nonumber \\
&&\hspace{0mm}
=e^{-k^{2}}
\!\left[2
\!\left({\bm k}{\bm k}\!-\!\frac{k^{2}}{3}\underline{1}\right)
\!:\!\hat{\bm\nabla}\hat{\bm v}
+\!\left(k^{2}\!-\!\frac{5}{2}\right)\!
{\bm k}\!\cdot\!\hat{\bm\nabla}\ln T\right] ,
\label{Boltzmann1b}
\end{eqnarray}
where $\hat{\bm\nabla}\!\equiv\!{\partial}/{\partial\hat{\bm r}}$, and we have redefined $\varphi^{(1)}$
as a function of ${\bm k} \!\equiv\!\bar{\bm p}/\sqrt{2mk_{\rm B}T}$. 
Similarly, Eq.\ (\ref{constraint}) is transformed into
\begin{equation}
\int d^{3}k\,e^{-k^{2}}{\bm k}^{n}\varphi^{(1)}_{{\bm k}}
=0. \hspace{10mm} (n=0,1,2) 
\label{constraint2}
\end{equation}
Equation (\ref{Boltzmann1b}) may be regarded as an integral equation for $\varphi^{(1)}_{{\bm k}}$ with
condition (\ref{constraint2}).

The right-hand side of Eq.\ (\ref{Boltzmann1b}) suggests that we may solve the equation by adopting the expression:\cite{CC90,HCB54}
\begin{equation}
\varphi^{(1)}_{\bm k}=-ln^{1/3}
\!\left[ A^{5/2}(k^{2})
\left({\bm k}{\bm k}-\frac{k^{2}}{3}\underline{1}\right)
:\hat{\bm\nabla}\hat{\bm v}
+A^{3/2}(k^{2})\,
{\bm k}\!\cdot\!\hat{\bm\nabla}\ln T
\,\right] ,
\label{varphi}
\end{equation}
where $A^{5/2}$ and $A^{3/2}$ are two unknown functions; they are distinguished here
by the superscripts $\alpha\!=\!5/2,3/2$ for the reason that will be clear shortly.
Substituting Eq.\ (\ref{varphi}) into Eq.\ (\ref{Boltzmann1b}), we obtain a couple of separate equations
for $A^{\alpha}$ as
\begin{eqnarray}
&&
\frac{2\sqrt{2}}{\sqrt{\pi}}\int\frac{d^{3}q}{4\pi}
\int\frac{d^{3}q'}{4\pi}\,\frac{\delta(q'\!-\!q)}{q}
e^{-k^{2}
-({\bm k}+{\bm q})^{2}} \left[\,T^{\alpha}_{{\bm k}}+
T^{\alpha}_{{\bm k}+{\bm q}}
-T^{\alpha}_{{\bm k}+({\bm q}-{\bm q}')/2}
-T^{\alpha}_{{\bm k}+({\bm q}+{\bm q}')/2}\right] 
\nonumber \\
&& =
e^{-k^{2}}{\cal R}_{{\bm k}}^{\alpha} ,
\label{Boltzmann1c}
\end{eqnarray}
\begin{table}[t]
\begin{center}
\caption{\label{tab:table1}Definitions of tensors in Eq.\ (\ref{Boltzmann1c}), where
$S_{0}^{\alpha}(\varepsilon)\!=\!1$ and
$S_{1}^{\alpha}(\varepsilon)\!=\!1\!+\!\alpha\!-\!\varepsilon$ are Sonine's polynomials.\cite{CC90,HCB54}}
\begin{tabular}{cccc}
\hline
$\alpha$ & ${\cal W}^{\alpha}_{\bm k}$ & ${\cal R}^{\alpha}_{\bm k}$ & 
$T^{\alpha}_{\bm k}$ \\
\hline
$5/2$ 
& $\displaystyle{\bm k}{\bm k}\!-\!({k}^{2}/{3})\underline{1}$ 
& $2S_{0}^{\alpha}({k}^{2}){\cal W}^{\alpha}_{\bm k}$ 
& $A^{\alpha}({k}^{2}){\cal W}^{\alpha}_{\bm k}$ \\
$3/2$  & ${\bm k}$ & 
$-S_{1}^{\alpha}({k}^{2}){\cal W}^{\alpha}_{\bm k}$ 
& $A^{\alpha}({k}^{2}){\cal W}^{\alpha}_{\bm k}$
\\
\hline
\end{tabular}   
\end{center}
\end{table}

\noindent
where ${\cal R}^{\alpha}_{\bm k}$ and $T^{\alpha}_{\bm k}$ are tensors given in Table \ref{tab:table1}.
The factor $e^{-k^{2}}$ on the right-hand side is identical 
as a function of $k^{2}$ with the weight function of Sonine's polynomial. \cite{CC90,HCB54}
Hence, we solve Eq.\ (\ref{Boltzmann1c}) by expanding
$A^{\alpha}(\varepsilon)$ in terms of Sonine's polynomials
$\{S_{\ell}^{\alpha}(\varepsilon)\}$ as
\begin{equation}
A^{\alpha}(\varepsilon)=\sum_{\ell=0}^{\infty}a_{\ell}^{\alpha}S_{\ell}^{\alpha}(\varepsilon) .
\label{A-exp}
\end{equation}
By using the orthogonality of 
$S_{\ell}^{\alpha}(\varepsilon)$ as well as Eqs.\ (\ref{varphi}) and (\ref{A-exp}), 
constraint (\ref{constraint2}) is then transformed into
\begin{equation}
a_{0}^{3/2}=0 .
\label{constraint3}
\end{equation}
Hence, we  will remove the $\ell=0$ term in expansion (\ref{A-exp}) for $\alpha\!=\!3/2$ hereafter.
Next, we take the tensor ($\alpha=5/2$) or the vector ($\alpha=3/2$) product of 
Eq.\ (\ref{Boltzmann1c}) with $S_{\ell}^{\alpha}(k^{2}){\cal W}^{\alpha}_{{\bm k}}/4\pi$ and subsequently
perform the integration over ${\bm k}$.
Equation (\ref{Boltzmann1c}) is thereby transformed into a set of linear algebraic equations for $a_{\ell}^{\alpha}$
as
\begin{equation}
\sum_{\ell'}T_{\ell\ell'}^{\alpha}a_{\ell'}^{\alpha}={\cal R}_{\ell}^{\alpha} .
\label{Boltzmann1d}
\end{equation}
Here, ${\cal R}_{\ell}^{\alpha}$ is given by
\begin{equation}
{\cal R}_{\ell}^{\alpha}\equiv \int\frac{d^{3}k}{4\pi}e^{-{k}^{2}}
S_{\ell}^{\alpha}({k}^{2})\,({\cal W}_{\bm k}^{\alpha},{\cal R}_{\bm k}^{\alpha})
=\left\{
\begin{array}{ll}
\vspace{2mm}
\displaystyle \frac{5}{4}\sqrt{\pi}\,\delta_{\ell 0} & :\alpha=5/2 \\
\displaystyle -\frac{15}{16}\sqrt{\pi}\,\delta_{\ell 1} & :\alpha=3/2
\end{array}
\right. \! ,
\end{equation}
with $({\cal W}_{\bm k}^{\alpha},{\cal R}_{\bm k}^{\alpha})$ denoting 
${\cal W}_{\bm k}^{\alpha}:{\cal R}_{\bm k}^{\alpha}$ and
${\cal W}_{\bm k}^{\alpha}\cdot{\cal R}_{\bm k}^{\alpha}$ for $\alpha\!=\!5/2$ and $3/2$, respectively.
Besides, $T_{\ell\ell'}^{\alpha}$ can be expressed with a change of variables
${\bm k}\rightarrow{\bm k}-{\bm q}/2$ as
\begin{eqnarray}
&&\hspace{-10mm}
T_{\ell\ell'}^{\alpha}
= \frac{2\sqrt{2}}{\sqrt{\pi}}\int_{0}^{\infty}\!dk\, 
e^{-2k^{2}}k^{2}
\int_{0}^{\infty}\!dq\, e^{-q^{2}/2}q^{3}
\int_{0}^{\infty}\!dq'\,\delta(q'\!-\! q)
\nonumber \\
&& \hspace{1mm}
\times \bigl[I_{\ell\ell'}^{\alpha}(k,q,q)
\!+\!I_{\ell\ell'}^{\alpha}(k,q,-q)
-2I_{\ell\ell'}^{\alpha}(k,q,q')\bigr] ,
\label{T_nn'}
\end{eqnarray}
where $I_{\ell\ell'}^{\alpha}(k,q,q')$ is defined by
\begin{equation}
I_{\ell\ell'}^{\alpha}(k,q,q')
\equiv 
\int \frac{d\Omega_{\bm q}}{4\pi}
\int \frac{d\Omega_{{\bm q}'}}{4\pi}
S^{\alpha}_{\ell}(|{\bm k}-{\bm q}/2|^{2})
S^{\alpha}_{\ell'}(|{\bm k}-{\bm q}'/2|^{2})
({\cal W}_{{\bm k}-{\bm q}/2}^{\alpha},
{\cal W}_{{\bm k}-{\bm q}'/2}^{\alpha}) ,
\label{I_pp'}
\end{equation}
and $I_{\ell\ell'}^{\alpha}(k,q,\pm q)$ is obtained from Eq.\
(\ref{I_pp'}) by the replacement  ${\bm k}-{\bm q}'/2\rightarrow{\bm k}\mp {\bm q}/2$.

It is easy to calculate the first few elements of Eq.\ (\ref{T_nn'}) analytically; we obtain
$T_{00}^{5/2}\!=\!T_{11}^{3/2}\!=\! 1$,
$T_{01}^{5/2}\!=\!T_{12}^{3/2}\!=\! -1/4$, 
$T_{11}^{5/2}\!=\! 205/48$, and $T_{22}^{3/2}\!=\! 45/16$.
Besides, those for a general $\ell\ell'$ may be evaluated numerically without difficulty.
Now, we can solve Eq.\ (\ref{Boltzmann1d}) approximately using $T_{\ell\ell'}^{(\alpha)}$ and ${\cal R}_{\ell}^{(\alpha)}$ of $\ell,\ell'\leq \ell_{\rm c}$,
and the errors can be estimated by increasing the cutoff $\ell_{\rm c}$.
Table \ref{tab:table2} shows the values of $a_{\ell}^{\alpha}$ thereby obtained.
The convergence in terms of $\ell_{\rm c}$ turns out to be quite rapid:
for example, the values of $a_{0}^{5/2}$ and $a_{1}^{3/2}$ listed in Table \ref{tab:table2} are different from their lowest-order values $5\sqrt{\pi}/4$ ($\ell_{\rm c}\!=\! 0$) and $-15\sqrt{\pi}/16$ ($\ell_{\rm c}\!=\! 1$), respectively, 
by only $\sim 2$\%. Thus, the expansion in Sonine's polynomial is quite efficient in obtaining the solution.
\begin{table}[t]
\begin{center}
\caption{\label{tab:table2}Values of $a_{\ell}^{\alpha}$ obtained by solving Eq.\ (\ref{Boltzmann1d}).}
\begin{tabular}{cccccc}
\hline
$\alpha$ & $a_{0}^{\alpha}$ & $a_{1}^{\alpha}$ & $a_{2}^{\alpha}$ & $a_{3}^{\alpha}$
 & $a_{4}^{\alpha}$ \\
 \hline
$5/2$  & $2.2511$ & $0.1390$ & $0.0233$ & $0.0058$ & $0.0018$ \\
$3/2$  & $0$ & $-1.7036$ & $-0.1626$ & $-0.0371$ & $-0.0117$\\
\hline
\end{tabular}   
\end{center}
\end{table}

Let us substitute Eqs.\ (\ref{Enskog}), (\ref{f_eq}), and (\ref{varphi}) into Eqs.\ (\ref{Pi-C}) and (\ref{j_Q-C}).
We then obtain expressions of the momentum flux density tensor and the heat flux density as
\begin{subequations}
\label{j_qPi^1}
\begin{eqnarray}
{\Pi}_{ij}&=& P\delta_{ij}-mn\nu\!\left(\frac{\partial v_{i}}{\partial r_{j}}+
\frac{\partial v_{j}}{\partial r_{i}}-\delta_{ij}\frac{2}{3}
{\bm\nabla}\!\cdot\!{\bm v}
\right) ,
\\
{\bm j}_{Q}&=& -n c_p \kappa\frac{\partial T}{\partial {\bm r}} .
\label{j_Q^1}
\end{eqnarray}
\end{subequations}
Here, $P=nk_{\rm B}T$ denotes the pressure for the monatomic gas,
$c_p=\frac{5}{2}k_{\rm B}$ is the specific heat per particle under constant pressure,
and $\nu$ and $\kappa$ are the kinematic viscosity and thermal diffusivity, respectively.
The latter two quantities are defined 
in terms of the mean-free path $l$ in Eq.\ (\ref{l}) as well as $a_{0}^{5/2}$ and $a_{1}^{3/2}$
in Table \ref{tab:table2} as
\begin{subequations}
\label{kappa-nu}
\begin{eqnarray}
\nu&=& \frac{1}{4}l\sqrt{\frac{2k_{\rm B}T}{m}}a_{0}^{5/2} ,
\\
\kappa&=& -\frac{1}{2}l\sqrt{\frac{2k_{\rm B}T}{m}}a_{1}^{3/2} ,
\label{kappa}
\end{eqnarray}
\end{subequations}
which clearly have the same dimension.
Their ratio $P_r\equiv\nu/\kappa$ constitutes the Prandtl number, \cite{CC90} i.e., one of the 
fundamental dimensionless quantities of fluid mechanics. 
Our theoretical estimate $P_r = 0.66$ from 
Table \ref{tab:table2}  and Eq.\ (\ref{kappa-nu}) agrees excellently with the experimental value $0.67$
of Ar at $T=273$ K and $P=1$ atm, and also those of many monatomic gases. \cite{HCB54}

Thus, we have succeeded in expressing $\underline{\Pi}$ and ${\bm j}_{Q}$ in terms of $(n,{\bm v},T)$
as Eq.\ (\ref{j_qPi^1}). 
With these expressions, Eq.\ (\ref{continuity-C}) now forms a closed set of equations for $(n,{\bm v},T)$.
Among them, Eq.\ (\ref{continuity-C-p}) is the standard Navier-Stokes equation,\cite{LL-FM}
and Eq.\ (\ref{continuity-C-E}) with ${\bm v}={\bm 0}$ is the heat equation.
It is worth emphasizing once again that these equations have been obtained by carrying out
Enskog's expansion from local equilibrium up to the first order.
Hence, it follows that, within their applicable range, the deviation of any solution to Eqs.\ (\ref{continuity-C}) and (\ref{j_qPi^1}) from the local equilibrium should be small to be expressed as Eq.\ (\ref{Enskog}) with Eqs.\ (\ref{f_eq}), (\ref{varphi}), (\ref{A-exp}), and Table \ref{tab:table2}.
Noting that those nonequilibrium phenomena are well described using the equations of fluid mechanics,
one may realize that most of the energy in the turbulence, pattern formation, etc., is still dominated by heat, which can be described appropriately by equilibrium statistical mechanics.
We finally point out once again that, using Eq.\ (\ref{Enskog}) thereby obtained, we can calculate
fluctuations in the fluid with the method described in \S  \ref{two-body}.

\section{Summary and future problems}

We have described the fundamentals of nonequilibrium quantum field theory on the basis of a many-body perturbation expansion with the Keldysh Green's function in a self-contained manner. The time-evolution operator and S-matrix were introduced concisely in \S 2; using the latter, we can calculate an expectation value of an arbitrary operator perturbatively with Eq.\ (\ref{<O_I3>}). It was shown in \S\S 3.1-- 3.3 that the perturbation expansion can be carried out with slight modifications of the equilibrium Matsubara formalism, specifically in the integration contour and Feynman rules. The formulation given here for a nonrelativistic two-body interaction can be applied easily to other many-body systems by merely changing the starting Hamiltonian so as to be suitable for the relevant problem. A special consideration was given in \S 3.4 to the self-consistent perturbation expansion, i.e., Baym's $\Phi$-derivable approximation, of satisfying various conservation laws. It consists of solving Eqs.\ (\ref{Dyson-check}) and (\ref{Sigma-Phi}) self-consistently for an appropriately chosen functional $\Phi$. These equations form a firm microscopic basis for tracing nonequilibrium time evolutions of various many-body systems. Indeed, active investigations in high-energy physics have been carried out over the last decade to solve those equations numerically without any further approximations.\cite{BC01,AB01,AABBS02,Berges02,Berges03,CDM03,JCG04,AST05,Berges06}  

On the other hand, coupled equations (\ref{Dyson-check}) and (\ref{Sigma-Phi}) may be simplified further into first-order differential equations in the phase space, i.e., quantum transport equations  (\S\S 4.1-- 4.4), by transforming them into Wigner representations and truncating the resultant gradient expansion at the first order. The procedure may be justified if the space-time inhomogeneity occurs over a length or time much larger than the microscopic length or time characteristic of the relevant system, e.g., average interparticle spacing. Indeed, this condition is met in a wide range of nonequilibrium phenomena. A wide applicability of transport equations was also confirmed by a numerical simulation.\cite{Berges06} Thus, quantum transport equations (\ref{G^R}) and (\ref{transport}) in the phase space may be used as a convenient alternative to Eqs.\ (\ref{Dyson-check}) and (\ref{Sigma-Phi}) in the full coordinate space. The reduced equations in the phase space also enabled us to obtain an expression of nonequilibrium entropy as Eq.\ (\ref{entropy-density}), which evolves as Eq.\ (\ref{continuity-s}) and also embraces the Boltzmann entropy for dilute classical gases as a high-temperature weak-coupling limit (\S 5.1). This may be regarded as a definite contribution to the basic issue of how and in what circumstances we can define nonequilibrium entropy. We also showed that the expression satisfies the law of increase in entropy within the self-consistent second-order perturbation expansion; it remains to be clarified whether the proof can be extended beyond the second order.
Concerning the law of increase in entropy, we presented an argument at the end of \S 3.1 on why thermalization is achieved in this formalism by only evolving the system mechanically.  

In \S 5, we performed further reductions of relevant variables with a couple of different approximation schemes, i.e., the quasiparticle and quasiclassical approximations. The quasiparticle approximation was shown to reproduce the Boltzmann equation.
Sections 6--8 were devoted to quantum transport equations for a charged system in electromagnetic fields, a derivation of the Bethe-Salpeter equation for two-body correlation functions in the $\Phi$-derivable approximation, and a proof of conservation laws in the $\Phi$-derivable approximation, respectively. Finally, in \S 9, we presented a derivation of the Navier-Stokes equation for a dilute classical gas
from the Boltzmann equation described in \S 5.1.

Thus, readers may have been convinced that the nonequilibrium quantum field theory provides us with a powerful theoretical tool for describing general nonequilibrium phenomena based on Hamiltonians. Particularly, the $\Phi$-derivable approximation enables us to handle one-particle, two-particle, and higher-order correlation functions on an equal footing for a given $\Phi$ so as to satisfy conservation laws. To put it another way, choosing an appropriate $\Phi$ amounts to fixing the whole BBGKY hierarchy.\cite{Cercignani88} Moreover, higher-order correlations and collisions beyond the two-particle ones can be incorporated systematically in the self-consistent perturbation expansion. These clear and simple structures of the approximation scheme are among the definite advantages of the $\Phi$-derivable approximation. Since no such schemes seem available in classical statistical mechanics, it may be used even for describing classical phenomena as a starting point to take the high-temperature limit eventually. Note also that the Boltzmann equation and Navier-Stokes equation can be derived from Eqs.\ (\ref{Dyson-check}) and (\ref{Sigma-Phi}) with successive reductions of relevant variables. Thus, those phenomena that have been discussed traditionally using the Boltzmann or Navier-Stokes equation can be handled microscopically and systematically by the $\Phi$-derivable approximation of nonequilibrium quantum field theory. They include two-body correlations of nonequilibrium fluids, especially near nonequilibrium phase transitions,\cite{CH93,CG09} which are clearly beyond the scope of the Navier-Stokes equation.\cite{LL-FM} Thus, typical nonequilibrium phenomena such as pattern formations, which have been discussed conventionally on the basis of deterministic evolution equations for averaged quantities (e.g., the Navier-Stokes equation for the average velocity),\cite{CH93,CG09} can also be treated using this approach so as to incorporate fluctuation effects with the Bethe-Salpeter equation. It also enables us to improve the Boltzmann equation systematically so as to include higher-order collisions and/or quantum effects. It is interesting to see how Enskog's expansion described in \S 9 may be extended to systems with strong correlations and/or quantum effects.

The approach may also be applied to superfluid phases,\cite{Rainer83} especially to Bose-Einstein condensates which have been actively investigated \cite{PS08,PS03,GNZ09} since the realization of Bose-Einstein condensation on an atomic gas in 1995.\cite{AEMWC95} In this context,
it is worth pointing out that an extension of the $\Phi$-derivable approximation to Bose-Einstein condensates has met a particular difficulty\cite{HM65,Griffin96,Kita05,Kita06} of not being able to reproduce the gapless Nambu-Goldstone boson\cite{PS95,Nambu61,Goldstone61} of broken $U(1)$ symmetry, which is the Bogoliubov mode in the weak-coupling regime.\cite{Bogoliubov47}
It is only very recently that the task has been successfully carried out.\cite{Kita09} Thus, applications of this approach to Bose-Einstein condensates remain mostly unexplored. They include a microscopic derivation of the two-fluid equations\cite{LL-FM} with a definite temperature dependence of the viscosity coefficient, etc. See Ref.\ \citen{GNZ09} for the present status of the issue.

Having derived an expression of nonequilibrium entropy in \S \ref{sec:QTE}, it is also interesting to seek its possible roles in nonequilibrium phenomena.
Much effort has been directed towards extending equilibrium statistical mechanics to nonequilibrium situations.\cite{GP71,Zubarev74,Haken75,Kuramoto84,Tsallis88,CH93,Jou93,Jarzynski97,ES02,ST06,Grandy08,CG09} One of its central issues has been whether there are any universal probability distributions in nonequilibrium systems corresponding to the microcanonical, canonical, and grand canonical ensembles of equilibrium statistical mechanics, or to put it another way, whether there are any thermodynamic functions that take minimum in nonequilibrium steady states. Various proposals have been presented on this issue, but no definite answer seems to have been obtained yet. 
In this context, it may be worth pointing out that entropy is the key quantity in equilibrium statistical mechanics. Indeed, every equilibrium statistical ensemble can be identified as the maximum of the Gibbs entropy (\ref{S-Gibbs}) under some constraints, as shown by Jaynes.\cite{Jaynes57} Thus, it may be worth asking (i) how entropy changes as we drive the system out of equilibrium and (ii) how far into the nonequilibrium region the maximum property of entropy holds true in the space of steady states. 
It should be noted at the same time that investigations of nonequilibrium states may be carried out by solving some evolution  equations, e.g., Eqs.\ (\ref{Dyson-check}) and (\ref{Sigma-Phi}) in this paper, without recourse to any thermodynamic function such as entropy. However, a thermodynamic function, if it does exist and takes minimum in nonequilibrium steady states, will be useful in clarifying and classifying nonequilibrium phase transitions.\cite{CH93,CG09}

The above questions (i) and (ii) on entropy were investigated using the Rayleigh-B\'enard transition of a dilute classical monatomic gas. \cite{Kita06b} It was shown that entropy as a function of mechanical variables becomes smaller under a heat flow than without it, in accordance with equilibrium statistical mechanics. Further, entropy was shown to be larger with convection than that without it in the region where the convection is stabilized. Thus, entropy as a function of mechanical variables apparently takes its largest possible value in the example, implying that the principle of maximum entropy may hold even in steady states beyond equilibrium.\footnote{It is worth pointing out that this consideration of nonequilibrium entropy differs substantially from that of extended irreversible thermodynamics developed by Jou et al.\ \cite{Jou93} in that: (i) we adopt an expression of nonequilibrium entropy with interaction; and (ii) we focus only on those steady states that are solutions to the Boltzmann equation (or more generally, Dyson's equation on the Keldysh contour). On the other hand, extended irreversible thermodynamics makes use of noninteracting entropy, adds variables characteristic of nonequilibrium to entropy, and directly maximizes the entropy in terms of those variables without asking whether the solution satisfies the Boltzmann equation. Thus, states identified with their procedure may not correspond to real steady states in nature.}
More investigations are required to establish the above conjecture, however.

\section*{Acknowledgements}

The author would like to thank Dr. H. Yamashita for a careful reading of the manuscript and resultant useful comments.

\appendix

\section{Second Quantization}

The second quantization method enables us to handle many-particle systems concisely and conveniently. 
It is completely equivalent to the description with many-body wave functions in the configuration space. 
However, this equivalence has not been explained adequately in the literature, thereby leaving students who want to learn many-body physics in an awkward situation. 
We explain here the above equivalence as concisely and clearly as possible starting from the symmetry in identical particles under permutation. This will be carried out without assuming that the $N$-body wave function can be expressed as a superposition of products of $N$ one-particle wave functions.

\subsection{Permutation\label{subsec:permutation}}

We first explain the permutation in the group theory.\cite{HK78}
A permutation signifies an action that $N$ persons sitting on $N$ chairs in a circle exchange their positions.
It can be expressed by the operator:
\begin{equation}
\hat{P} = \left(
\begin{array}{ccccc}
1 & 2 & 3 & \cdots & N
\\
p_1 & p_2 & p_3 & \cdots & p_N
\end{array}
\right),
\label{P-def}
\end{equation}
where $1\leq p_{i}\leq N$ with no duplication among $p_i$'s. Thus, the person previously on chair $i$ has moved onto chair $p_i$. 
The number of distinct permutations is easily identified to be $N!$.

Among the permutations, there is a special class called cyclic permutation.
It corresponds to each person moving onto an adjacent chair systematically as
\begin{equation}
\hat{C} = \left(
\begin{array}{cccccc}
1 & 2 & 3 & \cdots & k-1 & k
\\
2 & 3 & 4 & \cdots & k &1
\end{array}
\right)
\equiv (1\, 2\, 3\, \cdots \, k).
\end{equation}
A cyclic permutation of 2 elements is called transposition, which is expressed as
\begin{equation}
\hat{P}_{12} \equiv \left(
\begin{array}{cc}
1 & 2 \\ 2 & 1
\end{array}
\right)\equiv(1\, 2).
\label{P_12}
\end{equation}

We now summarize theorems on the permutation. See, e.g., Ref.\ \citen{HK78} for the proofs.
First, every permutation can be expressed as a product of cyclic permutations.
For example,
\begin{equation}
\hat{P}_{a} \equiv \left(
\begin{array}{cccccc}
1 & 2 & 3 & 4 & 5 & 6
\\
2 & 5 & 6 & 4 & 1 &3
\end{array}
\right)
= 
(4) \, (3 \, 6) \, (1 \, 2 \, 5) \,  \equiv \, (3 \, 6) \, (1 \, 2 \, 5) .
\label{P_a}
\end{equation}
The identity permutation is often dropped as $(4)$ above.
Second, every cyclic permutation can be expressed as a product of transpositions as
\begin{eqnarray*}
\hspace{-10mm} \;\;\; \left(
\begin{array}{cccccc}
1 & 2 &  \cdots & k\!-\!2 & k\!-\!1 & k
\\
2 & 3 &  \cdots & k\!-\!1 & k & 1
\end{array}
\right)
=(1\, 2) \,
(1 \, 3) \,\cdots\,  (1 \, k\!-\!1)(1 \, k),
\end{eqnarray*}
where operations on the right-hand side proceed from right to left.
Combining the two statements, we realize that every permutation can be expressed as a product of transpositions.
For example, Eq.\ (\ref{P_a}) may be expressed as
\begin{eqnarray*}
\hat{P}_{a}= 
(3 \, 6) \, (1 \, 2)\, (1 \, 5) .
\end{eqnarray*}
Another example is given by
\begin{eqnarray*}
\hat{P}_{b}\equiv\left(
\begin{array}{cccccc}
1 & 2 & 3 & 4 & 5 & 6
\\
2 & 5 & 6 & 3 & 1 & 4
\end{array}
\right)
= 
(3 \, 6 \, 4) \, (1 \, 2 \, 5) = (3 \, 6) \, (3 \, 4) \, (1 \, 2)\, (1 \, 5) .
\end{eqnarray*}
An odd (even) permutation is a permutation that can be expressed as a product of an odd (even) number of transpositions as $\hat{P}_{a}$ ($\hat{P}_{b}$) above. It can be shown that every permutation is either odd or even.

\subsection{Permutation symmetry in systems of identical particles}

Many-particle systems composed of identical particles have a special symmetry under permutations.
Consider an $N$-particle system described by the Hamiltonian:
\begin{equation}
\hat{H}  = \sum_{j=1}^{N} \left[ \frac{\hat{\bm p}_{j}^{2}}{2m}
+U({\bm r}_{j}) \right] + \sum_{i<j}V(|{\bm r}_{i}-{\bm r}_{j}|) 
\equiv \sum_{j=1}^{N}\hat{h}^{(1)}_{j} +\sum_{i<j} \hat{h}^{(2)}_{ij},
\label{H-many}
\end{equation}
where $\hat{\bm p}_{j}$ is the momentum operator, and $U$ and $V$ denote the one-body and two-body potentials, respectively.
This Hamiltonian commutes with Eq.\ (\ref{P-def}) as
\begin{equation}
\hat{P}\hat{H}\hat{P}^{-1} = \hat{H} ,
\label{PH-HP}
\end{equation}
which is exemplified for the case of $N=2$ as follows:
\begin{eqnarray*}
\hat{P}_{12}\hat{H}\hat{P}_{12}^{-1}
= \left[ \frac{\hat{\bm p}_{2}^{2}}{2m}+\frac{\hat{\bm p}_{1}^{2}}{2m} +U({\bm r}_{2})+U({\bm r}_{1})
+ V(|{\bm r}_{2}-{\bm r}_{1}|)\right]\hat{P}_{12}\hat{P}_{12}^{-1} = \hat{H}.
\end{eqnarray*}
Thus, any permutation upon $\hat{H}$  only changes the order of summation to keep $\hat{H}$ itself invariant.
It follows from Eq.\ (\ref{PH-HP}) that $\hat{P}$ and $\hat{H}$ can be diagonalized simultaneously.\cite{Sakurai} 
In cases where $\hat{H}$ has no time dependence, the expectation value of $\hat{P}$ does not change in time.

To begin with, let us determine eigenvalues of transposition (\ref{P_12}).
Noting that 
$\hat{P}_{12}^{2}$ is the identity permutation, we conclude that an eigenvalue $\varepsilon_{P_{12}}$ of $\hat{P}_{12}$ satisfies $\varepsilon_{P_{12}}^{2}=1$.
Hence, it follows that $\varepsilon_{P_{12}}$ can be either $1$ or $-1$.
To put it another way, a wave function of identical particles either remains invariant or changes its sign upon a transposition. 
It was Pauli in 1940, who identified the connection between the eigenvalue $\varepsilon_{P_{12}}$ and the spin, which is summarized as
\begin{eqnarray*}
\varepsilon_{P_{12}}=\left\{\begin{array}{ll} +1 &(\mbox{boson})\\
-1 &(\mbox{fermion})\end{array}\right. \hspace{3mm} 
\longleftrightarrow\hspace{3mm} 
\mbox{spin:}\,\,\left\{ \begin{array}{ll} 
\vspace{2mm}0,1,2,\cdots &(\mbox{integer})\\
\displaystyle\frac{1}{2},\frac{3}{2},\frac{5}{2},\cdots &(\mbox{half integer})
\end{array}\right. .
\end{eqnarray*}
For example, the electron, proton, and neutron are all fermions with spin $1/2$, whereas the photon is a boson with spin $1$.
The rule applies also to composite particles.
Thus, the hydrogen composed of one electron and one proton, whose total spin can be either $0$ or $1$, behaves as a boson.

The above consideration can be generalized easily to the case of general permutations.
Remembering ``every permutation can be expressed as a product of transpositions'' and 
``every permutation is either odd or even'' in \S \ref{subsec:permutation}, we obtain the eigenvalue $\varepsilon_{P}\equiv(\pm 1)^{P}$ of 
a permutation $\hat{P}$ as
\begin{equation}
(\pm 1)^{P}=\left\{\begin{array}{ll} 1 & \mbox{: even permutation} \\
\pm 1 & \mbox{: odd permutation}
\end{array}\right. ,
\end{equation}
where the upper (lower) sign corresponds to bosons (fermions).

We finally write down explicitly how $\hat{P}$ operates on an $N$-body wave function in the configuration space.
Let us express the wave function as
\begin{equation}
\Psi_{\nu}(x_{1},x_{2},\cdots,x_{N}),
\label{Psi_nu}
\end{equation}
where $\nu$ signifies a set of quantum numbers to specify the $N$-body wave function,
and $x_{j}\equiv ({\bm r}_{j},\alpha_{j})$ ($j=1,2,\cdots ,N$) with ${\bm r}_{j}$ and 
$\alpha_{j}$ denoting the space and spin coordinates, respectively. 
The operation of Eq.\ (\ref{P-def}) on $\Psi_N$ is defined by
\begin{subequations}
\label{P-Psi}
\begin{equation}
\hat{P}\Psi_{\nu}(x_{1},x_{2},\cdots,x_{N})\equiv\Psi_{\nu}(x_{p_{1}},x_{p_{2}},\cdots,x_{p_{N}}) .
\end{equation}
On the other hand, the fact of $\Psi_{\nu}$ being an eigenfunction of $\hat{P}$ with the eigenvalue $(\pm 1)^{P}$
is expressed as
\begin{equation}
\hat{P}\Psi_{\nu}(x_{1},x_{2},\cdots,x_{N})=(\pm 1)^{P} \Psi_{\nu}(x_{1},x_{2},\cdots,x_{N}) .
\label{P-Psi2}
\end{equation}
\end{subequations}

\subsection{Constructing the eigenspace of permutation}

Equation (\ref{P-Psi}) tells us that an eigenfunction of identical particles should be an eigenstate of every permutation $\hat{P}$, i.e., it should be either symmetrized (bosons) or antisymmetrized (fermions). We construct here the eigenspace of $\hat{P}$ where many-body wave functions of identical particles belong to.

Following the creation and annihilation operators of the harmonic oscillator,\cite{Sakurai}
we first  introduce the operators $\hat{\psi}$ and $\hat{\psi}^{\dagger}$, which satisfy the commutation relations:
\begin{subequations}
\label{psi-commute}
\begin{equation}
[\hat{\psi}(x),
\hat{\psi}^{\dagger}(x')]_{\mp}
\equiv
\hat{\psi}(x)\hat{\psi}^{\dagger}(x')\mp
\hat{\psi}^{\dagger}(x')\hat{\psi}(x)
=\delta(x,x'),
\end{equation}
\begin{equation}
[\hat{\psi}(x),\hat{\psi}(x')]_{\mp}=
[\hat{\psi}^{\dagger}(x),
\hat{\psi}^{\dagger}(x')]_{\mp}= 0,
\end{equation}
\end{subequations}
with $\delta(x,x')\equiv\delta({\bm r}-{\bm r}')\delta_{\alpha\alpha'}$.
We next define the brackets $|0\rangle$ and $\langle 0 |$ through
\begin{equation}
\hat{\psi}(x)|0\rangle = 0 , \hspace{10mm}
0= (\hat{\psi}(x)|0\rangle)^{\dagger}=\langle 0 | \hat{\psi}^{\dagger}(x) ,
 \hspace{10mm}
\langle 0 | 0\rangle = 1  .
\label{|0>}
\end{equation}
These are basic ingredients to construct the eigenspace of $\hat{P}$.

We next introduce the state vector $|x_{1}x_{2}\cdots x_{N}\rangle$ and its Hermitian conjugate
as
\begin{subequations}
\label{Fock}
\begin{eqnarray}
&&\hspace{-10mm}
|x_{1}x_{2}\cdots x_{N}\rangle
\equiv \frac{1}{\sqrt{N!}}
\hat{\psi}^{\dagger}(x_{1})\hat{\psi}^{\dagger}(x_{2})
\cdots \hat{\psi}^{\dagger}(x_{N})|0\rangle  ,
\label{Fock1}
\\
&&\hspace{-10mm}
\langle x_{1}x_{2}\cdots x_{N}|\equiv
|x_{1}x_{2}\cdots x_{N}\rangle^{\dagger}=
\frac{1}{\sqrt{N!}}\langle 0 |
\hat{\psi}(x_{N})\cdots \hat{\psi}(x_{2})\hat{\psi}(x_{1})  .
\label{Fock2}
\end{eqnarray}
\end{subequations}
The space spanned by Eq.\ (\ref{Fock1}) naturally forms the eigenspace of $\hat{P}$.
Indeed, using $\hat{\psi}^{\dagger}(x_{2})\hat{\psi}^{\dagger}(x_{1})
=\pm \hat{\psi}^{\dagger}(x_{1})\hat{\psi}^{\dagger}(x_{2})$, one can show
$|x_{2}x_{1}x_{3}\cdots x_{N}\rangle=\pm 
|x_{1}x_{2}x_{3}\cdots x_{N}\rangle$ easily.
The commutation relation may be used repeatedly to prove 
\begin{equation}
|x_{p_1}x_{p_2}\cdots x_{p_N}\rangle = (\pm 1)^{P}|x_{1}x_{2}\cdots x_{N}\rangle .
\label{P|x_j>}
\end{equation}
The vector $|x_{1}x_{2}\cdots x_{N}\rangle$ satisfies the normalization condition:
\begin{equation}
\langle x'_{1}x'_{2}\cdots x'_{N'}|x_{1}x_{2}\cdots x_{N}\rangle
=\frac{\delta_{N'N}}{N!}\sum_{\hat{P}}(\pm 1)^{P}\delta(x'_{1},x_{p_{1}})
\delta(x'_{2},x_{p_{2}})\cdots \delta(x'_{N},x_{p_{N}}).
\label{ortho}
\end{equation}
To prove it, we make use of the identity:
\begin{eqnarray}
\hat{\psi}(x'_{1})\hat{\psi}^{\dagger}(x_{1})
\hat{\psi}^{\dagger}(x_{2})\cdots
\hat{\psi}^{\dagger}(x_{N})
&=&
\delta(x'_{1},x_{1})
\hat{\psi}^{\dagger}(x_{2})\cdots
\hat{\psi}^{\dagger}(x_{N}) 
\nonumber \\
& &
+(\pm 1)^{1}
\delta(x'_{1},x_{2})
\hat{\psi}^{\dagger}(x_{1})\hat{\psi}^{\dagger}(x_{3})\cdots
\hat{\psi}^{\dagger}(x_{N})
\nonumber \\
& &
+\cdots
\nonumber \\
& &
+(\pm 1)^{N-1} 
\delta(x'_{1},x_{N})
\hat{\psi}^{\dagger}(x_{1})\cdots\hat{\psi}^{\dagger}(x_{N-1})
\nonumber \\
& &
 +(\pm 1)^{N} 
\hat{\psi}^{\dagger}(x_{1})\hat{\psi}^{\dagger}(x_{2})\cdots
\hat{\psi}^{\dagger}(x_{N})\hat{\psi}(x'_{1}),
\label{psi-ident}
\end{eqnarray}
to place the annihilation operator $\hat{\psi}(x_{j}')$ to the left of $|0\rangle$ and subsequently use
$\hat{\psi}(x_{j}')|0\rangle=0$.
For example, Eq.\ (\ref{ortho}) in the case of $N=N'=2$ is shown as
\begin{eqnarray*}
\langle x'_{1}x'_{2}| x_{1}x_{2}\rangle
&=&\frac{1}{2!}
\langle 0|\hat{\psi}(x'_{2})\hat{\psi}(x'_{1})
\hat{\psi}^{\dagger}(x_{1})\hat{\psi}^{\dagger}(x_{2})|0\rangle
\\
&=&\frac{1}{2!}
\langle 0|\hat{\psi}(x'_{2})\bigl[
\delta(x'_{1},x_{1})\hat{\psi}^{\dagger}(x_{2})\pm \delta(x'_{1},x_{2})\hat{\psi}^{\dagger}(x_{1})
\nonumber \\
& &
+(\pm 1)^{2}\hat{\psi}^{\dagger}(x_{1})\hat{\psi}^{\dagger}(x_{2})\hat{\psi}(x'_{1})\bigr]|0\rangle
\nonumber \\
&=&\frac{1}{2!}\bigl[
\delta(x'_{1},x_{1})\delta(x'_{2},x_{2})
\pm\delta(x'_{1},x_{2})\delta(x'_{2},x_{1})\bigr] .
\end{eqnarray*}
Thus, we have constructed the eigenspace of $\hat{P}$.

\subsection{Bracket vectors for many-body wave functions}

We now introduce the bracket vector for wave function (\ref{Psi_nu}) with symmetry (\ref{P-Psi}) by
\begin{subequations}
\label{Psi-bracket}
\begin{eqnarray}
&&\hspace{-10mm}
|\Psi_{\nu}\rangle\equiv \int d x_{1} \int d x_{2} \cdots
\int d x_{N}\,
|x_{1}x_{2}\cdots x_{N}\rangle \Psi_{\nu}(x_{1},x_{2},\cdots,x_{N}),
\label{Psi-cket}
\\
&&\hspace{-10mm}
\langle\Psi_{\nu}|\equiv \int d x_{1} \int d x_{2} \cdots \int d x_{N}\,
\langle x_{1}x_{2}\cdots x_{N}| \Psi_{\nu}^{*}(x_{1},x_{2},\cdots,x_{N})=
|\Psi_{\nu}\rangle^{\dagger},
\label{Psi-bra}
\end{eqnarray}
\end{subequations}
where the integration over $x_{j}=({\bm r}_{j},\alpha_{j})$ is composed of that over ${\bm r}_{j}$ and the summation over $\alpha_{j}$.
The bracket vector has the properties:
\begin{subequations}
\label{|Psi>-Prop}
\begin{equation}
\hat{\psi}(x_{1})|\Psi_{\nu}\rangle
=\sqrt{N}
\!\int\! d x'_{2} \cdots \int d x'_{N}\,
|x'_{2}\cdots x'_{N}\rangle \Psi_{\nu}(x_{1},x'_{2},\cdots,x'_{N}),
\label{|Psi>-Prop1}
\end{equation}
\begin{equation}
\hat{\psi}(x_{2})\hat{\psi}(x_{1})|\Psi_{\nu}\rangle
=\sqrt{N(N-1)}
\!\int\! d x'_{3} \cdots \int d x'_{N}\,
|x'_{3}\cdots x'_{N}\rangle \Psi_{\nu}(x_{1},x_{2},x'_{3},\cdots,x'_{N}),
\label{|Psi>-Prop2}
\end{equation}
\begin{equation}
\hat{\psi}(x_{N})\cdots\hat{\psi}(x_{2})\hat{\psi}(x_{1})|\Psi_{\nu}\rangle
=\sqrt{N!}\,|0\rangle \Psi_{\nu}(x_{1},x_{2},\cdots,x_{N}).
\label{|Psi>-Prop3}
\end{equation}
\end{subequations}
Thus, an operation of $\hat{\psi}(x)$ on $|\Psi_{\nu}\rangle$ has the effect of extracting the argument $x$ from $|\Psi_{\nu}\rangle$.
The proof of Eq.\ (\ref{|Psi>-Prop}) may be carried out straightforwardly using Eqs.\ (\ref{P-Psi}) and (\ref{psi-ident})
together with changes of integration variables.
It follows from Eqs.\ (\ref{|0>}), (\ref{Fock2}), and (\ref{|Psi>-Prop3}) that
\begin{equation}
\langle x_{1}x_{2}\cdots x_{N}|\Psi_{\nu}\rangle
=\Psi_{\nu}(x_{1},x_{2},\cdots,x_{N}) \, .
\label{|Psi>-Prop3'}
\end{equation}
Substituting Eq.\ (\ref{|Psi>-Prop3'}) into Eq.\ (\ref{Psi-cket}), we obtain
$$
|\Psi_{\nu}\rangle\equiv \int d x_{1} \int d x_{2} \cdots
\int d x_{N}\,
|x_{1}x_{2}\cdots x_{N}\rangle 
\langle x_{1}x_{2}\cdots x_{N}|\Psi_{\nu}\rangle, 
$$
which is equivalent to
\begin{equation}
\int d x_{1} \int d x_{2} \cdots\int d x_{N}\,
|x_{1}x_{2}\cdots x_{N}\rangle 
\langle x_{1}x_{2}\cdots x_{N}| =1  .
\label{complete}
\end{equation}
Thus, the vectors $\{|x_{1}x_{2}\cdots x_{N}\rangle \}$ form a complete set for the eigenspace of $\hat{P}$.

Equation (\ref{Psi-bracket}) can also be used to symmetrize or antisymmetrize any wave function $\tilde{\Psi}(x_{1},x_{2},\cdots,x_{N})$ which is not an eigenstate of $\hat{P}$.
Indeed, we only need to construct
\begin{equation}
|\Psi\rangle =A_{N} \int d x'_{1}\int d x'_{2}\cdots \int d x'_{N}
|x'_{1}x'_{2}\cdots x'_N\rangle \tilde{\Psi}(x'_{1},x'_{2},\cdots,x'_{N}) ,
\label{|Psi'>}
\end{equation}
with $A_{N}$ denoting some normalization constant.
Then, it follows from Eq.\ (\ref{P|x_j>}) that integrations over $x_{1}'$, $\cdots$, $x_{N}'$ in Eq.\ (\ref{|Psi'>}) with $|x'_{1}x'_{2}\cdots x'_N\rangle$ in the integrand extracts  from
$\tilde{\Psi}(x'_{1},x'_{2},\cdots,x'_{N})$ only the symmetric or antisymmetric contribution.
Equation (\ref{|Psi'>}) may be regarded as projecting $\tilde{\Psi}(x_{1},x_{2},\cdots,x_{N})$ onto the eigenspace of 
$\hat{P}$.
The wave function of Eq.\ (\ref{|Psi'>}) in the configuration space is obtained 
with Eqs.\ (\ref{ortho}) and (\ref{|Psi>-Prop3'}) as
\begin{equation}
\Psi(x_{1},x_{2},\cdots,x_{N})\equiv \langle x_{1},x_{2},\cdots,x_{N} |\Psi\rangle
=\frac{A_{N}}{N!}\sum_{\hat{P}}(\pm 1)^{P}\tilde{\Psi}(x_{p_1},x_{p_{2}},\cdots,x_{p_N}) ,
\label{Psi-tildePsi}
\end{equation}
which indeed satisfies Eq.\ (\ref{P-Psi}).

\subsection{Orthonormality and completeness of bracket vectors}

Let us assume that the wave function $\Psi_{\nu}(x_{1},x_{2},\cdots,x_{N})$ with symmetry (\ref{P-Psi})
satisfies the orthonormality and completeness given by
\begin{subequations}
\label{oc1}
\begin{equation}
\int\!d x_{1}\int d x_{2}\cdots \int d x_{N} 
\Psi_{\nu'}^{*}(x_{1},x_{2},\cdots,x_{N})
\Psi_{\nu}(x_{1},x_{2},\cdots,x_{N})=\delta_{\nu'\nu}  ,
\label{ortho1}
\end{equation}
\begin{eqnarray}
\sum_{\nu}\Psi_{\nu}(x_{1},x_{2},\cdots,x_{N})
\Psi_{\nu}^{*}(x_{1}',x_{2}',\cdots,x_{N}')
&=& 
\frac{1}{N!}\sum_{\hat{P}}(\pm 1)^{P}\delta(x_{1}',x_{p_{1}})
\delta(x_{2}',x_{p_{2}})\cdots 
\nonumber \\
& &\times \delta(x_{N}',x_{p_{N}}) ,
\label{compl1}
\end{eqnarray}
\end{subequations}
respectively.
These relations can be expressed alternatively in terms of the bracket vectors in Eq.\ (\ref{Psi-bracket}) as
\begin{equation}
\langle \Psi_{\nu'}|
\Psi_{\nu}\rangle =\delta_{\nu'\nu}  ,
\hspace{10mm}
\sum_{\nu}|\Psi_{\nu}\rangle\langle\Psi_{\nu}|=1 .
\label{oc2}
\end{equation}
The proof proceeds straightforwardly by substituting Eq.\ (\ref{Psi-bracket}) into Eq.\ (\ref{oc2}) and making use of Eqs.\ (\ref{P-Psi}), (\ref{ortho}), (\ref{complete}), and (\ref{oc1}).

\subsection{Matrix elements of operators}

We now see that matrix elements of operators for identical particles can be expressed concisely in terms of $|\Psi_{\nu}\rangle$ and $\hat{\psi}(x)$.

Equation (\ref{H-many}) exemplifies that every one-operator $\hat{H}^{(1)}$ and two-particle operator $\hat{H}^{(2)}$ for systems of identical particles may be expressed generally as 
\begin{equation}
\hat{H}^{(1)}\equiv\sum_{j=1}^{N}\hat{h}_{j}^{(1)},\hspace{10mm}
\hat{H}^{(2)}\equiv\sum_{i<j}\hat{h}_{ij}^{(2)}.
\end{equation}
The matrix elements of these operators between $\Psi_{\nu'}^{*}$ and $\Psi_{\nu}$ can be expressed equivalently with $\langle\Psi_{\nu'}|$ and $|\Psi_{\nu}\rangle$ as
\begin{subequations}
\label{MatEll}
\begin{eqnarray}
&&
\int d x_{1}\cdots  \int d x_{N}
\Psi_{\nu'}^{*}(x_{1},\cdots,x_{N})\hat{H}^{(1)}
\Psi_{\nu}(x_{1},\cdots,x_{N})
\nonumber \\
&&\hspace{20mm}
=\int d x_{1}\langle\Psi_{\nu'}|\hat{\psi}^{\dagger}(x_{1})\hat{h}_{1}^{(1)}
\hat{\psi}(x_{1})|\Psi_{\nu}\rangle ,
\label{f_1}
\\
&&\int d x_{1}\cdots \int d x_{N}
\Psi_{\nu'}^{*}(x_{1},\cdots,x_{N})\hat{H}^{(2)}
\Psi_{\nu}(x_{1},\cdots,x_{N})
\nonumber \\
&& \hspace{20mm}
=\frac{1}{2}
\int d x_{1} \int d x_{2}\langle\Psi_{\nu'}|\hat{\psi}^{\dagger}(x_{1})
\hat{\psi}^{\dagger}(x_{2})\hat{h}_{12}^{(2)}
\hat{\psi}(x_{2})\hat{\psi}(x_{1})|\Psi_{\nu}\rangle  .
\label{f_2}
\end{eqnarray}
\end{subequations}
The equalities can be shown to hold in an elementary way by substituting Eq.\ (\ref{|Psi>-Prop}) into the right-hand sides of Eqs.\ (\ref{f_1}) and (\ref{f_2}) and using Eqs.\ (\ref{P-Psi}) and (\ref{ortho}).

\subsection{Second quantization as an equivalent alternative to first quantization}

The preceding consideration has clarified that there are a couple of equivalent descriptions 
for many-body systems composed of identical particles. 

The first one is called the method of first quantization that makes use of Hamiltonian (\ref{H-many}) and the wave function (\ref{Psi_nu}) in the configuration space. Wave functions have to be symmetrized or antisymmetrized as Eq.\ (\ref{P-Psi}). They may be constructed so as to satisfy 
Eq.\ (\ref{oc1}).

An alternative description is called the second quantization method, where the Hamiltonian is expressed concisely as
\begin{subequations}
\label{2nd-Q}
\begin{eqnarray}
&&
\hat{\cal H}  = \int d x_{1}\hat{\psi}^{\dagger}(x_{1})
\!\left[ \frac{\hat{\bm p}_{1}^{2}}{2m}
+U({\bm r}_{1}) \right]\!\hat{\psi}(x_{1})
\nonumber \\
&&\hspace{8mm}
+ \frac{1}{2}\!\int\! d x_{1}\!\int\! d x_{2}
\hat{\psi}^{\dagger}(x_{1})\hat{\psi}^{\dagger}(x_{2})V(|{\bm r}_{1}-{\bm r}_{2}|)
\hat{\psi}(x_{2})\hat{\psi}(x_{1}) .
\label{2nd-Q1}
\end{eqnarray}
The corresponding state vector is given in terms of Eq.\ (\ref{Fock}) by
\begin{equation}
|\Psi_{\nu}\rangle=
\!\int\! d x_{1}\cdots \int d x_{N} 
|x_{1}\cdots x_{N}\rangle
\Psi_{\nu}(x_{1},\cdots,x_{N}),
\label{|Psi>}
\end{equation}
\end{subequations}
which satisfies Eq.\ (\ref{oc2}).
It follows from  Eq.\ (\ref{MatEll}) that Eq.\ 
(\ref{2nd-Q}) of second quantization is equivalent to the description of first quantization.

\subsection{Second quantization in the absence of interaction}
 
We now express the Hamiltonian and brackets of noninteracting systems in terms of one-particle eigenstates.
They also form a convenient starting point for a perturbation expansion with respect to the interaction.

Consider the Hamiltonian:
\begin{equation}
\hat{\cal H}_{0} \equiv \int \hat{\psi}^{\dagger}(x_{1}) \left[ \frac{\hat{\bm p}_{1}^{2}}{2m}
+U({\bm r}_{1}) \right]\hat{\psi}(x_{1}) d x_{1},
\label{H_0App}
\end{equation}
and assume that the following one-particle eigenvalue problem has been solved:
\begin{equation}
\left[\frac{\hat{\bm p}_{1}^{2}}{2m}
+U({\bm r}_{1})\right]\varphi_{k}(x_{1})=\epsilon_{k}\varphi_{k}(x_{1}) ,
\label{eigen1}
\end{equation}
where $k$ specifies the one-particle eigenstate and $\epsilon_{k}$ is the eigenvalue.
We also assume that the eigenfunctions $\varphi_{k}(x)=\langle x |k\rangle$ satisfy the orthonormality relations:
\begin{subequations}
\label{varphi-oc}
\begin{equation}
\langle k| k'\rangle \equiv \int  \varphi_{k}^*(x_{1})\varphi_{k'}(x_{1}) d x_{1}
=\delta_{kk'} ,
\label{varphi-oc1}
\end{equation}
\begin{equation}
\sum_{k}\varphi_{k}(x_{1})\varphi_{k}^*(x_{2})=\delta(x_{1},x_{2}) .
\label{varphi-oc2}
\end{equation}
\end{subequations}
We next expand the operator $\hat{\psi}(x)$ in terms of $\varphi_{k}(x_{1})$ as
\begin{equation}
\hat{\psi}(x) = \sum_{k}\hat{c}_{k}\varphi_{k}(x),\hspace{10mm}
\hat{\psi}^{\dagger}(x) = \sum_{k}\hat{c}_{k}^{\dagger}\varphi_{k}^{*}(x).
\label{psi-exp}
\end{equation}
Using Eqs.\ (\ref{psi-commute}) and (\ref{varphi-oc}), one can show easily that $\hat{c}_{k}$ and $\hat{c}_{k}^{\dagger}$
above obey the commutation relations:
\begin{equation}
[\hat{c}_{k},\hat{c}_{k'}^{\dagger}]_{\mp}=\delta_{kk'},\hspace{10mm}
[\hat{c}_{k},\hat{c}_{k'}]_{\mp}=[\hat{c}_{k}^{\dagger},\hat{c}_{k'}^{\dagger}]_{\mp}=0.
\label{c-commute}
\end{equation}
Let us substitute Eq.\ (\ref{psi-exp}) into Eq.\ (\ref{H_0App}) and take Eqs.\ (\ref{eigen1}) and (\ref{varphi-oc1}) into account.
Then, Eq.\ (\ref{H_0App}) is transformed into
\begin{equation}
\hat{\cal H}_{0}=\sum_{k}\epsilon_{k}\hat{c}_k^{\dagger}\hat{c}_{k} .
\label{H_0App-diag}
\end{equation}
Thus, we have obtained the diagonalized expression of $\hat{\cal H}_{0}$.

We next express bracket vectors for $\hat{\cal H}_{0}$ in the same representation.
To this end, we first derive an $N$-particle eigenfunction of $\hat{\cal H}_{0}$ in the configuration space.
It may be constructed from a product of $N$ one-particle wave functions $\prod_{j=1}^{N}\langle x_{j}|k_{j}\rangle\equiv \tilde{\Psi}_{\nu}(x_1,x_2,\cdots,x_N)$ by Eq.\ (\ref{Psi-tildePsi}), i.e.,
\begin{equation}
\Psi_{\nu}(x_{1},x_{2},\cdots,x_{N})=\frac{A_{N}}{N!}\sum_{\hat{P}}(\pm 1)^{P}
\langle x_{1}|k_{p_1}\rangle\langle x_{2}|k_{p_2}\rangle\cdots \langle x_{N}|k_{p_N}\rangle , 
\label{Psi-NonInt}
\end{equation}
where $\nu$ denotes $\nu=(k_{1},k_{2},\cdots,k_{N})$.
In the case of fermions with the factor $(-1)^{P}$, the summation over permutations in Eq.\ (\ref{Psi-NonInt}) is equivalently expressed as\cite{Aitken}
\begin{equation}
\Psi_{\nu}^{({\rm F})}(x_{1},x_{2},\cdots,x_{N})=\frac{A_{N}^{({\rm F})}}{N!}\det
\left[
\begin{array}{ccc}
\langle x_{1} | k_{1}\rangle & \cdots &\langle x_{1} | k_{N}\rangle
\\
\vdots & &\vdots
\\
\langle x_{N} | k_{1}\rangle & \cdots &\langle x_{N} | k_{N}\rangle
\end{array}
\right] ,
\label{Slater}
\end{equation}
which is known as the Slater determinant.
It follows from the properties of the determinant that $\Psi_{\nu}^{({\rm F})}$ vanishes when two of the columns or rows are identical. We thereby arrive at the Pauli exclusion principle that no two identical fermions can occupy the same one-particle state or the same coordinate (including spin).
Thus, the Pauli exclusion principle naturally results from the permutation symmetry of identical particles.
By noting that all the  $k_{j}$'s are different, the normalization condition is transformed as
\begin{eqnarray*}
&&\hspace{0mm}
1=\int\!d x_{1}\cdots \int d x_{N}|
\Psi_{\nu}^{({\rm F})}(x_{1},\cdots,x_{N})|^2
= \frac{\bigl(A_N^{({\rm F})}\bigr)^2}{(N!)^{2}}\sum_{\hat{P}}\sum_{\hat{P}'}(-1)^{P'+P}\prod_{j=1}^{N}
\langle  k_{p_{j}'}| k_{p_{j}}\rangle
\\
&&\hspace{3mm}
=\frac{\bigl(A_N^{({\rm F})}\bigr)^2}{(N!)^{2}}\sum_{\hat{P}}\sum_{\hat{P}'}
(-1)^{P'+P}\prod_{j=1}^{N}\delta_{p_{j}'p_{j}}
=\frac{\bigl(A_N^{({\rm F})}\bigr)^2}{(N!)^{2}}\sum_{\hat{P}}\sum_{\hat{P}'}(-1)^{P'+P}\delta_{\hat{P}'\hat{P}}
=\frac{\bigl(A_N^{({\rm F})}\bigr)^2}{N!}.
\end{eqnarray*}
Hence, we obtain $A_N^{({\rm F})}$ as
\begin{equation}
A_N^{({\rm F})}=\sqrt{N!} .
\label{A^F}
\end{equation}

For bosons, on the other hand, every state $k_{j}$ can accommodate multiple particles.
Consider specifically the case where there are $n_{j}$ particles in the state $k_{j}$ ($j=1,2,\cdots,\ell; \,\,\ell\leq N$), i.e., 
\begin{equation}
\nu=(\,\underbrace{k_{1},\cdots,k_{1}}_{n_{1}},\underbrace{k_{2},\cdots,k_{2}}_{n_{2}},\cdots,\underbrace{k_{\ell},\cdots,k_{\ell}}_{n_{\ell}}\,),\hspace{10mm}
\sum_{j=1}^{\ell}n_{j}=N .
\label{nu-Bose}
\end{equation}
The corresponding wave function is given by Eq.\ (\ref{Psi-NonInt}) with the upper sign, where
$k_{p_{j}}$ denotes a permutation of $\nu$ in Eq.\ (\ref{nu-Bose}).
The normalization condition is transformed as
\begin{eqnarray*}
&&\hspace{0mm}
1=\int d x_{1}\cdots \int d x_{N}|
\Psi_{\nu}^{({\rm B})}(x_{1},\cdots,x_{N})|^{2}
= \frac{\bigl(A_N^{({\rm B})}\bigr)^2}{(N!)^2}
\sum_{\hat{P}'}\sum_{\hat{P}}\prod_{j=1}^{N}
\langle k_{p_{j}'}| k_{p_{j}}\rangle 
\\
&&\hspace{3mm}
= \frac{\bigl(A_N^{({\rm B})}\bigr)^2}{(N!)^2}N! \sum_{\hat{P}} \prod_{j=1}^{N}
\langle  k_{j}| k_{p_{j}}\rangle
= \frac{\bigl(A_N^{({\rm B})}\bigr)^2}{N!} n_{1}!n_{2}!\cdots n_{\ell}! ,
\end{eqnarray*}
where the third equality originates from multiplying the result for $\hat{P}'=$(identity permutation) by $N!$.
Hence, we obtain the normalization constant as
\begin{equation}
A_N^{({\rm B})}=\frac{\sqrt{N!}}{\sqrt{n_{1}!n_{2}!\cdots n_{\ell}!}} .
\label{A^B}
\end{equation}

Now that we have obtained $N$-body wave functions in the configuration space, we write down the corresponding 
bra-vector $|\Psi_{\nu}\rangle$.
Consider first the case of fermions. 
Let us substitute Eq.\ (\ref{A^F}) into Eq.\ (\ref{Psi-NonInt}) with the sign $(-1)^{P}$, put the resultant expression into Eq.\ (\ref{|Psi>}), and transform it further  with Eqs.\ (\ref{varphi-oc})--(\ref{c-commute}).
We thereby obtain 
$|\Psi_{\nu}^{({\rm F})}\rangle$ as
\begin{subequations}
\label{|Psi^FB>}
\begin{equation}
|\Psi_{\nu}^{({\rm F})}\rangle=\hat{c}_{k_{1}}^{\dagger}\hat{c}_{k_{2}}^{\dagger}\cdots
\hat{c}_{k_{N}}^{\dagger}|0\rangle.
\label{|Psi^F>}
\end{equation}
This expression is much simpler than Eq.\ (\ref{Slater}) in the configuration space, and 
the sign change of the latter upon the exchange of columns $k_i\leftrightarrow k_j$ is algebraically realized in $|\Psi_{\nu}^{({\rm F})}\rangle$ through the exchange of positions between $\hat{c}_{k_{i}}^{\dagger}$ and $\hat{c}_{k_{j}}^{\dagger}$ with Eq.\ (\ref{c-commute}).
Specifically,  the $N$-body ground state corresponds to the state composed of $N$ lowest one-particle-energy states, which is sometimes called Fermi vacuum.

As for the case of bosons, we substitute Eq.\ (\ref{A^B}) into Eq.\ 
(\ref{Psi-NonInt}) with the sign $(+1)^{P}$ and transform the resultant expression in the same way as the case of fermions above.
We thereby obtain 
$|\Psi_{\nu}^{({\rm B})}\rangle$ corresponding to $\nu$ in Eq.\ (\ref{nu-Bose}) as
\begin{equation}
|\Psi_{\nu}^{({\rm B})}\rangle
=\frac{(\hat{c}_{k_{1}}^{\dagger})^{n_{1}}}{\sqrt{n_{1}!}}\cdots
\frac{(\hat{c}_{k_{\ell}}^{\dagger})^{n_{\ell}}}
{\sqrt{n_{\ell}!}}|0\rangle .
\label{|Psi^B>}
\end{equation}
\end{subequations}
This state is identical in expression to that of a harmonic oscillator with multiple frequencies.\cite{Sakurai}

It follows from Eq.\ (\ref{c-commute}) that $\hat{c}_{k}^{2}|0\rangle =0$ holds for fermions.
Taking this fact into account, we can put Eqs.\ (\ref{|Psi^F>}) and (\ref{|Psi^B>}) into a single expression as 
\begin{equation}
|\Psi_{\nu}\rangle
=\frac{(\hat{c}_{k_{1}}^{\dagger})^{n_{1}}}{\sqrt{n_{1}!}}\cdots
\frac{(\hat{c}_{k_{\ell}}^{\dagger})^{n_{\ell}}}
{\sqrt{n_{\ell}!}}|0\rangle .
\label{|Psi-nonint>}
\end{equation}
One may now realize that the second quantization method enables us to describe many-particle systems with identical particles concisely and conveniently.

\subsection{Density matrix in second quantization\label{subsec:DM}}

The density matrix is undoubtedly one of the most fundamental concepts  in quantum statistical mechanics.
We express here equilibrium density matrices in second quantization.

Consider first the canonical ensemble. 
Its density matrix can be expressed in terms of Hamiltonian (\ref{2nd-Q1}), its eigenstate (\ref{|Psi>}), and the eigenvalue $E_{\nu}$ as 
\begin{equation}
\hat{\rho} = 
\sum_{\nu}e^{\beta(F-\hat{\cal H})}|\Psi_{\nu}\rangle \langle \Psi_{\nu}|
, \hspace{10mm} F\equiv -\frac{1}{\beta}\ln\sum_{\nu}e^{-\beta E_{\nu}} ,
\label{rho-2nd}
\end{equation}
where $\beta\equiv 1/k_{\rm B}T$ with $k_{\rm B}$ denoting the Boltzmann constant and $T$ the temperature.
By using Eq.\ (\ref{rho-2nd}), the expectation value of an Hermitian operator $\hat{{\cal O}}$ is evaluated as
\begin{equation}
\langle\hat{\cal O}\rangle \equiv \sum_{\nu}e^{\beta(F-E_{\nu})}\langle \Psi_{\nu}|\hat{{\cal O}}|\Psi_{\nu}\rangle 
={\rm Tr}\,\hat{\rho}\, \hat{{\cal O}}.
\end{equation}
For example, the expectation value of the density operator 
$\hat{n}(x)=\hat{\psi}^{\dagger}(x)\hat{\psi}(x)$ is expressed as
\begin{equation}
\langle \hat{n}(x)\rangle =\sum_{\nu}e^{\beta(F-E_{\nu})}\langle \Psi_{\nu}|
\hat{\psi}^{\dagger}(x)\hat{\psi}(x)|\Psi_{\nu}\rangle
= {\rm Tr}\,\hat{\rho}\, \hat{\psi}^{\dagger}(x)\hat{\psi}(x).
\end{equation}

Next, the density matrix of the grand canonical ensemble is given by
\begin{equation}
\hat{\rho} = 
\sum_{\nu N}e^{\beta(\Omega-\hat{\cal H}+\mu\hat{N})}|\Psi_{\nu N}\rangle \langle \Psi_{\nu N}|,
\hspace{10mm}\Omega\equiv -\frac{1}{\beta}\ln\sum_{\nu N}e^{-\beta (E_{\nu}-\mu N)},
\label{rho-GC}
\end{equation}
where $\mu$ is the chemical potential, and 
\begin{equation}
\hat{N}\equiv \int \hat{\psi}^{\dagger}(x)\hat{\psi}(x)\,d  x
=\sum_{k}\hat{c}_{k}^{\dagger}\hat{c}_{k}
\label{N-opp}
\end{equation}
denotes the particle-number operator.
The second expression of Eq.\ (\ref{N-opp}) is obtained by making use of Eqs.\ (\ref{varphi-oc1}) and (\ref{psi-exp}).
We now focus specifically on the noninteracting case with Hamiltonian (\ref{H_0App-diag}). Substituting Eqs.\ (\ref{H_0App-diag}), (\ref{|Psi-nonint>}), and (\ref{N-opp}) into Eq.\ (\ref{rho-GC}),
we obtain the grand canonical density matrix without interaction as
\begin{equation}
\hat{\rho}_{0}=\prod_{k}\frac{\displaystyle
\sum_{n} e^{-\beta(\epsilon_{k}-\mu)n}|n\rangle_{\hspace{-0.3mm} k}
 \hspace{0.3mm}{_{k}\hspace{-0.3mm}\langle n|}}
{\displaystyle\sum_{n}e^{-\beta(\epsilon_{k}-\mu)n}},
\hspace{15mm}|n\rangle_{k} \equiv \frac{(\hat{c}_{k}^{\dagger})^{n}}{\sqrt{n!}}
|0\rangle_{k} ,
\label{rho_GC0}
\end{equation}
where $|0\rangle_{k}$ denotes the vacuum of the one-particle state $k$ defined by $\hat{c}_{k}|0\rangle_{k}=0$, and the $n$ summation runs over $n=0,1,2,\cdots$ ($n=0,1$) for bosons (fermions).

Equation (\ref{2nd-Q1}) tells us that it may be useful to introduce
\begin{subequations}
\label{rho^12}
\begin{eqnarray}
\rho^{(1)}(x_{1},x_{1}')&\equiv&{\rm Tr}\,\hat{\rho}\, \hat{\psi}^{\dagger}(x_{1}')\hat{\psi}(x_{1})
=\langle \hat{\psi}^{\dagger}(x_{1}')\hat{\psi}(x_{1})\rangle,
\label{rho^1}
\\
\rho^{(2)}(x_{1},x_{2};x_{1}',x_{2}')&\equiv&\langle \hat{\psi}^{\dagger}(x_{1}')
\hat{\psi}^{\dagger}(x_{2}')\hat{\psi}(x_{2})\hat{\psi}(x_{1})\rangle,
\label{rho^2}
\end{eqnarray}
\end{subequations}
which are called reduced density matrices. These expressions apply to both the canonical and grand canonical ensembles with either Eq.\ (\ref{rho-2nd}) or (\ref{rho-GC}) used as the density matrix.
The expectation value of every one-body (two-body) operator can be expressed with $\rho^{(1)}$ 
($\rho^{(2)}$). 
For example, the statistical average of Eq.\ (\ref{2nd-Q1}) can be written in terms of Eq.\ 
(\ref{rho^12}) as
\begin{eqnarray}
&&
\langle \hat{\cal H}\rangle  = \!\int\! d x_{1}\!\left[ \frac{\hat{\bm p}_{1}^{2}}{2m}
+U({\bm r}_{1}) \right]\!\rho^{(1)}(x_{1},x_{1}')\biggr|_{x_1'=x_1}
\nonumber \\
&&\hspace{10.5mm}
+ \frac{1}{2}\!\int\! d x_{1}\!\int\! d x_{2}
V(|{\bm r}_{1}-{\bm r}_{2}|)\rho^{(2)}(x_{1},x_{2};x_{1},x_{2}) .
\label{<calH>}
\end{eqnarray}
It is worth pointing out that Eq.\
(\ref{rho^12}) can also be expressed with Eqs.\ (\ref{|Psi>-Prop}) and (\ref{rho-2nd}) or (\ref{rho-GC}) in terms of wave functions in the configuration space. In the case of the canonical ensemble, for example, they are given by
\begin{subequations}
\begin{equation}
\rho^{(1)}(x_{1},x_{1}')=\sum_{\nu}e^{\beta (F-E_{\nu})}N \int d x_{2}
\cdots\int d x_{N} \Psi_\nu(x_1,x_2,\cdots,x_N)\Psi_\nu^*(x_1',x_2,\cdots,x_N),
\end{equation}
\begin{eqnarray}
&&
\rho^{(2)}(x_{1},x_{2};x_{1}',x_{2}')=
\sum_{\nu}e^{\beta (F-E_{\nu})}N (N-1)\int d x_{3}
\cdots\int d x_{N} \Psi_\nu(x_1,x_2,x_3,\cdots,x_N)
\nonumber \\
&&\hspace{34mm}
\times\Psi_\nu^*(x_1',x_2',x_3,\cdots,x_N).
\end{eqnarray}
\end{subequations}
These reduced density matrices were used by Yang\cite{Yang62}
to discuss the concept of off-diagonal long-range order in 
superfluids and superconductors.

\section{Self-Consistent Perturbation Expansion\label{sec:SCPE}}

We explain here the self-consistent perturbation expansion in the equilibrium Matsubara formalism.
Its advantages are summarized below Eq.\ (\ref{checkSigma-symm}).
The contents here also form the basis to understand nonequilibrium quantum field theory developed in \S\S \ref{sec:rep} and \ref{sec:noneq}.

\subsection{Thermodynamic potential and Green's function}

We consider a system described by the Hamiltonian:
\begin{equation}
\hat{\cal H} = \hat{\cal H}_{0}+\hat{\cal H}_{\rm int}  ,
\label{H-M}
\end{equation}
where $\hat{\cal H}_{0}$ and $\hat{\cal H}_{\rm int}$ are defined by Eqs.\ (\ref{H_0}) and (\ref{H_int}), respectively,
with no time dependence in $U$ of Eq.\ (\ref{K_1}). Note that Eq.\ (\ref{K_1}) includes the chemical potential.
We drop here the spin degrees of freedom for simplicity once again.
The thermodynamic potential is given by\cite{LW60,AGD63,FW71}
\begin{equation}
\Omega = -\frac{1}{\beta}\ln {\rm Tr}\,e^{-\beta\hat{\cal H}},
\label{Omega}
\end{equation}
with $\beta\equiv 1/k_{\rm B}T$. 
To calculate $\Omega$ perturbatively in terms of $\hat{\cal H}_{\rm int}$,
let us introduce the scattering matrix $\hat{\cal S}(\beta)$ as
\begin{equation}
\hat{\cal S}(\beta)\equiv e^{\beta\hat{\cal H}_{0}}e^{-\beta\hat{\cal H}} .
\label{S(b)}
\end{equation}
It obeys the equation of motion:
\begin{equation}
\frac{d  \hat{\cal S}(\beta)}{d  \beta}
=-\hat{\cal H}_{\rm int}(\beta)\hat{\cal S}(\beta),\hspace{10mm}\hat{\cal S}(0)=1,
\label{dSdb}
\end{equation}
with
\begin{equation}
\hat{\cal H}_{\rm int}(\beta)\equiv e^{\beta\hat{\cal H}_{0}}
\hat{\cal H}_{\rm int}e^{-\beta\hat{\cal H}_{0}}.
\label{H_int(b)}
\end{equation}
Looking back at Eqs.\ (\ref{U-def}) and (\ref{U-eq-motion1}), we can solve Eq.\ (\ref{dSdb}) as
\begin{eqnarray}
&&\hspace{0mm}
\hat{\cal S}(\beta)
=1
+\sum_{n=1}^{\infty}(-1)^{n}
\int_{0}^{\beta}d \tau_{n}\cdots
\int_{0}^{\beta}d \tau_{1}\hat{\cal H}_{\rm int}(\tau_{n})\cdots
\hat{\cal H}_{\rm int}(\tau_{1})
\nonumber \\
&&\hspace{9mm}
={\cal T}_{\tau}\exp\left[-\int_{0}^{\beta}d \tau\,\hat{\cal H}_{\rm int}(\tau)\right] ,
\label{S(b)-explicit2}
\end{eqnarray}
where ${\cal T}_{\tau}$ orders the Hamiltonians into the ascending sequence of the arguments from right to left
with an extra sign change upon every permutation for fermion field operators.\cite{AGD63,FW71}
One may easily see that this expression with Eq.\ (\ref{H_int}) is equivalent to Eq.\ (\ref{S-M}).
We next introduce the noninteracting Matsubara Green's function:
\begin{equation}
G_{0}(1,2)=-\langle {\cal T}_{\tau} \hat{\psi}(1)\hat{\psi}^{\dagger}(2)\rangle_{0},
\end{equation}
where $\hat{\psi}(1)\equiv e^{\tau_{1}\hat{\cal H}_{0}}\hat{\psi}({\bm r}_{1})e^{-\tau_{1}\hat{\cal H}_{0}}$
and $\hat{\psi}^{\dagger}(1)\equiv e^{\tau_{1}\hat{\cal H}_{0}}\hat{\psi}^{\dagger}({\bm r}_{1})e^{-\tau_{1}\hat{\cal H}_{0}}$
with $1\equiv {\bm r}_{1}\tau_{1}$, and $\langle\cdots\rangle_{0}$ denotes the average
in terms of the density matrix $\hat{\rho}_0\equiv e^{-\beta\hat{\cal H}_{0}}/{\rm Tr}e^{-\beta\hat{\cal H}_{0}}$.
Note that $\hat{\psi}^{\dagger}(1)$ is not a Hermitian conjugate of $\hat{\psi}(1)$ in the equilibrium Matsubara formalism.\cite{AGD63,FW71}
Writing Eq.\ (\ref{S(b)}) as $e^{-\beta\hat{\cal H}}=e^{-\beta\hat{\cal H}_{0}}\hat{\cal S}(\beta)$, we can transform Eq.\ (\ref{Omega}) into
\begin{equation}
\Omega=\Omega_{0}-\frac{1}{\beta}\ln\langle\hat{\cal S}(\beta)\rangle_{0}
=\Omega_{0}-\frac{1}{\beta}\bigl[\langle\hat{\cal S}(\beta)\rangle_{0{\rm c}} -1\bigr],
\label{Omega2}
\end{equation}
where $\Omega_{0}$ is the noninteracting thermodynamic potential, and the subscript $_{\rm c}$ denotes retaining only connected Feynman diagrams in the expansion of $\langle\hat{\cal S}(\beta)\rangle_{0}$ with $G_0$.
See, e.g., Ref.\ \citen{AGD63} for the proof of the second equality that taking the logarithm of $\langle\hat{\cal S}(\beta)\rangle_{0}$ amounts to retaining only connected diagrams of $\langle \hat{\cal S}(\beta)\rangle_{0}-1$. 

We focus next on the one-particle properties. They can be evaluated with the Matsubara Green's function:
\begin{subequations}
\label{G-Matsubara}
\begin{equation}
G(1,2)=-\langle {\cal T}_{\tau} \hat{\psi}_{\rm H}(1)\hat{\psi}^{\dagger}_{\rm H}(2)\rangle,
\label{G-Matsubara1}
\end{equation}
where $\hat{\psi}_{\rm H}(1)\equiv e^{\tau_{1}\hat{\cal H}}\hat{\psi}({\bm r}_{1})e^{-\tau_{1}\hat{\cal H}}$,
$\hat{\psi}_{\rm H}^{\dagger}(1)\equiv e^{\tau_{1}\hat{\cal H}}\hat{\psi}^{\dagger}({\bm r}_{1})e^{-\tau_{1}\hat{\cal H}}$, and $\langle\cdots\rangle$ denotes the average
in terms of the density matrix $\hat{\rho}\equiv e^{-\beta\hat{\cal H}}/{\rm Tr}e^{-\beta\hat{\cal H}}$. 
Indeed, Eq.\ (\ref{G-Matsubara1}) is a direct extension of the one-particle density matrix (\ref{rho^1}) to incorporate the $\tau$ dependence.
The Matsubara Green's function can also be expressed with $\hat{\cal S}(\beta)$ as\cite{AGD63,FW71}
\begin{equation}
G(1,2)=-\langle {\cal T}_{\tau} \hat{\cal S}(\beta)\hat{\psi}(1)\hat{\psi}^{\dagger}(2)\rangle_{0{\rm c}},
\label{G-Matsubara2}
\end{equation}
\end{subequations}
which can be calculated perturbatively with $G_{0}$.\cite{AGD63,FW71}

To derive an equation of motion for $G$, we
differentiate $\hat{\psi}_{\rm H}(1)\equiv e^{\tau_{1}\hat{\cal H}}\hat{\psi}({\bm r}_{1})e^{-\tau_{1}\hat{\cal H}}$ in terms of $\tau_{1}$ to obtain
$$
\frac{\partial \hat{\psi}_{\rm H}(1)}{\partial \tau_{1}}=-e^{\tau_{1}\hat{\cal H}}[\hat{\psi}({\bm r}_{1})\hat{\cal H}-
\hat{\cal H}\hat{\psi}({\bm r}_{1})]e^{-\tau_{1}\hat{\cal H}} .
$$
We then substitute Eq.\ (\ref{H-M}) with Eqs.\ (\ref{H_0}) and (\ref{H_int}) into the above equation and  calculate the commutator.
The result can be expressed as
\begin{equation}
\frac{\partial \hat{\psi}_{\rm H}(1)}{\partial \tau_{1}}
=-\hat{K}_{1}\hat{\psi}_{\rm H}(1)
-\int d 2\, \bar{V}(1-2)\hat{\psi}_{\rm H}^{\dagger}(2)\hat{\psi}_{\rm H}(2)\hat{\psi}_{\rm H}(1),
\label{eq-motion-psi}
\end{equation}
where $\hat{K}_{1}$ is defined by Eq.\ (\ref{K_1}) with no time dependence in $U$, and $\bar{V}(1-2)\equiv \delta(\tau_{1}-\tau_{2})V({\bm r}_1-{\bm r}_2)$.
Let us multiply Eq.\ (\ref{eq-motion-psi}) by $-\hat{\psi}_{\rm H}^{\dagger}(1')$ from the right, operate  ${\cal T}_{\tau}$,
average the resultant expression with $\hat{\rho}\equiv e^{-\beta\hat{\cal H}}
/{\rm Tr}\,e^{-\beta\hat{\cal H}}$, and make use of $
\langle {\cal T}_{\tau}\frac{\partial \hat{\psi}_{\rm H}(1)}{\partial \tau_{1}}
\hat{\psi}_{\rm H}^{\dagger}(1')\rangle=\frac{\partial \langle {\cal T}_{\tau}\hat{\psi}_{\rm H}(1)\hat{\psi}_{\rm H}^{\dagger}(1')\rangle}{\partial \tau_{1}}-\delta(1,1')$.
The final result can be written  in terms of Eq.\ (\ref{G-Matsubara}) as
\begin{equation}
\left(-\frac{\partial }{\partial \tau_{1}}-\hat{K}_{1}\right) G(1,1')
+\int d 2\, \bar{V}(1-2)\langle {\cal T}_{\tau}\hat{\psi}_{\rm H}^{\dagger}(2)\hat{\psi}_{\rm H}(2)\hat{\psi}_{\rm H}(1)
\hat{\psi}_{\rm H}^{\dagger}(1')\rangle 
=\delta(1,1').
\label{eq-motion-G}
\end{equation}
The equation of motion for $G_{0}$ is obtained from above by setting $\bar{V}=0$, which may be expressed alternatively as
\begin{equation}
G_{0}^{-1}(1,1')=\left(-\frac{\partial}{\partial \tau_{1}}-\hat{K}_{1}\right)\delta(1,1') .
\label{G_0^-1-M}
\end{equation}

On the other hand, inspection of the perturbation expansion of Eq.\ (\ref{G-Matsubara2}) shows that $G(1,2)$ satisfies Dyson's equation:\cite{LW60,AGD63,FW71}
\begin{subequations}
\label{Dyson-M}
\begin{equation}
G(1,1')=G_{0}(1,1')+\int d 2 \int d 2' G_{0}(1,2)\Sigma(2,2')G(2',1') ,
\label{Dyson1}
\end{equation}
where $\Sigma(2,2')$ is the (proper) self-energy, which cannot be separated by cutting any single-particle line. 
The first few Feynman diagrams in the perturbation expansion for $\Sigma$ are given in Fig.\ \ref{fig:self-energy},
where straight and wavy lines denote $G_{0}$ and $\bar{V}$, respectively.
\begin{figure}[t]
\begin{center}
\includegraphics[height=15mm]{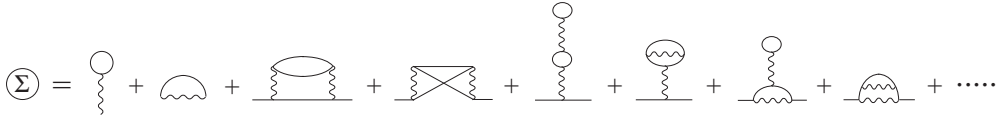}
\end{center}
\caption{First few Feynman diagrams for the proper self-energy.\label{fig:self-energy}}
\end{figure}
Let us multiply Eq.\ (\ref{Dyson1}) by $G_{0}^{-1}(3,1)$, carry out an integration over $1$, and make use of Eq.\ (\ref{G_0^-1-M}). We thereby obtain an alternative expression of Eq.\ (\ref{Dyson1}) as
\begin{equation}
\left(-\frac{\partial}{\partial \tau_{1}}-\hat{K}_{1}\right)
G(1,1')-\int\Sigma(1,2)G(2,1')\, d 2=\delta(1,1').
\label{Dyson2}
\end{equation}
\end{subequations}

Comparing Eq.\ (\ref{Dyson2}) with Eq.\ (\ref{eq-motion-G}), we arrive at the identity:
\begin{equation}
\int d 2\, \Sigma(1,2)G(2,1')=-
\int d 2\, \bar{V}(1-2)\langle {\cal T}_{\tau}\hat{\psi}_{\rm H}^{\dagger}(2)
\hat{\psi}_{\rm H}(2)\hat{\psi}_{\rm H}(1)\hat{\psi}_{\rm H}^{\dagger}(1')\rangle.
\label{Sigma-V}
\end{equation}
We now set $1'=1_+$ in Eq.\ (\ref{Sigma-V}), where the subscript $_+$ denotes the presence of an extra infinitesimal positive constant in $\tau_{1}$ to place the creation operator $\hat{\psi}_{\rm H}^{\dagger}(1')$ to the left of the annihilation operators. We then multiply the resultant equation by
$\mp 1/2\beta$ and perform an integration over $1$.
The result can be expressed in terms of Eq.\ (\ref{H_int}) as
\begin{equation}
\langle\hat{\cal H}_{\rm int}\rangle=\mp\frac{1}{2\beta}
\int d 1\int d 2\, \Sigma(1,2)G(2,1')\bigr|_{1'=1_+},
\label{<H_int>}
\end{equation}
which will play an important role below.

\subsection{Renormalization of self-energy}

Let us focus on the last four diagrams of the second order in Fig.\ \ref{fig:self-energy}. We then notice that the parts enclosed by squares below have the effect of inserting the first-order self-energy into $G_{0}$ to renormalize $G_{0}\rightarrow G$ ($G$ is denoted by a thick line).
\vspace{2mm}
\begin{center}
\includegraphics[height=20mm]{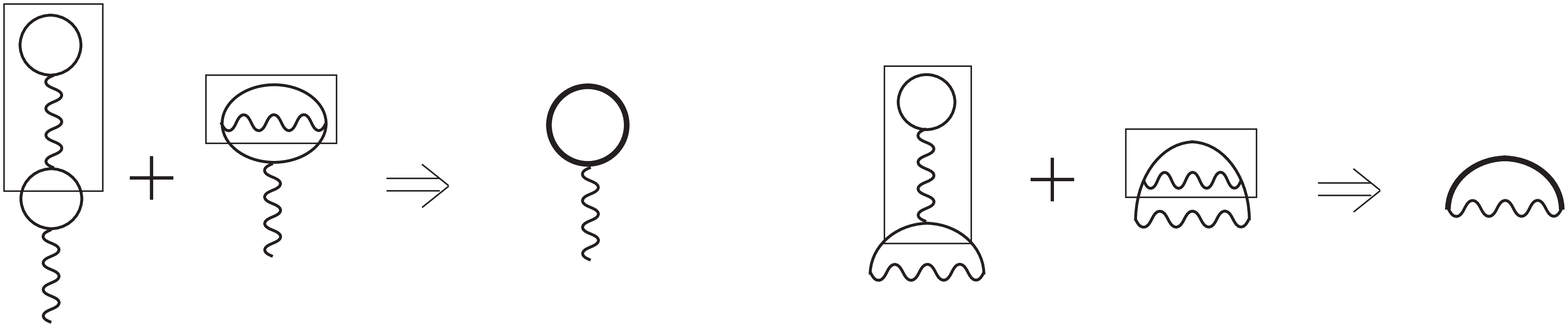}
\end{center}
\vspace{2mm}
The observation suggests an alternative way to obtain the self-energy:
consider only skeleton diagrams, i.e., diagrams without self-energy insertions, in the expansion for the self-energy
and put $G_{0}\rightarrow G$.
We call this expansion the renormalized skeleton diagram expansion for the self-energy,
which can be expressed graphically as follows.
\vspace{2mm}
\begin{center}
\includegraphics[height=10mm]{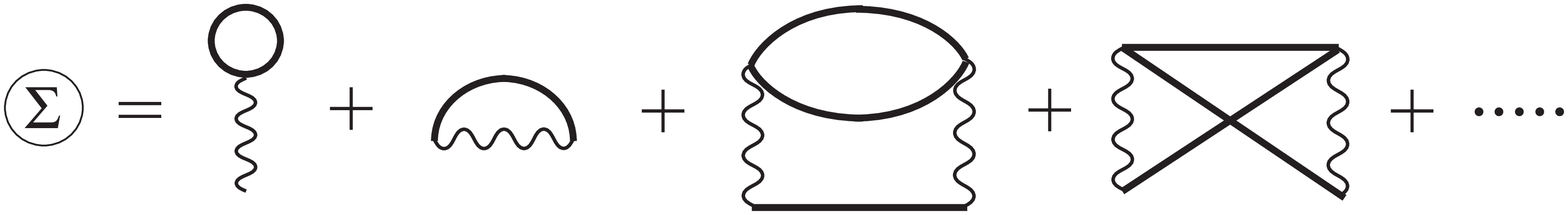}
\end{center}
\vspace{2mm}
The corresponding analytic expression is given by
\begin{eqnarray}
&&\hspace{0mm}
\Sigma(1,1')=\mp \delta(1,1')\int d 2\,\bar{V}(1-2)G(2,2)-\bar{V}(1-1')G(1,1')
\nonumber \\
&&\hspace{18mm}
+\int d 2\int d 2'\,
\bar{V}(1-2)\bar{V}(1'-2')[\pm G(1,1')G(2,2')G(2',2)
\nonumber \\
&&\hspace{18mm}
+G(1,2')G(2',2)G(2,1')]
+\cdots ,
\label{Sigma-exp}
\end{eqnarray}
where $G(1,1')$ for $\tau_{1}=\tau_{1}'$ means $G(1,1'_{+})$.

The renormalized skeleton diagram expansion is a self-consistent perturbation expansion.
Indeed, $\Sigma$ here is given as a functional of $G$ as 
$\Sigma=\Sigma[G]$, whereas 
$G$ obeys Dyson's equation:
\begin{figure}[h]
\begin{center}
\hspace{10mm}
\includegraphics[height=5.5mm]{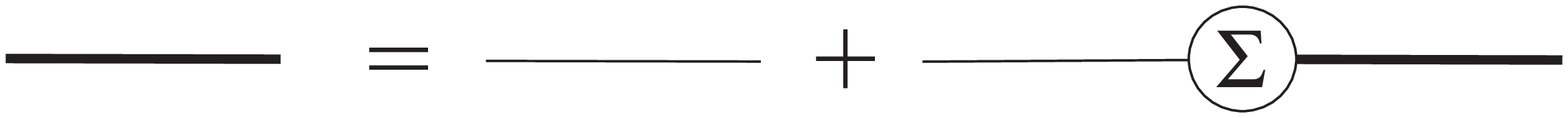}
\end{center}
\end{figure}

\vspace{0mm}
\noindent
so that it necessarily depends on $\Sigma$.
Hence, it follows that 
$G$ and $\Sigma$ have to be determined self-consistently in the renormalized skeleton diagram expansion.
The first-order self-energy, denoted by $\Sigma^{\rm HF}$, is expressed graphically as follows.
\vspace{2mm}
\begin{center}
\includegraphics[height=12mm]{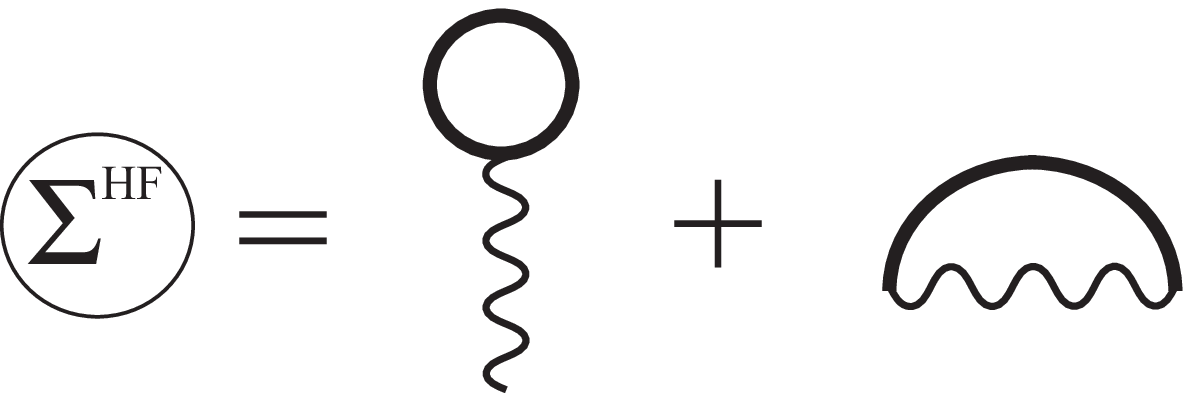}
\end{center}
Using $\Sigma^{\rm HF}$ in Dyson's equation (\ref{Dyson-M}) reproduces the self-consistent
Hartree-Fock approximation, as first pointed out by Luttinger.\cite{Luttinger60}
He also suggested more general approximations of incorporating higher-order contributions in Eq.\ (\ref{Sigma-exp})
into the self-energy of Dyson's equation (\ref{Dyson-M}).\cite{Luttinger60}

\subsection{Self-energy as a functional derivative of $\Phi$}

The renormalized self-energy in Eq.\ (\ref{Sigma-exp}) can also be obtained as follows. Let us introduce the functional $\Phi=\Phi[G]$ as the contribution of the skeleton diagrams in the perturbation expansion for $\Omega$
with the replacement $G_{0}\rightarrow G$.
It may be expressed graphically as follows.
\begin{center}
\includegraphics[height=10mm]{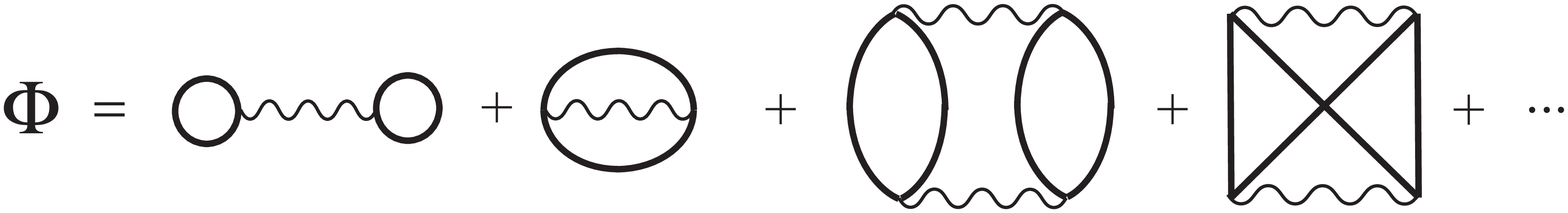}
\end{center}
The corresponding analytic expression is obtained from Eq.\ (\ref{Omega2}) as
\begin{eqnarray}
&&\hspace{0mm}
\Phi \equiv-\frac{1}{\beta}
[\langle \hat{\cal S}(\beta)\rangle_{0{\rm c}}-1]_{{\rm skeleton},G_{0}\rightarrow G}
\nonumber \\
&&\hspace{4mm}
=\frac{1}{2\beta}\int d 1\int d 1'\,\bar{V}(1-1')
[G(1,1)G(1',1')\pm G(1,1')G(1',1)]
\nonumber \\
&&\hspace{8.5mm}
-\frac{1}{4\beta}\int d 1\int d 1'\int d 2\int d 2'\,
\bar{V}(1-1')\bar{V}(2-2') [G(1,2)G(2,1)
\nonumber \\
&&\hspace{8.5mm}
\times G(1',2')G(2',1')
\pm G(1,2)G(2,1')G(1',2')G(2',1)]
+\cdots .
\label{Phi}
\end{eqnarray}
Comparing Eq.\ (\ref{Phi}) with Eq.\ (\ref{Sigma-exp}), we notice that there is an additional $G$ in each order
with an extra factor $-(\pm 1)/2n\beta$ in the $n$th-order terms.
The factor $1/2n$ reflects the fact that $2n$ $G$'s are equivalent in the $n$th-order closed diagrams for $\Phi$, and $\pm 1$ originates from the fact that there is an extra closed particle line. With these observations, one may convince oneself order by order that the self-energy of Eq.\ (\ref{Sigma-exp}) is also obtained from $\Phi$ above by
\begin{equation}
\Sigma(1,1')= \mp \beta \frac{\delta\Phi}{\delta G(1',1)} .
\label{dPhi/dG}
\end{equation}
This is an exact relation with $\Phi$ defined by Eq.\ (\ref{Phi}). 
Equation (\ref{dPhi/dG}) also provides us with a systematic self-consistent approximation scheme called ``$\Phi$-derivable approximation'' by Baym.\cite{Baym62} It consists of (i) retaining only part of the series in Eq.\ (\ref{Phi}) and (ii) determining 
$G$ and $\Sigma$ self-consistently using Eqs.\ (\ref{Dyson-M}) and (\ref{dPhi/dG}).
Although it is equivalent to considering some partial series in Eq.\ (\ref{Sigma-exp}),
the $\Phi$-derivable approximation has a clear advantage that giving $\Phi$ also enables us to calculate two-particle and higher-order correlations, as shown in \S \ref{two-body}. It is also with $\Phi$ that one can prove conservation laws concisely; see \S \ref{sec:conserv} on this point.

\section{Luttinger-Ward Functional\label{LW}}

The thermodynamic potential is defined by Eq.\  (\ref{Omega}), which can also be expressed as Eq.\ (\ref{Omega2}).
Below, we provide it with a couple of alternative expressions. The latter expression, i.e., the Luttinger-Ward expression,\cite{LW60} is quite useful for clarifying its general properties as well as for carrying out practical calculations.

Let us introduce a parameter $\lambda'$ into $\hat{\cal H}=\hat{\cal H}_0+\hat{\cal H}_{\rm int}$ of Eq.\ (\ref{Omega}) as $\Omega_{\lambda'} = -\beta^{-1}\ln {\rm Tr}\,e^{-\beta(\hat{\cal H}_0+\lambda'\hat{\cal H}_{\rm int})}$.
We next differentiate $\Omega_{\lambda'}$ in terms of $\lambda'$
to obtain
$$
\frac{\partial \Omega_{\lambda'}}{\partial \lambda'}=
\frac{{\rm Tr}\,e^{-\beta \hat{\cal H}_{\lambda'}}\hat{\cal H}_{\rm int}}
{{\rm Tr}\,e^{-\beta \hat{\cal H}_{\lambda'}}}=\frac{\langle \lambda'\hat{\cal H}_{\rm int}\rangle_{\lambda'}}
{\lambda'},
$$
where $\langle \cdots\rangle_{\lambda'}$ denotes the grand canonical average with respect to
$\hat{\cal H}_{\lambda'}\equiv\hat{\cal H}_0+\lambda'\hat{\cal H}_{\rm int}$. 
Let us integrate the above equation in terms of $\lambda'$ from $0$ to $\lambda$ with 
$\hat{\cal H}_{\lambda'=0}=\hat{\cal H}_{0}$ in mind.
We thereby obtain
\begin{equation}
\Omega_{\lambda}=\Omega_{0}+\int_{0}^{\lambda}\frac{\langle \lambda'\hat{\cal H}_{\rm int}\rangle_{\lambda'}}
{\lambda'}d \lambda' .
\label{Omega-3}
\end{equation}
We finally substitute Eq.\ (\ref{<H_int>}) into Eq.\ (\ref{Omega-3}) to arrive at
\begin{equation}
\Omega_{\lambda}=\Omega_0\mp\frac{1}{2\beta}\int_{0}^{\lambda}\frac{d \lambda'}{\lambda'}
\int d 1\int d 2\, \Sigma_{\lambda'}(1,2)G_{\lambda'}(2,1_{+}).
\label{Omega_lambda}
\end{equation}
We need $\Sigma_{\lambda}$ and $G_{\lambda}$ over $0\leq\lambda\leq 1$ before using this expression.

Luttinger and Ward\cite{LW60} found an alternative expression for $\Omega$ as a functional of the self-energy $\Sigma$.
It is more convenient\cite{Baym62} to rewrite their result as a functional of $G$ as
\begin{equation}
\Omega = \pm \frac{1}{\beta}{\rm Tr}[\ln (-\underline{G}_{0}^{-1}+\underline{\Sigma} )
+\underline{\Sigma}\,\underline{G}]+\Phi .
\label{Omega-4}
\end{equation}
Here, $\underline{G}$ denotes the matrix with the element $G(1,1')$, the symbol ${\rm Tr}$ signifies
\begin{equation}
{\rm Tr}\underline{A}\equiv\int A(1,1_{+})d 1,
\label{Tr-def}
\end{equation}
and $\Phi$ is defined by Eq.\ (\ref{Phi}).
This expression enables us to calculate $\Omega$ directly from $G$ and $\Sigma$, and also 
to obtain some exact results on thermodynamic quantities.

Expression (\ref{Omega-4}) has an important property of being stationary with respect to a variation of $G$ as
\begin{eqnarray}
&&\hspace{0mm}
\frac{\delta\Omega}{\delta G(1',1)}
=\pm 
\frac{1}{\beta}{\rm Tr}\!\left[\!- 
\frac{\delta\underline{\Sigma}}{\delta G(1',1)}(\underline{G}_{0}^{-1}\!-\underline{\Sigma})^{-1}
+\frac{\delta\underline{\Sigma}}{\delta G(1',1)}\underline{G}
+\underline{\Sigma}\frac{\delta\underline{G}}{\delta G(1',1)}\!\right]
+\frac{\delta\Phi}{\delta G(1',1)}
\nonumber \\
&&\hspace{16.2mm}
=0,
\label{dOmega/dG}
\end{eqnarray}
where we have made use of Eq.\ (\ref{Dyson2}), i.e.\ $\underline{G}^{-1}=\underline{G}_{0}^{-1}-\underline{\Sigma}$, and Eq.\ (\ref{dPhi/dG}).
Thus, $\Omega$ is invariant through an infinitesimal variation in $G$.
See Ref.\ \citen{dDM64} by de Dominicis and Martin for a more general proof of the stationarity as well as its extensions to the superfluid phase.

Using Eq.\ (\ref{dOmega/dG}), we can prove the equivalence of Eq.\ (\ref{Omega-4}) with Eq.\ (\ref{Omega_lambda}) for $\lambda=1$. 
To this end, we express as $\tilde{\Omega}_\lambda$ the quantity obtained from the right-hand side of Eq.\ (\ref{Omega-4}) by the replacement $\hat{\cal H}_{\rm int}\rightarrow\lambda\hat{\cal H}_{\rm int}$, i.e.,
\begin{equation}
\tilde{\Omega}_\lambda \equiv \pm\frac{1}{\beta}{\rm Tr}[\ln (-\underline{G}_{0}^{-1}+\underline{\Sigma}_{\lambda} )
+\underline{\Sigma}_{\lambda}\underline{G}_{\lambda}]+\Phi_\lambda .
\label{Y}
\end{equation}
There are two types of $\lambda$ dependences in 
$\tilde{\Omega}_\lambda$. The first one originates from the $n$ interaction lines in the $n$th-order contribution to $\Phi_\lambda$, and the other from the implicit dependence through $G_{\lambda}$.
Let us differentiate $\tilde{\Omega}_\lambda$ with respect to $\lambda$.
Due to Eq.\ (\ref{dOmega/dG}), however, we need not consider the latter dependence in the differentiation.
Noting $d \lambda^{n}/d \lambda=n\lambda^{n}/\lambda$ in the explicit dependence as well as Eqs.\  (\ref{dPhi/dG}) and (\ref{Omega_lambda}), we obtain
\begin{equation}
\frac{\partial \tilde{\Omega}_\lambda}{\partial\lambda}
=\mp\frac{1}{2\beta\lambda}
{\rm Tr}\underline{\Sigma}_{\lambda}\underline{G}_{\lambda}
=\frac{\partial\Omega_{\lambda}}{\partial\lambda}.
\label{dY/dlambda}
\end{equation}
Thus, $\tilde{\Omega}_\lambda$ and $\Omega_{\lambda}$ satisfy the same first-order differential equation in terms of $\lambda$.

\begin{figure}[t]
\begin{center}
\includegraphics[width=50mm]{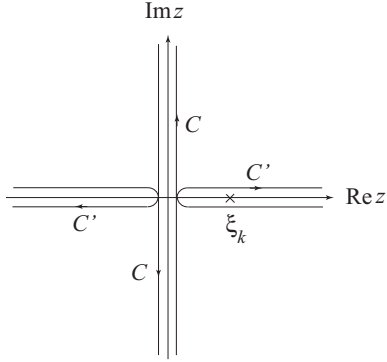}
\end{center}
\caption{Contours on the complex $z$ plane.\label{Contours-z}}
\end{figure}

To complete our proof of $\tilde{\Omega}_\lambda=\Omega_{\lambda}$, we now show that the initial value $\tilde{\Omega}_{0}=\pm {\beta}^{-1}{\rm Tr}\ln (-\underline{G}_{0}^{-1})$ satisfies $\tilde{\Omega}_{0}=\Omega_{0}$.
To this end, we adopt the representation of diagonalizing 
Eq.\ (\ref{G_0^-1-M}) to expand it as
$$
{G}_{0}^{-1}(1,1')=\frac{1}{\beta}\sum_{\ell=-\infty}^{\infty}\sum_{k}
(i\varepsilon_{\ell}-\xi_{k})\varphi_{k}({\bm r}_{1})\varphi_{k}^{*}({\bm r}_{1}')e^{-i\varepsilon_{\ell}(\tau_1-\tau_1')} ,
$$
where $\varepsilon_{\ell}$ denotes $2\ell\pi/\beta$ and $(2\ell+1)\pi/\beta$ for bosons and fermions, and $\xi_{k}$ and $\varphi_{k}({\bm r}_{1})$ are the eigenvalue and eigenfunction of the operator
$\hat{K}_{1}\equiv -({\hbar^{2}}/{2m})\nabla_{1}^{2}+U({\bm r}_{1})-\mu$, respectively.
Then, $\tilde{\Omega}_{0}$ is expressed as
\begin{equation}
\tilde{\Omega}_{0}=\pm \frac{1}{\beta}\sum_{\ell}\sum_{k}e^{i\varepsilon_{\ell} 0_+}\ln(\xi_{k}-i\varepsilon_{\ell}).
\label{Y_0-0}
\end{equation}
The factor $e^{i\varepsilon_{\ell} 0_+}$ originates from the subscript $_+$ in Eq.\ (\ref{Tr-def}).
By noting that $i\varepsilon_{\ell}$ is a pole of the function
\begin{equation}
f(z)=\frac{1}{e^{\beta z}\mp 1} ,
\label{f(z)}
\end{equation}
with the residue $\pm \beta^{-1}$,
the summation over $\ell$ can be transformed into an integration on the complex $z$ plane as
$$
\tilde{\Omega}_{0}=\int_{C}\frac{d z}{2\pi i}\sum_{k}\frac{e^{z 0_+}}{e^{\beta z}\mp 1}\ln(\xi_{k}-z),
$$
where $C$ denotes the contour in Fig.\ \ref{Contours-z}.
With the factor $e^{z 0_+}$, we can further make use of  Jordan's lemma to deform $C$ into $C'$ in Fig.\ \ref{Contours-z}.
We then perform an integration by parts with $(e^{\beta z}\mp 1)^{-1}=
\pm\beta^{-1}\frac{d }{d z}\ln(1\mp e^{-\beta z})$.
It then turns out that there is null contribution from the edges of $C'$.
Finally, using the residue theorem, we obtain
\begin{equation}
\tilde{\Omega}_{0}=\pm\frac{1}{\beta}\sum_{k}\ln(1\mp e^{-\beta \xi_{k}})=\Omega_{0}.
\label{Y_0}
\end{equation}
Equations (\ref{dY/dlambda}) and (\ref{Y_0}) enable us to conclude $\tilde{\Omega}_\lambda=\Omega_\lambda$; setting $\lambda=1$, we arrive at Eq.\ (\ref{Omega-4}).

\section{Expression of Equilibrium Entropy\label{sec:S_eq}}

We derive here an exact expression of equilibrium entropy $S=-{\partial \Omega}/{\partial T}$ from Eq.\ 
(\ref{Omega-4}).\cite{Kita99,Kita06a} 

Let us expand Green's function and the self-energy as
\begin{subequations}
\label{GS-exp}
\begin{eqnarray}
G(1,2)&=&\frac{1}{\beta}\sum_{\ell=-\infty}^{\infty}G({\bm r}_{1},{\bm r}_{2}; i\varepsilon_{\ell})\,e^{-i\varepsilon_{\ell}(\tau_{1}-\tau_{2})},
\label{G-exp}
\\
\Sigma(1,2)&=&\frac{1}{\beta}\sum_{\ell=-\infty}^{\infty}\Sigma({\bm r}_{1},{\bm r}_{2};i\varepsilon_{\ell})\,e^{-i\varepsilon_{\ell}(\tau_{1}-\tau_{2})},
\label{S-exp}
\end{eqnarray}
\end{subequations}
where $\varepsilon_{\ell}=2\ell\pi/\beta$ and $(2\ell+1)\pi/\beta$ for bosons and fermions, respectively.
Substituting these expressions into it and carrying out Tr over $\tau$, we can transform Eq.\ (\ref{Omega-4}) into
\begin{equation}
\Omega = \pm \frac{1}{\beta}\sum_{\ell}
{\rm Tr}\bigl\{\ln \bigl[\underline{K}+\underline{\Sigma}(i\varepsilon_{\ell} )
-i\varepsilon_{\ell}\underline{1}\bigr]
+\underline{\Sigma}(i\varepsilon_{\ell} )\,\underline{G}(i\varepsilon_{\ell} )\bigr\}
e^{i\varepsilon_{\ell}0_{+}}
+\Phi .
\label{Omega-4b}
\end{equation}
Here, $\underline{K}$, $\underline{G}(i\varepsilon_{\ell})$, and $\underline{\Sigma}(i\varepsilon_{\ell})$ signify matrices with the elements $$K({\bm r}_{1},{\bm r}_{1}')\equiv \left[-({\hbar^{2}}/{2m})\nabla_{1}^{2}+U({\bm r}_{1})-\mu\right]\delta({\bm r}_{1}-{\bm r}_{1}'),$$
$G({\bm r}_1,{\bm r}_1';i\varepsilon_{\ell})$, and $\Sigma({\bm r}_1,{\bm r}_1';i\varepsilon_{\ell})$, respectively,
and the symbol ${\rm Tr}$ is defined by Eq.\ (\ref{Tr-def}) without the integration over $\tau_{1}$.
We then express $\underline{G}(i\varepsilon_{\ell})$ in the Lehmann representation\cite{AGD63,FW71} as
\begin{equation}
G({\bm r}_1,{\bm r}_1';i\varepsilon_{\ell})= \frac{1}{2\pi}\int_{-\infty}^{\infty}\frac{A({\bm r}_1,{\bm r}_1';\varepsilon)}{i\varepsilon_{\ell}-\varepsilon}d\varepsilon,
\label{Lehmann}
\end{equation}
where $A({\bm r}_1,{\bm r}_1';\varepsilon)$ is the spectral function.
Substituting Eq.\ (\ref{Lehmann}) into it, we may regard Eq.\ (\ref{Omega-4b}) as a functional of  $\underline{A}(\varepsilon)$ with the property:
\begin{equation}
\frac{\delta\Omega}{\delta A({\bm r}_1',{\bm r}_1;\varepsilon)}=0,
\label{dOmega/dA}
\end{equation}
which originates from Eq.\ (\ref{dOmega/dG}).
Using the procedure to obtain Eq.\ (\ref{Y_0}) from Eq.\ (\ref{Y_0-0}),
we further transform Eq.\ (\ref{Omega-4b}) into
\begin{eqnarray}
&&\hspace{-8mm}
\Omega =-{\rm P}\int_{-\infty}^{\infty}\frac{d \varepsilon}{\pi}f(\varepsilon)
{\rm Tr}\biggl\{ {\rm Im} \ln \bigl[\underline{K}+\underline{\Sigma}(\varepsilon_{-})
-\varepsilon_{-}\underline{1}\bigr]
\nonumber \\
&&\hspace{0mm}
+{\rm Im}\underline{\Sigma}(\varepsilon_{-}){\rm Re}\underline{G}(\varepsilon_{-})
+{\rm Re}\underline{\Sigma}(\varepsilon_{-}) {\rm Im}\underline{G}(\varepsilon_{-})
\biggr\}
+\Phi ,
\label{Omega-4c}
\end{eqnarray}
where P denotes Cauchy principal value
to remove the pole of the Bose distribution function in Eq.\ (\ref{f(z)}) at $z=0$, 
$\varepsilon_{\pm}\equiv \varepsilon\pm i0_+$ with $0_+$ as an infinitesimal positive constant,
and ${\rm Re}\underline{G}(\varepsilon_{-})$ and ${\rm Im}\underline{G}(\varepsilon_{-})$ are defined generally as
\begin{equation}
{\rm Re}\underline{G}(\varepsilon_{-})\equiv \frac{\underline{G}(\varepsilon_{-})+\underline{G}(\varepsilon_{+})}{2},
\hspace{5mm}
{\rm Im}\underline{G}(\varepsilon_{-})\equiv \frac{\underline{G}(\varepsilon_{-})-\underline{G}(\varepsilon_{+})}{2i},
\label{ReIm}
\end{equation}
respectively. Note ${\rm Im}\underline{G}(\varepsilon_{-})=\frac{1}{2}\underline{A}(\varepsilon)$ from Eq.\ (\ref{Lehmann}). 

We next express $\Phi$ of Eq.\ (\ref{Phi}) with $\underline{A}$ and $f$.
Let us substitute Eq.\ (\ref{G-exp}) with Eq.\ (\ref{Lehmann}) into the first-order term of Eq.\ (\ref{Phi})
and carry out integrations over $\tau$'s and summations over $\ell$'s with Eq.\ (\ref{f(z)}).
We thereby obtain an expression of the Hartree-Fock contribution to $\Phi$ as
\begin{eqnarray}
&&\hspace{-10mm}
\Phi^{({\rm HF})}=\frac{1}{2}\int d {\bm r}_{1}\int d {\bm r}_{1}'
V({\bm r}_{1}-{\bm r}_{1}')
\int_{-\infty}^{\infty}\frac{d \varepsilon}{2\pi}
\int_{-\infty}^{\infty}\frac{d \varepsilon'}{2\pi}
f(\varepsilon)f(\varepsilon')
\nonumber \\
&&\hspace{5mm}
\times
[A({\bm r}_{1},{\bm r}_{1};\varepsilon)A({\bm r}_{1}',{\bm r}_{1}';\varepsilon')\pm
A({\bm r}_{1},{\bm r}_{1}';\varepsilon)A({\bm r}_{1}',{\bm r}_{1};\varepsilon')].
\end{eqnarray}
Similarly, the second-order contribution in Eq.\ (\ref{Phi}) is transformed into
\begin{eqnarray}
&&\hspace{0mm}
\Phi^{(2)}=-\frac{1}{4}\int d {\bm r}_{1}\int d {\bm r}_{1}'\int d {\bm r}_{2}\int d {\bm r}_{2}'
\,V({\bm r}_{1}-{\bm r}_{1}')V({\bm r}_{2}-{\bm r}_{2}')
\int_{-\infty}^{\infty}\!\frac{d \varepsilon_{1}}{2\pi}
\int_{-\infty}^{\infty}\!\frac{d \varepsilon_{1}'}{2\pi}
\nonumber \\
&&\hspace{12mm}\times
\int_{-\infty}^{\infty}\!\frac{d \varepsilon_{2}}{2\pi}
\int_{-\infty}^{\infty}\!\frac{d \varepsilon_{2}'}{2\pi}
\bigl[A({\bm r}_{1},{\bm r}_{2};\varepsilon_{1})A({\bm r}_{2},{\bm r}_{1};\varepsilon_{2})
A({\bm r}_{1}',{\bm r}_{2}';\varepsilon_{1}')A({\bm r}_{2}',{\bm r}_{1}';\varepsilon_{2}')
\nonumber \\
&&\hspace{12mm}
\pm
A({\bm r}_{1},{\bm r}_{2};\varepsilon_{1})A({\bm r}_{2},{\bm r}_{1}';\varepsilon_{2})
A({\bm r}_{1}',{\bm r}_{2}';\varepsilon_{1}')A({\bm r}_{2}',{\bm r}_{1};\varepsilon_{2}')\bigr]J(\varepsilon_{1},\varepsilon_{1}',
\varepsilon_{2},\varepsilon_{2}'),
\end{eqnarray}
where $J(\varepsilon_{1},\varepsilon_{1}',
\varepsilon_{2},\varepsilon_{2}')$ 
signifies
\begin{subequations}
\label{Jdef}
\begin{eqnarray}
&&\hspace{0mm}
J(\varepsilon_{1},\varepsilon_{1}',\varepsilon_{2},\varepsilon_{2}')
\equiv \frac{(1\pm f_{1})(1\pm f_{1}')f_{2}f_{2}'}{\beta}
\int_{0}^{\beta}d \tau_{1}\int_{0}^{\tau_{1}}d \tau_{2}\,e^{-(\tau_{1}-\tau_{2})
(\varepsilon_{1}+\varepsilon_{1}'-\varepsilon_{2}-\varepsilon_{2}')}
\nonumber \\
&&\hspace{29.5mm}
+(1\leftrightarrow 2)
\nonumber \\
&&\hspace{25.5mm}
= \frac{(1\pm f_{1})(1\pm f_{1}')f_{2}f_{2}'}
{\varepsilon_{1}+\varepsilon_{1}'-\varepsilon_{2}-\varepsilon_{2}'}\left[1+
\frac{e^{-\beta(\varepsilon_{1}+\varepsilon_{1}'-\varepsilon_{2}-\varepsilon_{2}')}-1}
{\beta(\varepsilon_{1}+\varepsilon_{1}'-\varepsilon_{2}-\varepsilon_{2}')}\right]
+(1\leftrightarrow 2)
\nonumber \\
&&\hspace{25.5mm}
= \frac{(1\pm f_{1})(1\pm f_{1}')f_{2}f_{2}'-f_{1}f_{1}'(1\pm f_{2})(1\pm f_{2}')}
{\varepsilon_{1}+\varepsilon_{1}'-\varepsilon_{2}-\varepsilon_{2}'},
\label{Jdef1}
\end{eqnarray}
with $f_{1}=f(\varepsilon_{1})$, $f_{1}'=f(\varepsilon_{1}')$, etc.
The final expression of Eq.\ (\ref{Jdef1}) has been obtained with 
$$
(1\pm f_{1})(1\pm f_{1}')f_{2}f_{2}'
e^{-\beta(\varepsilon_{1}+\varepsilon_{1}'-\varepsilon_{2}-\varepsilon_{2}')}
=f_{1}f_{1}'(1\pm f_{2})(1\pm f_{2}').
$$
Note that $J(\varepsilon_{1},\varepsilon_{1}',\varepsilon_{2},\varepsilon_{2}')$ is analytic at
$\varepsilon_{1}+\varepsilon_{1}'-\varepsilon_{2}-\varepsilon_{2}'=0$, as is evident from the first expression.
Hence, it follows that $J$ can be expressed alternatively as
\begin{equation}
J(\varepsilon_{1},\varepsilon_{1}',\varepsilon_{2},\varepsilon_{2}')={\rm P}
\frac{(1\pm f_{1})(1\pm f_{1}')f_{2}f_{2}'}
{\varepsilon_{1}+\varepsilon_{1}'-\varepsilon_{2}-\varepsilon_{2}'}-{\rm P}
\frac{f_{1}f_{1}'(1\pm f_{2})(1\pm f_{2}')}
{\varepsilon_{1}+\varepsilon_{1}'-\varepsilon_{2}-\varepsilon_{2}'},
\label{Jdef2}
\end{equation}
\end{subequations}
with P denoting the principal value.

Let us differentiate $\Phi^{({\rm HF})}$ and $\Phi^{(2)}$ above in terms of $f(\varepsilon)$
and compare the resultant expressions with Eq.\ (\ref{Sigma-exp}) in the same representation.
We then observe that the following relation holds order by order:
\begin{equation}
\frac{\delta \Phi}{\delta f(\varepsilon)}
=\frac{1}{2\pi}{\rm Tr} \underline{A}(\varepsilon){\rm Re}\underline{\Sigma}(\varepsilon_{-}),
\label{dPhi/df}
\end{equation}
which apparently originates from Eq.\ (\ref{dPhi/dG}).\cite{Kita99,Kita06a}

With these preliminaries, we now carry out the differentiation of $S=-\partial\Omega/\partial T$.
Equation (\ref{dOmega/dA}) tells us that we only need to consider the $T$ dependence of $f(\varepsilon)$ for this purpose. It then follows from Eq.\ (\ref{dPhi/df}) that the contribution from $\Phi$ exactly cancels the one from the third term in the curly brackets of Eq.\ (\ref{Omega-4c}).
Hence, we obtain
\begin{eqnarray*}
&&
S={\rm P}\int_{-\infty}^{\infty}\frac{d \varepsilon}{\pi}\,
\frac{\partial f}{\partial T}
{\rm Tr}\bigl\{ {\rm Im}
\ln \bigl[\underline{K}\!+\!\underline{\Sigma}(\varepsilon_-)
\!-\!\varepsilon_-\underline{1}\bigr]
+{\rm Im}\underline{\Sigma}(\varepsilon){\rm Re} \underline{G}^{\rm R}(\varepsilon)\bigr\}.
\end{eqnarray*}
Next, we express ${\partial f}/{\partial T}=
-({\partial }/{\partial \varepsilon})k_{\rm B}[-f\ln f\pm (1\pm f)\ln (1\pm f)]$ and subsequently carry out an integration by parts.
We thereby obtain
\begin{equation}
S=k_{\rm B}
\int_{-\infty}^{\infty}\frac{d \varepsilon}{2\pi} 
\sigma[f(\varepsilon)]\,{\rm Tr}
\left[\underline{A}(\varepsilon)\!
\left(\underline{1}-
\frac{\partial {\rm Re}
\underline{\Sigma}(\varepsilon_{-})}{\partial \varepsilon}\right)
+2{\rm Im} \underline{\Sigma}(\varepsilon_{-})\frac{\partial {\rm Re}\underline{G}(\varepsilon_{-})}{\partial \varepsilon}
\right] ,
\label{Seq}
\end{equation}
where $\sigma$ and $f$ are defined by Eqs.\ (\ref{sigma-def}) and (\ref{f(z)}), respectively,
and we have removed P from the integration based on the observation that $\sigma[f(\varepsilon)]$ only has a logarithmic singularity at $\varepsilon=0$ for the relevant Bose distribution function.

We now consider the case $U({\bm r})=0$ in Eq.\ (\ref{H_0}). It then follows that 
every matrix such as $A({\bm r}_{1},{\bm r}_{1}';\varepsilon)$ can be expanded in plane waves as
$$A({\bm r}_{1},{\bm r}_{1}';\varepsilon)=\frac{1}{\cal V}\sum_{{\bm p}}A_{\bm p}(\varepsilon)e^{i{\bm p}\cdot({\bm r}_{1}-{\bm r}_{1}')/\hbar},
$$
with ${\cal V}$ the volume.
Substituting it into Eq.\ (\ref{Seq}) and carrying out an integration over ${\bm r}$, we obtain 
\begin{equation}
\frac{S}{{\cal V}}=\hbar k_{\rm B}
\int\frac{d^{3}p d\varepsilon}{(2\pi\hbar)^{4}}
\sigma[f(\varepsilon)]
\biggl\{A_{\bm p}(\varepsilon)\!
\left[1-
\frac{\partial {\rm Re}\Sigma_{\bm p}(\varepsilon_{-})}{\partial \varepsilon}\right]
+2{\rm Im}\Sigma_{\bm p}(\varepsilon_{-})\frac{\partial {\rm Re}G_{\bm p}(\varepsilon_{-})}{\partial \varepsilon}
\biggr\} .
\label{Seq2}
\end{equation}
Noting ${\rm Re}G_{\bm p}(\varepsilon_{-})={\rm Re}G^{\rm A}({\bm p}\varepsilon)={\rm Re}G^{\rm R}({\bm p}\varepsilon)$ and $2{\rm Im}\Sigma_{\bm p}(\varepsilon_{-})=A_{\Sigma}({\bm p}\varepsilon)$, we realize that this expression is also obtained from Eq.\ (\ref{entropy-density}) with $\phi\rightarrow f$.
Thus, we have confirmed that Eq.\ (\ref{entropy-density}) is compatible with equilibrium entropy.


\begin{thebibliography}{99}
 
 
\bibitem{Keldysh64}L. V. Keldysh, Zh. Eksp. Teor. Fiz. {\bf 47} (1964), 1515 [Sov. Phys. JETP {\bf 20} (1965), 1018].

\bibitem{Bonitz00}M. Bonitz (Ed.), {\em Progress in Nonequilibrium Green's Functions} (World Scientific, Singapore, 2000).

\bibitem{BS03}M. Bonitz and D. Semkat (Eds.), {\em Progress in Nonequilibrium Green's Functions II} (World Scientific, Singapore, 2003).

\bibitem{BF06}M. Bonitz and A. Filinov (Eds.), {\em Progress in Nonequilibrium Green's Functions III},
Journal of Physics: Conference Series {\bf 35} (2006).

\bibitem{Langreth76}D. C. Langreth, in {\em Linear and Nonlinear Electron Transport in Solids}, eds. J. T. Devreese and V. E. van Doren (Plenum Press, New York, 1976), p. 3.

\bibitem{Danielewicz84}P. Danielewicz, Ann. of Phys. {\bf 152} (1984), 239. 

\bibitem{CSHY85}K. C. Chou, Z. B. Su, B. L. Hao and L. Yu, Phys. Rep. {\bf 118} (1985), 1.

\bibitem{RS86}J. Rammer and H. Smith, Rev. Mod. Phys. {\bf 58} (1986), 323.

\bibitem{Berges04}J. Berges, hep-ph0409233.

\bibitem{HJ98} H. Haug and A. -P. Jauho, {\em Quantum Kinetics in Transport and Optics of Semiconductors} (Springer-Verlag, Berlin, 1998).

\bibitem{Rammer07}J. Rammer, {\em Quantum Field Theory of Nonequilibrium States} (Cambridge University Press, Cambridge, 2007).

\bibitem{LP}E. M. Lifshitz and L. P. Pitaevskii, {\em Physical Kinetics} (Pergamon, New York, 1981).

\bibitem{Mahan}G. D. Mahan, {\em Many-Particle Physics} (Kluwer Academic/Plenum, New York, 2000).

\bibitem{LW60}J. M. Luttinger and J. C. Ward, Phys. Rev. {\bf 118} (1960), 1417.

\bibitem{Luttinger60}J. M. Luttinger, Phys. Rev. {\bf 119} (1960), 1153.

\bibitem{Baym62}G. Baym, Phys. Rev. {\bf 127} (1962), 1391.

\bibitem{Cercignani88}C. Cercignani, {\em The Boltzmann Equation and Its Applications} (Springer-Verlag, New York, 1988).

\bibitem{CH93}M. C. Cross and P. C. Hohenberg, Rev. Mod. Phys. {\bf 65} (1993), 851.

\bibitem{CG09}M. C. Cross and H. Greenside, {\em 	
Pattern Formation and Dynamics in Nonequilibrium Systems} (Cambridge University Press, Cambridge, 2009).

\bibitem{Jaynes57}E. T. Jaynes, Phys. Rev. {\bf 106} (1957), 620.

\bibitem{LL80}L. D. Landau and E. M. Lifshitz, {\em Statistical Physics} (Pergamon, Oxford, 1980).

\bibitem{Boltzmann72} L. Boltzmann, Wiener Berichte {\bf 63} (1872), 275.

\bibitem{Resibois78}P. Resibois, J. Stat. Phys. {\bf 19} (1978), 593.

\bibitem{Lebowitz04}S. Goldstein and J. L. Lebowitz, Physica D {\bf 193} (2004), 53; 
P. L. Garrido, S. Goldstein and J. L. Lebowitz,
Phys. Rev. Lett. {\bf 92} (2004), 050602.

\bibitem{Ivanov00}Yu. B. Ivanov, J. Knoll and D. N. Voskresensky,
Nucl. Phys. A {\bf 672} (2000), 313.

\bibitem{Matsubara55}T. Matsubara, Prog. Theor. Phys. {\bf 14} (1955), 351.

\bibitem{Wick50} G. C. Wick, Phys. Rev. {\bf 80} (1950),  268.

\bibitem{BdD59}C. Bloch and C. de Donimicis, Nucl. Phys. {\bf 10} (1959), 509.

\bibitem{Gaudin60}M. Gaudin, Nucl. Phys. {\bf 15} (1960), 89.

\bibitem{Landau56}L. D. Landau, Zh. Eksp. Teor. Fiz. {\bf 30} (1956), 1058 
[Sov. Phys. JETP {\bf 3} (1957), 920]; Zh. Eksp. Teor. Fiz. {\bf 32} (1957), 59 
[Sov. Phys. JETP {\bf 5} (1957), 101]. 

\bibitem{AGD63}
A. A. Abrikosov, L. P. Gor'kov and I. E. Dzyaloshinski, {\em Methods of Quantum Field Theory in Statistical Physics} (Prentice-Hall, Englewood Cliffs, N.J., 1963).

\bibitem{Bogoliubov62}N. N. Bogoliubov, in {\em Studies in Statistical Mechanics}, eds. J. de Boer and G. E. Uhlenbeck (North-Holland, Amsterdam, 1962) vol. 1, p. 1.

\bibitem{KB62}L. P. Kadanoff and G. Baym, {\em Quantum Statistical Mechanics} (Benjamin, New York, 1962).

\bibitem{Schwinger61}J. Schwinger, J. Math. Phys. {\bf 2} (1961), 407.

\bibitem{comment1}L. V. Keldysh, article on pp. 4--17 of Ref.\ \citen{BS03}.

\bibitem{CCNS71}C. Caroli, R. Combescot, P. Nozi\`eres and D. Saint-James, J. of Phys. C: Solid St. Phys. {\bf 4} (1971), 916. 


\bibitem{AG75}A. G. Aronov and V. L. Gurevich, Fiz. Tverd. Tela {\bf 16} (1974), 2656 [Sov. Phys. Solid State {\bf 16} (1975), 1722].

\bibitem{LO75}A. I. Larkin and Yu. B. Ovchinnikov, Zh. Eksp. Teor. Fiz. {\bf 68} (1975), 1915 [Sov. Phys. JETP {\bf 41} (1975), 960].

\bibitem{FW71}A. L. Fetter and J. D. Walecka, {\em Quantum Theory of Many-Particle Systems} (McGraw-Hill, New York, 1971).

\bibitem{OB67}J. Orban and A. Bellemans, Phys. Lett. {\bf 24}A (1967), 620.

\bibitem{Lebowitz93}See, e.g., J. Lebowitz, Phys. Today {\bf 46}  (1993), 32.

\bibitem{dDM64}C. de Dominicis and P. C. Martin, J. Math. Phys. {\bf 5}  (1964), 14.

\bibitem{Kita96} T. Kita, J. Phys. Soc. Jpn. {\bf 65} (1996), 1355; J. Phys. Soc. Jpn. {\bf 65} (1996), 1373.

\bibitem{Kita09} T. Kita, Phys. Rev. B {\bf 80} (2009), 214502.

\bibitem{CJT74}J. M. Cornwall, R. Jackiw and E. Tomboulis, Phys. Rev. D 
{\bf 10} (1974), 2428.

\bibitem{Kita06a}T. Kita, J. Phys. Soc. Jpn.  {\bf 75} (2006), 114005.

\bibitem{BC01}J. Berges and J. Cox, Phys. Lett. B {\bf 517} (2001), 369.

\bibitem{AB01}G. Aarts and J. Berges, Phys. Rev. D {\bf 64} (2001), 105010.

\bibitem{AABBS02}G. Aarts, D. Ahrensmeier, R. Baier, J. Berges and J. Serreau, Phys. Rev. D {\bf 66} (2002),
045008.

\bibitem{Berges02}J. Berges, Nucl. Phys. A {\bf 699} (2002), 847.

\bibitem{Berges03}J. Berges, S. Bors\'anyi and J. Serreau, Nucl. Phys. B {\bf 660} (2003), 51.


\bibitem{CDM03}F. Cooper, J. F. Dawson and B. Mihaila, Phys. Rev. D {\bf 67} (2003),
056003.

\bibitem{JCG04}S. Juchem, W. Cassing and C. Greiner, Phys. Rev. D {\bf 69} (2004),
025006.

\bibitem{AST05} A. Arrizabalaga, J. Smit and A. Tranberg, Phys. Rev. D {\bf 72} (2005), 025014.

\bibitem{Berges06}J. Berges and S. Bors\'anyi, Phys. Rev. D 74 (2006), 045022.

\bibitem{Wigner32}E. P. Wigner, Phys. Rev. {\bf 40} (1932), 749.

\bibitem{HOSW84}M. Hillery, R. F. O'Connell, M. O. Scully
and E. P. Wigner, Phys. Rep. {\bf 106} (1984), 121.

\bibitem{Weyl31}H. Weyl, {\em The Theory of Groups and Quantum Mechanics}
(Dover, New York, 1931).

\bibitem{Moyal49}J. E. Moyal, Proc. Cambridge Philos. Soc. {\bf 45} (1949), 99.

\bibitem{Groenewold46}H. J. Groenewold, Physica {\bf 12} (1946), 405.

\bibitem{BM90}W. Botermans and R. Malfliet, Phys. Rep. {\bf 198} (1990), 115.

\bibitem{Kita99}T. Kita, J. Phys. Soc. Jpn. {\bf 68} (1999), 3740.

\bibitem{CP75}G. M. Carneiro and C. J. Pethick, Phys. Rev. B {\bf 11} (1975), 1106.

\bibitem{PK64}R. E. Prange and L. P. Kadanoff, Phys. Rev. {\bf 134} (1964), A566.

\bibitem{Eilenberger68} G. Eilenberger, Z. Phys. {\bf 214}  (1968), 195.

\bibitem{Rainer83}J. W. Serene and D. Rainer, Phys. Rep. {\bf 101} (1983), 221.

\bibitem{LO86}A. I. Larkin and Y. N. Ovchinnikov,
in {\em Nonequilibrium Superconductivity}, Vol. 12,
eds. D. N. Langenberg and A. I. Larkin (Elsevier, Amsterdam, 1986), p. 493.

\bibitem{Kita01}T. Kita, Phys. Rev. B {\bf 64} (2001), 054503.

\bibitem{KY08}T. Kita and H. Yamashita, J. Phys. Soc. Jpn. {\bf 77} (2008), 024711.

\bibitem{Parks69}See, e.g. {\em Superconductivity}, ed. by R. D. Parks
(Dekker, New York, 1969).

\bibitem{Stratonovich56} R. L. Stratonovich, Dok. Akad. Nauk SSSR {\bf 1} (1956), 72 [Sov. Phys. Dolkady {\bf 1} (1956), 414].

\bibitem{Fujita66}S. Fujita, {\em Introduction to Non-Equilibrium Quantum
Statistical Mechanics} (W. B. Saunders, Philadelphia, 1966).

\bibitem{LF01}M. Levanda and V. Fleurov, Ann. of Phys. {\bf 292} (2001), 199.

\bibitem{Kubo64}R. Kubo, J. Phys. Soc. Jpn. {\bf 19} (1964), 2127.

\bibitem{Onoda06}S. Onoda, N. Sugimoto and N. Nagaosa, Prog. Theor. Phys. {\bf 116} (2006), 61.

\bibitem{SJV86}O. T. Serimaa, J. Javanainen and S. Varr\'o, Phys. Rev. A {\bf 33} (1986), 2913.
See its Appendix.


\bibitem{MS59}P. C. Martin and J. Schwinger, Phys. Rev. {\bf 115} (1959), 1342.

\bibitem{BK61}G. Baym and L. Kadanoff, Phys. Rev. {\bf 124} (1961), 287.

\bibitem{SB51}E. E. Salpeter and H. A. Bethe, Phys. Rev. {\bf 84} (1951), 1232.

\bibitem{LL-FM}L. D. Landau and E. M. Lifshitz, {\em Fluid Mechanics} (Pergamon, Oxford, 1987).

\bibitem{CC90}S. Chapman and T. G. Cowling,
{\em The Mathematical Theory of Non-uniform Gases} (Cambridge University Press,
Cambridge, 1990).

\bibitem{HCB54}J. O. Hirschfelder, C. F. Curtiss and R. B. Bird,
{\em Molecular Theory of Gases and Liquids}
(Wiley, New York, 1954).

\bibitem{Kita06b} T. Kita, J. Phys. Soc. Jpn. {\bf 75} (2006), 124005.

\bibitem{LL-M}L. D. Landau and E. M. Lifshitz, {\em Mechanics} (Pergamon, Oxford, 1969)  \S 18.

\bibitem{PS08}C. J. Pethick and H. Smith, {\em Bose-Einstein Condensation in Dilute Gases}
(Cambridge University Press, Cambridge, 2008).

\bibitem{PS03}L. Pitaevskii and S. Stringari, {\em Bose-Einstein Condensation}
(Oxford University Press, Oxford, 2003).

\bibitem{GNZ09}A. Griffin, T. Nikuni and E. Zaremba, {\em Bose-Condensed Gases at Finite Temperatures} (Cambridge University Press, Cambridge, 2009).

\bibitem{AEMWC95}M. R. Anderson, J. R. Ensher, M. R. Matthews, C. E. Wieman and E. A. Cornell,
Science {\bf 269} (1995), 198.


\bibitem{HM65}P. C. Hohenberg and P. C. Martin, Ann. of Phys.
{\bf 34} (1965), 291.

\bibitem{Griffin96}A. Griffin, Phys. Rev. B {\bf 53} (1996), 9341.

\bibitem{Kita05}T. Kita, J. Phys. Soc. Jpn. {\bf 74} (2005), 1891 [Errata: {\bf 74} (2005), 3397].

\bibitem{Kita06}T. Kita, J. Phys. Soc. Jpn. {\bf 75} (2006), 044603.

\bibitem{PS95}M. E. Peskin and D. V. Schroeder, 
{\em An Introduction to Quantum Field Theory}
(Westview Press, Boulder, 1995).

\bibitem{Nambu61}Y. Nambu and G. Jona-Lasinio, Phys. Rev. {\bf 122} (1961), 345.

\bibitem{Goldstone61}J. Goldstone, Nuovo Cim. {\bf 19} (1961), 154.
\\
J. Goldstone, A. Salam and S. Weinberg, Phys. Rev. {\bf 127} (1962), 965.

\bibitem{Bogoliubov47}N. N. Bogoliubov, J. Phys. (USSR) {\bf 9} (1947), 23.

\bibitem{GP71}P. Glansdorff and I. Prigogine, {\em
Thermodynamic Theory of Structure, Stability and Fluctuations}
(Wiley, New York, 1971).

\bibitem{Zubarev74}D. N. Zubarev, {\em Nonequilibrium Statistical Thermodynamics} (Consultants Bureau, New York, 1974).

\bibitem{Haken75}H. Haken, Rev. Mod. Phys. {\bf 47} (1975), 67.

\bibitem{Kuramoto84}Y. Kuramoto: {\em
Chemical Oscillations, Waves, and Turbulence} (Springer-Verlag, Berlin, 1984).

\bibitem{Tsallis88}C. Tsallis, J. Stat. Phys. {\bf 52} (1988), 479.


\bibitem{Jou93}D. Jou, J. Casas-V\'azquez and G. Lebon, {\em Extended Irreversible Thermodynamics} 
(Springer, Berlin, 1993).

\bibitem{Jarzynski97}C. Jarzynski, Phys. Rev. Lett. {\bf 78} (1997), 2690.

\bibitem{ES02}D. J. Evans and D. J. Searles, Adv. Phys. {\bf 51} (2002), 1529.

\bibitem{ST06}S. Sasa and H. Tasaki, J. Stat. Phys. {\bf 125} (2006), 125.

\bibitem{Grandy08}W. T. Grandy Jr., {\em Entropy and the Time Evolution of Macroscopic Systems} (Oxford University Press, New York, 2008).

\bibitem{HK78}I. N. Herstein and I. Kaplansky, {\em Matters Mathematical} (Chelsea Pub., New York, 1978)

\bibitem{Sakurai}J. J. Sakurai, {\em Modern Quantum Mechanics} (Reading, MA, 1994).

\bibitem{Aitken}A. C. Aitken, {\em Determinants and Matrices} (Oliver and Boyd, Edinburgh, 1956).

\bibitem{Yang62}C. N. Yang, Rev. Mod. Phys. {\bf 34} (1962), 694.




\end{thebibliography}
\end{document}